\documentclass[10pt,journal,compsoc]{IEEEtran}
\usepackage{enumitem}
\usepackage{amsfonts}
\usepackage{algorithmic}
\usepackage{array}
\usepackage[caption=false,font=normalsize,labelfont=sf,textfont=sf]{subfig}
\usepackage{textcomp}
\usepackage{stfloats}
\usepackage{cite}
\usepackage{verbatim}
\usepackage{graphicx}
\usepackage{balance}
\usepackage[scr=rsfs]{mathalfa}
\usepackage{amsmath, amssymb, bm}
 \usepackage{hyperref}
\usepackage{cleveref}
\usepackage{algorithm}
 \usepackage{setspace}
 \usepackage{xcolor}
\usepackage{tikz}
\usepackage{bbm}
\usepackage{lipsum}
\usepackage[acronym]{glossaries}
\glsdisablehyper

\newacronym{name}{SigNova}{{}}

\newtheorem{theorem}{Theorem}[section]
\newtheorem{definition}[theorem]{Definition}
\newtheorem{example}{Example}
\newtheorem{remark}[theorem]{Remark}
\usepackage{cleveref}
\usepackage{mathtools}

\Crefformat{figure}{#2Fig.~#1#3}
\Crefmultiformat{figure}{Figs.~#2#1#3}{ and~#2#1#3}{, #2#1#3}{ and~#2#1#3}
\Crefformat{section}{#2Sec.~#1#3}
\Crefformat{algorithm}{#2Alg.~#1#3}
\Crefformat{theorem}{#2Thm.~#1#3}
\Crefformat{example}{#2Example~#1#3}
\Crefformat{definition}{#2Def.~#1#3}
\let\Algorithm\algorithm
\renewcommand\algorithm[1][]{\Algorithm[#1]\setstretch{1.5}}

\newglossaryentry{corpus}{name={corpus},
description={},
symbol={\ensuremath{\mathcal{D}_n}}
}
\newglossaryentry{ES}{name={Expected signature},
description={},
symbol={\ensuremath{\varphi}}
}
\newglossaryentry{cov}{name={same covariance matrix},
description={},
symbol={\ensuremath{\Sigma}}
}
\newglossaryentry{covn}{name={same covariance matrix},
description={},
symbol={\ensuremath{\Sigma_n}}
}

\usepackage[]{footmisc} 

\renewcommand{\footnoterule}{%
   
    \kern -3pt
    \hspace{1.5cm}\rule{0.25\textwidth}{.05pt}
    \kern -3pt
}

\begin{document}

\title{Novelty Detection on Radio Astronomy Data using Signatures}

\author{%
  \IEEEauthorblockN{%
Paola Arrubarrena\IEEEauthorrefmark{1}, 
    Maud Lemercier\IEEEauthorrefmark{1}, 
    Bojan Nikolic, 
    Terry Lyons, 
    and Thomas Cass
  }%
}

\IEEEtitleabstractindextext{%
\begin{abstract}
  We introduce \acrshort{name}, a new semi-supervised framework for detecting anomalies in streamed data. While our initial examples focus on detecting radio-frequency interference (RFI) in digitized signals within the field of radio astronomy, it is important to note that \acrshort{name}'s applicability extends to any type of streamed data. The framework comprises three primary components. Firstly, we use the signature
    transform to extract a canonical collection of summary statistics
    from observational sequences. This allows us to represent
    variable-length visibility samples as finite-dimensional feature
    vectors. Secondly, each feature vector is assigned a novelty
    score, calculated as the Mahalanobis distance to its nearest
    neighbor in an RFI-free training set. By thresholding these scores
    we identify observation ranges that deviate from the expected
    behavior of RFI-free visibility samples without relying on
    stringent distributional assumptions. Thirdly, we integrate this
    anomaly detector with Pysegments, a segmentation algorithm, to
    localize consecutive observations contaminated with RFI, if
    any. This approach provides a compelling alternative to classical
    windowing techniques commonly used for RFI detection. Importantly,
    the complexity of our algorithm depends on the RFI pattern rather
    than on the size of the observation window. We demonstrate how
    \acrshort{name} improves the detection of various types of RFI
    (e.g., broadband and narrowband) in time-frequency visibility
    data. We validate our framework on the
    Murchison Widefield Array (MWA) telescope and simulated data and the Hydrogen Epoch of Reionization Array (HERA).
\end{abstract}

\begin{IEEEkeywords}
Novelty detection, anomaly detection, semi-supervised learning, radio frequency interference.
\end{IEEEkeywords}}

\maketitle

\IEEEdisplaynontitleabstractindextext
\IEEEpeerreviewmaketitle

\begingroup\renewcommand\thefootnote{}

\footnotetext{\textit{
 \begin{itemize}[leftmargin=0.3cm]
     \item Paola Arrubarrena$^{*}$ is with the Department of Mathematics at Imperial College London, London SW7 2AZ, UK, and also with The Alan Turing Institute, London NW1 2DB, UK. E-mail: p.arrubarrena@imperial.ac.uk.
     \item Maud Lemercier$^{*}$ is with the Mathematical Institute at University of Oxford, Oxford OX2 6GG, UK, and also with The Alan Turing Institute, London NW1 2DB, UK. E-mail: maud.lemercier@maths.ox.ac.uk.
     \item Bojan Nikolic is with The Cavendish Laboratory of the Department of Physics at University of Cambridge, Cambridge CB3 0HE, UK. E-mail: bn204@cam.ac.uk.
     \item Terry Lyons is with the Mathematical Institute at University of Oxford, Oxford OX2 6GG, UK, and also with The Alan Turing Institute, London NW1 2DB, UK. E-mail: terry.lyons@maths.ox.ac.uk.
     \item Thomas Cass is with the Department of Mathematics at Imperial College London, London SW7 2AZ, UK, also with The Alan Turing Institute, London NW1 2DB, UK, and with the Institute for Advance Study, New Jersey 08540, USA. E-mail: thomas.cass@imperial.ac.uk.
 \end{itemize}
This work was supported in part by EPSRC (NSFC) under Grant EP/S026347/1, in part by The Alan Turing Institute under the EPSRC grant EP/N510129/1, the Data Centric Engineering Programme (under the Lloyd’s Register Foundation grant G0095), the Defence and Security Programme (funded by the UK Government) and the Office for National Statistics \& The Alan Turing Institute (strategic partnership) and in part by the Hong Kong Innovation and Technology Commission (InnoHK Project CIMDA), in part by the Programme Grant, and in part by the Erin Ellentuck Fellowship at the Institute for Advanced Study.}}

\endgroup

\begingroup\renewcommand\thefootnote{}
\footnotetext{\textit{$^{*}$Equal contribution and corresponding author.}}
\endgroup

\IEEEraisesectionheading{\section{Introduction}\label{sec:introduction}}

\IEEEPARstart{R}{adio} astronomy provides a unique perspective on the universe by observing the celestial radiation at radio frequencies (50\,MHz up to 950\,GHz). Lower parts of the radio spectrum (50\,MHz to about 5\,GHz) are of particular current scientific interest and with the Square Kilometer Array (SKA)~\cite{SKA} there is substantial international investment in making it possible to make very sensitive, large-field observations at these frequencies. Amongst the goals are investigation of the very earliest galaxies through the impact they have on the primordial intergalactic neutral hydrogen by observing the red-shifted hyperfine transition of hydrogen ($\lambda_{\text{rest}}=21$ cm).

\hspace{0pt}\\One of the major challenges for observations at these lower
frequencies is radio-frequency interference (RFI), which refers to any
unwanted electromagnetic signal that contaminates the radio
observations. RFI can seriously degrade the quality of radio
observations and can even render them unusable. In this paper we
present a technique to identify sections of data from interferometric
telescopes that is contaminated by RFI. Since radio observations
combine information from a range of frequencies, times and antennas,
if only sections of data are contaminated by RFI they can be excluded
from further combination allowing high fidelity and sensitivity final
measurements or images.

\hspace{0pt}\\Radio interferometers work by measuring the correlated signal in the
electric field received by pairs of separated antennas,
$\gamma_{i,j}= \langle E_i(t) E^{*}_j(t) \rangle$, which because of
similarity to optical interferometry is called the
\emph{visibility}. If there are $N_A$ antennas the telescope's digital
correlator will calculate correlations between each pair of antennas
(including each antenna with itself), giving $N_B:=N_A(N_A+1)/2$
measurements of visibilities.  Each visibility between antenna pair
$(i,j)$ is a complex-valued signal indexed on a time-frequency domain
$I\times \Omega$,
$$\gamma_{i,j}:I\times \Omega\to \mathbb{C}, ~(t,\nu)\mapsto \gamma_{i,j}(t,\nu).$$
The digital correlator performs this correlation at regular time
intervals. At each time the interferometer outputs a sample of the
visibility in the frequency domain. Much research has gone into the
design of automated RFI detection techniques to flag corrupted
visibility samples $\gamma_{i,j}(t_m,\nu_n)$.

\subsection{Related Work}
\textsc{AOFlagger} \cite{offringa2012morphological} is a widely adopted set of flagging strategies for detecting RFI in visibility data. It includes a range of pipelines customized for different radio telescopes. These pipelines flag the RFI to detect line-structured RFI in the time and frequency directions. One of the key components of these flagging strategies is the \textsc{SumThreshold} algorithm~\cite{Offringa_2010} which operates on the amplitudes of time-frequency visibilities in a given polarisation and baseline. This algorithm flags a consecutive sub-sequence of visibilities with $M$ samples, if the average of the sub-sequence exceeds a threshold. This procedure is applied iteratively to the data by sliding a window of increasing size, starting with $M=1$ which corresponds to analysing single samples. The threshold is chosen to be a decreasing function of $M$. While the computational complexity of the iterative algorithm is quadratic in the length of the base sequence to analyse, it is common practice to consider windows with exponentially increasing size to obtain a loglinear complexity.

\hspace{0pt}\\The Sky- Subtracted Incoherent Noise Spectra (\textsc{ssins}) \cite{wilensky2019absolving} is another flagger that detects RFI in time-frequency visibility data. A sky-subtraction technique is applied and it is shown that the resulting visibility data can be modeled as a circular complex Gaussian process. The flagging strategy of \textsc{ssins} consists of averaging the visibility amplitudes over all baselines, and leveraging the Central Limit Theorem to derive an asymptotic distribution for the baseline-averaged amplitudes (standardized in each frequency channel). Each visibility sample is then assigned a z-score and flagged if the score exceeds a certain threshold. The average-per-baseline~\cite{wilensky2019absolving} design may be dependent on a large number of antennas to achieve accurate performance and loses single-antenna information.

\hspace{0pt}\\\textbf{Statistical bias} Both \textsc{AOFlagger} and \textsc{ssins} are unsupervised flagging procedures that rely on estimating expectations. While in \textsc{AOFlagger}, the average of visibility samples in a window is computed and thresholded, in \textsc{ssins}, the average over all observations in time is computed to standardize the data and obtain a z-score for each sample. In both cases, RFI contamination will bias the estimation of the mean causing false positives and false negatives. Therefore, several iterations are conducted by removing previously flagged samples before estimating the mean, which increases the computational complexity of the proposed algorithms. Furthermore, \textsc{ssins} faces robustness issues in the presence of narrowband RFI that persists through the entire observation since the mean will be consistently corrupted. If the RFI contamination is constant over time, the flagger will miss it as the RFI is removed by the sky-subtraction step. All these challenges arise from the fact that both \textsc{AOFlagger} and \textsc{ssins} are \textit{unsupervised} anomaly detection techniques that do not leverage available RFI-free data.

\hspace{0pt}\\\textbf{Incomplete flagging and faint RFI} The strategies implemented in \textsc{AOFlagger} typically combine \textsc{SumThreshold} with additional steps such as the application of the SIR operator to extend the flags in time and frequency and fill the gaps in the flag mask. As discussed in \cite{offringa2023interference}, the SIR operator is prone to certain pitfalls, especially in the presence of invalid data (e.g. when a correlator fails), and several modifications might be considered to avoid over- and under-flagging. In the \textsc{ssins} framework, an integrated baseline approach is adopted to boost the sensitivity to faint RFI. However, despite this improvement, the authors highlight that, as they use a single-sample analysis, faint RFI might still be missed. To mitigate this issue, they propose to take as a test statistic the average of the visibilities at a given time over a sub-band of frequency channels. 

\subsection{Contributions} 
This paper presents \acrshort{name}\footnote{SigNova's Github \href{https://github.com/datasig-ac-uk/SigNova}{https://github.com/datasig-ac-uk/SigNova}.}, a novelty detection framework for streamed data, offering significant advantages in terms of sensitivity and accuracy for detecting RFI. Our framework can efficiently localize RFI anomalies and enables the identification of very faint RFI that might have been missed by other detection methods. These advantages stem from the combination of key principles from semi-supervised anomaly detection and rough path theory. The main components are outlined below.

\hspace{0pt}\\\textbf{Signature transform} RFI often contaminates consecutive time or frequency samples (e.g. narrowband or broadband RFI). Using relevant time series analysis tools becomes advantageous in such scenarios. In recent years, the signature transform, grounded in rough path theory, has proved to be a powerful tool for extracting valuable insights from streamed data in diverse real-world applications. In this work, we propose to use the signature to represent complex-valued visibility data over any interval into a finite-dimensional feature vector. The signature features, which correspond to a collection of moments of the underlying visibility signal, retain phase information, contrary to amplitude-based techniques. The ability to encode visibility streams into fixed-dimensional feature vectors, allows us to apply advanced anomaly detection techniques for multivariate data. 

\hspace{0pt}\\\textbf{Data-driven anomaly score} The vast amount of visibility data generated by modern radio telescopes, combined with a significant labelling effort, provides an opportunity to use semi-supervised anomaly detection techniques. We propose to leverage the availability of archival data labelled as clean to determine the presence of RFI in new visibility data. We assign to each new data instance an anomaly score, given by the Mahalanobis distance to its nearest neighbor in a training dataset of uncontaminated visibility sequences. By thresholding these scores, we identify observations that deviate from the expected behavior of clean data. Importantly, our approach does not rely on a particular statistical model of normal behavior, unlike \textsc{SSINS} or spectral kurtosis techniques \cite{nita2010generalized}.

\hspace{0pt}\\\textbf{Segmentation} Using the signature as a feature map allows us to run an anomaly detector on arbitrary time intervals. In order to determine precisely the start and the end position of RFI contaminations (if any) in time-frequency data, we use an efficient segmentation algorithm called Pysegments. This algorithm provides a compelling alternative to the traditional windowing techniques commonly used to identify RFI patterns, as the complexity of our algorithm depends on the RFI pattern rather than on the size of the observation window.

\begin{figure*}[!t]
    \centering
\includegraphics[width=\textwidth]{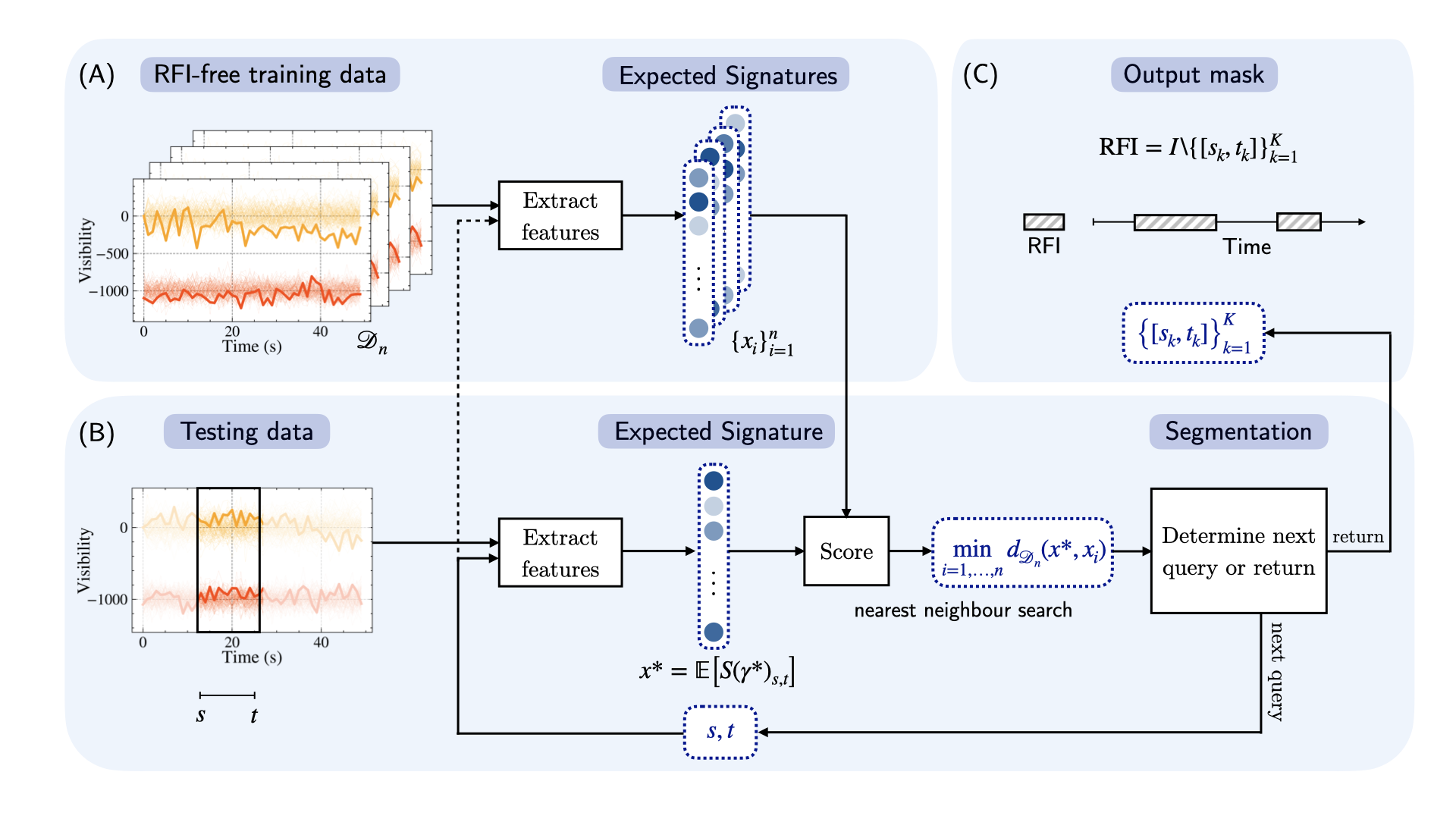}
    \caption{Schematic of \acrshort{name}. Panel (A) represents the training dataset (corpus). It corresponds to visibility data labeled as "clean". Each datum is itself an ensemble of $N_A$ streams whose expected signature can be queried on any time interval. Panel (B) illustrates how the RFI-detection framework operates on new visibility data, that is, a new ensemble of $N_A$ streams (associated with one antenna in a frequency channel). The segmentation algorithm determines dynamically on which interval $[s,t]$ one should analyse the signal, that is, test whether it is RFI-free. At every step, the analysis consists of computing an anomaly score on $[s,t]$. This score is obtained by computing the minimum of the (Mahalanobis) distances between the new data and every datum in the corpus (Panel (A)). Panel (C) shows the output of the framework: a collection of disjoint intervals which have been marked as ``clean", that is RFI-free. Based on this output, one can determine the RFI localisation.}
    \label{fig:enter-label}
\end{figure*}

\hspace{0pt}\\\textbf{Group anomaly detection} \acrshort{name} can be used to flag RFI in individual baselines and can be easily adapted to flag simultaneously all baselines associated with one antenna. To conduct an \emph{antenna-oriented} analysis for antenna $i$, we consider the set of visibilities $\{\gamma_{i,j}:j=1,\ldots,N_A\}$, and tackle a \textit{group anomaly detection} problem \cite{muandet2013one,xiong2011hierarchical}, by computing the anomaly score with expected signatures as variables. This approach somewhat sits in between the baseline analysis of \textsc{AOFlagger} and the integrated baseline analysis of \textsc{SSINS}. The rationale is that when RFI or a defect occurs at one antenna, it might be reflected in several baselines containing the antenna. A scheme of \acrshort{name} is shown in~\Cref{fig:enter-label}.

\hspace{0pt}\\We present our analysis of data from a cutting-edge radio telescope, the Murchison Widefield Array (MWA)~\cite{MWA}, that is being used to study key phenomena in the early universe, including the observing the red-shifted hyperfine transition
of hydrogen ($\lambda_{\rm rest} = 21\,$cm). 
The MWA, which features 127 antennas, observes in the frequency range of 72-231 MHz. It is worth noting that this methodology is applicable to any telescope, as it is not experimentally prone. It can also be effectively implemented with telescopes such as the HERA telescope~\cite{HERA} and others.

\hspace{0pt}\\The rest of this paper is organized as follows.In \Cref{sec:background}, we present the key mathematical and algorithmic components of our framework. Specifically, in \Cref{ssec:signature}, we introduce the signature and explain its role in the feature extraction process. \Cref{ssec:conformance} outlines how we define a novelty score to identify anomalies in streamed data. \Cref{ssec:pysegments} presents Pysegments, a segmentation algorithm that we leverage for localizing and characterizing RFI. In \Cref{sec:method}, we apply our method to streamed data, and we present our results in \Cref{sec:results} using simulated data, real data, and both. We present a discussion in~\Cref{sec:discussion}, where to find the data in~\Cref{sec:data_availability}, and finally we conclude the paper in \Cref{sec:conclusion}.

\section{Background}\label{sec:background}
We begin with some background on the mathematical and algorithmic components of our framework. 

\subsection{The signature}\label{ssec:signature}
Rough path theory \cite{lyons1998differential} is an area of stochastic analysis that provides a rigorous mathematical framework to describe the interaction of a stream with a physical control system. Recently, this mathematical tool-set has found numerous applications in machine learning. In particular, the signature transform possesses powerful properties which position it as an effective feature map for streamed data \cite{levin2013learning, lyons2014rough, arribas2018signature, moore2019using, kidger2019deep, morrill2020generalised, shao2023dimensionless, lemercier2021distribution, lyons2022signature, fermanian2023new}.

\hspace{0pt}\\Let $I=[t_L,t_U]$ be a closed interval with $0\leq t_L<t_U$. Let $\gamma$ be a path, that is a continuous function from $I$ to $\mathbb{R}^d$ with $d>0$. In the sequel, we consider paths of bounded variation \cite[Definition 1.5]{lyons2007differential}. 
The signature maps any path into the following space of sequences of tensors
\begin{equation}
 \resizebox{0.44\textwidth}{!}{$   T((\mathbb{R}^d)) = \left\{ \mathcal{A}=(A_0, A_1,  \ldots, A_k, \ldots)~|~\forall k\geq 0, A_k\in(\mathbb{R}^d)^{\otimes k}\right\} $}
\end{equation}
\noindent where $\otimes$ denotes the tensor product of vector spaces. By convention $(\mathbb{R}^d)^{\otimes 0}=\mathbb{R}$, therefore $A_0$ is a scalar, $(\mathbb{R}^d)^{\otimes 1}=\mathbb{R}^d$, and $A_1$ is a vector, then $(\mathbb{R}^d)^{\otimes 2} = \mathbb{R}^d \otimes \mathbb{R}^d$ can be identified with the space of $d \times d$ matrices and $(\mathbb{R}^d)^{\otimes 3}$ the space of $d \times d \times d$ tensors.\\ 

\begin{definition}[Signature]
The signature of a path $\gamma$ over the interval $[s,t]$, denoted by $S(\gamma)_{s,t}$, is the element of $T((\mathbb{R}^d))$,
\begin{equation}
    S(\gamma)_{s,t}=(1,S_1(\gamma)_{s,t}, \ldots, S_m(\gamma)_{s,t}, \ldots)
\end{equation}
where for any $m>0$, the $m^{th}$ term is given by the following iterated (Riemann–Stieltjes) integral 
\begin{align*}    S_m(\gamma)_{s,t}=\underset{s< t_1<\ldots<t_m<t}{\int \ldots \int} d\gamma(t_1)\otimes \ldots \otimes d\gamma(t_m).
\end{align*}
\end{definition}

\noindent The $m^{th}$ term can be interpreted as a sequentially ordered moment. More precisely, we have 
\begin{equation}\label{eqn:sig}
S_m(\gamma)_{s,t} =  (t-s)^m\mathbb{E}_{t_1<\ldots<t_m}[\dot{\gamma}(t_1)\otimes\ldots\otimes\dot{\gamma}(t_m)]/m!
\end{equation} 
where the expectation is taken over $t_1,\ldots,t_n\sim\mathrm{Law}(U_{(1)}, \ldots, U_{(m)})$, that is, $m$ random times distributed as the $m$ order statistics $U_{(1)}, \ldots, U_{(m)}$ of a uniform random variable on $[s,t]$. To simplify notation we drop the time indices for the signature over the full interval, that is, we write $S(\gamma):=S(\gamma)_{t_L,t_U}$.\\

\begin{example} For a $2$-dimensional path $\gamma:t\mapsto(\gamma_1(t),\gamma_2(t))$ from $[0,1]$ to $\mathbb{R}^2$, we have

\begin{align*}
    \resizebox{0.46\textwidth}{!}{$   S(\gamma)=\left(1,\begin{pmatrix}\mathbb{E}_t[\dot{\gamma}_1(t)]\\\mathbb{E}_t[\dot{\gamma}_2(t)]\end{pmatrix},\frac{1}{2}\begin{pmatrix}\mathbb{E}_{s<t}[\dot{\gamma}_1(s)\dot{\gamma}_1(t)] & \mathbb{E}_{s<t}[\dot{\gamma}_1(s)\dot{\gamma}_2(t)]\\\mathbb{E}_{s<t}[\dot{\gamma}_2(s)\dot{\gamma}_1(t)]&\mathbb{E}_{s<t}[\dot{\gamma}_2(s)\dot{\gamma}_2(t)]\end{pmatrix}, ~\ldots\right) 
    $}
\end{align*}
\end{example}
\noindent There are fundamental differences
between the signature and classical signal
processing transforms such as the Fourier transform. Importantly, the Fourier transform is a linear transformation which treats the channels in a multimodal stream independently. Furthermore, the signature transform has the universal approximation property, which guarantees that any continuous function on paths can be accurately approximated 
by linear combinations of their iterated integrals. This property is particularly important when carrying regression analysis tasks. 

\hspace{0pt}\\The range of the signature is included in a subset of $T((\mathbb{R}^d))$ which is the Hilbert space defined by
\begin{align*}
\resizebox{0.48\textwidth}{!}{$
    H=\Big\{\mathcal{A}=(1,A_1,\ldots, A_k, \ldots)\in T((\mathbb{R}^d))~\Big|~1+\sum_{k=1}^{\infty}\|A_k\|_{(\mathbb{R}^d)^{\otimes k}}^2<\infty\Big\}, $}
\end{align*}
with inner product $\langle \cdot, \cdot\rangle_H$ defined for any $\mathcal{A},\mathcal{B}\in H$ by
\begin{align*}
    \langle \mathcal{A},\mathcal{B}\rangle_H = 1 + \sum_{k=1}^{\infty}\langle A_k, B_k\rangle_{(\mathbb R^d)^{\otimes k}}.
\end{align*}
Therefore the signature maps paths to infinite-dimensional vectors which are easier to handle numerically. In view of numerical applications the signature is truncated at some level $n$ and we will denote the feature vector of $D=1+d+d^2+\ldots+d^n=(d^{n+1}-1)/(d^n-1)$ signature features by
\begin{align}
    \mathrm{sig}(\gamma)=\mathrm{vec}(1,S_1(\gamma), \ldots, S_n(\gamma))\in \mathbb{R}^{D}.
\end{align} 
In practice, the input data often corresponds to a sample $\gamma(t_1), \ldots, \gamma(t_M)$ of an underlying continuous path. The signature (the iterated integrals) of the piecewise linear interpolation of such time series can be efficiently computed using existing highly optimized Python packages \cite{esig,reizenstein2018iisignature,kidger2020signatory,roughpy}.

\hspace{0pt}\\\textbf{Invariances and Equivariances} In machine learning tasks, such as anomaly detection, when one has prior knowledge that certain transformations (e.g. translations or scaling) should not impact the prediction or decision, using features that are insensitive to these transformations becomes very advantageous. Remarkably, the signature is invariant to time reparameterization, and can be seen as a filter that removes an infinite
dimensional group of symmetries. It is also invariant to translation. The signature is also equivariant with respect to different groups, which will become relevant in the next subsection.

\subsection{Novelty scores}\label{ssec:conformance}
Having introduced a feature map to encode streamed data into finite-dimensional feature vectors, we address the problem of novelty detection in multivariate data. 

\hspace{0pt}\\Several anomaly detection methods are based on comparing distances between data instances. In semi-supervised anomaly detection, a classical technique consists in defining the anomaly score of a new data point as its distance to its nearest neighbour in a dataset of instances labelled as normal. A threshold is then applied to the anomaly score to determine if a test instance is anomalous or not. Different distance metrics can be used to handle different data types. A powerful approach for anomaly detection on sequential data has been proposed in \cite{shao2023dimensionless}, where the authors make use of the Mahalanobis distance defined on the range of the signature map. In the sequel, we provide a brief account of the formalism introduced in \cite{shao2023dimensionless}.

\begin{definition}[$\mu$-variance norm] Let $\mu$ be a probability measure on a real Hilbert space $H$ with finite second-order moments, and let $X$ denote a random variable with law $\mu$. Denote by $H^*$ the set of linear functionals from $H$ to $\mathbb{R}$. The covariance operator
defined for all $\phi,\psi\in H^*$ by 
\begin{align*}
    \mathscr{C}(\psi,\phi) := \mathrm{Cov}(\phi(X),\psi(X)), 
\end{align*}
induces a norm on $H$, called the $\mu$-variance norm, defined for $x \in H$ by
\begin{align}
    \|x\|_{\mu} := \sup_{\phi:\mathscr{C}(\phi,\phi)\leq 1}\phi(x),
\end{align}
The $\mu$-variance norm is finite on the linear span of the support of $\mu$, and infinite outside of it. 
\end{definition}
\begin{definition}[$\mu$-variance distance] Let $\mu$ be a centred probability measure on a real Hilbert space $H$ with finite second-order moments. The $\mu$-variance distance of $x\in H$ to $\mu$ is defined by
\begin{align}\label{eq:conformance} \alpha(x;\mu)=\inf_{y\in\mathrm{supp}(\mu)}\|x-y\|_{\mu}.
\end{align}
\end{definition}
Let $\mu$ be a probability measure on $H=\mathbb{R}^d$ with finite second-order second moments and denote its covariance matrix by $\glossentrysymbol{cov}$. Let $X$ be a random variable with law $\mu$. In this finite-dimensional case, the covariance matrix $\glossentrysymbol{cov}$ coincides with the covariance operator $\mathscr{C}$, in the sense that for all $\phi,\psi\in\mathbb{R}^D$, 
\begin{align*}
    \mathscr{C}(\phi,\psi)= \mathrm{Cov}(\phi^\top X, \psi^\top X)=\phi^\top \glossentrysymbol{cov} \psi
\end{align*}
the $\mu$-variance norm is given for all $x\in\mathbb{R}^D$ by
\begin{align}
    \|x\|_{\mu}=(x^\top \glossentrysymbol{cov}^{-1}x)^{\frac{1}{2}}
\end{align}
and the $\mu$-variance distance of $x\in\mathbb{R}^D$ to $\mu$ is given by
\begin{align}  \alpha(x;\mu)=\inf_{y\in\mathrm{supp}(\mu)}\left((x-y)^\top \glossentrysymbol{cov}^{-1}(x-y)\right)^{\frac{1}{2}}
\end{align}
where $d_{\mu}(x,y):=\left((x-y)^\top \glossentrysymbol{cov}^{-1}(x-y)\right)^{\frac{1}{2}}$ is the Mahalanobis distance between $x$ and $y$. 

\hspace{0pt}\\In practice, the measure is an empirical measure $\mu_n$ on $\mathbb{R}^D$ associated with $\glossentrysymbol{corpus}=\{x_1, \ldots, x_n\}$ with covariance matrix $\glossentrysymbol{covn}$. The $\mu_n$-variance distance of $x$ to $\mu_n$ is the Mahalanobis distance to the nearest neighbor in $\glossentrysymbol{corpus}$
\begin{align}
    d_{\glossentrysymbol{corpus}}(x,x_i)=\left((x-x_i)^\top \glossentrysymbol{covn}^{-1}(x-x_i)\right)^{\frac{1}{2}}
\end{align}
\hspace{0pt}The Mahalanobis distance is well-known in statistics and machine learning and presents advantages compared to the Euclidean distance. The effect of $\glossentrysymbol{covn}^{-1}$ is to decorrelate and standardize the data before taking the Euclidean distance. It is invariant to non-degenerate linear transformations, in the sense that $\forall x,y\in \mathbb{R}^D$
\begin{align}
    d_{\glossentrysymbol{corpus}}(x,y) = d_{\{Ax|x\in\glossentrysymbol{corpus}\}}(Ax,Ay).
\end{align}
\begin{remark} We note that the Mahalanobis distance is often computed between a point and the empirical mean of a distribution, in which case, it is thought of as a notion of distance between a point and a probability measure. In our notations, this corresponds to $d_{\glossentrysymbol{corpus}}(x,\bar{x})$ where $\bar{x}=\frac{1}{n}\sum_{i=1}^{n}x_i$. However, this idea breaks down in high dimensions.
For example, in the case where $\mu$ is the standard Gaussian measure on $H=\mathbb{R}^D$ (and $d_\mu$ becomes the Euclidean distance), points near the mean are extremely unlikely to be sampled, as nearly all the probability is concentrated in an annulus at radius $\sqrt{D}$~\cite{shao2023dimensionless}.
\end{remark}
\hspace{0pt}\\\cite{shao2023dimensionless} proposed to use $\mu$-variance distance of $x=\mathrm{sig}(\gamma)$ defined in \Cref{eq:conformance} as an outlierness score for any path $\gamma$. In other words, the score of a path $\gamma$ is given by the Mahalanobis distance between its signature and its nearest neighbor in a given set of uncontaminated paths. Compared to other distance metrics, the Mahalanobis distance is invariant under linear transformations of the data. This property positions the Mahalanobis distance as an ideal metric for novelty detection.

\hspace{0pt}\\\textbf{Invariances}  The Mahalanobis distance (hence the anomaly score) is invariant under linear transformations of the data. There are several operations on paths $\gamma$ which correspond to linear operations on $S(\gamma)$. For example, scaling all paths $\gamma$ in our dataset by a scalar $\theta$ results in a linear transformation 
 of their initial signatures $S(\theta\gamma)=(1,\theta S_1(\gamma), \theta^2 S_2(\gamma), \ldots)$. If all paths $\gamma$ in our dataset are pre-concatenated (or post-concatenated) with another given path $\gamma_0$, then $S(\gamma_0*\gamma)=S(\gamma_0)\otimes S(\gamma)$ (or $S(\gamma*\gamma_0)=S(\gamma)\otimes S(\gamma_0)$) and $S(\gamma_0)\otimes$ is a linear operation on $S(\gamma)$. These properties ensure that when the streams are altered by a change of coordinate system, their anomaly scores remain unchanged. 
\subsection{Pysegments: an efficient segmentation algorithm}\label{ssec:pysegments}
Several problems in time series analysis consist of searching for specific patterns within a sequence of observations. A classical approach consists in sliding a window of fixed length over the data and analyzing the data within the window at each step to determine whether the pattern is present. These techniques suffer from a number of challenges, including the choice of the window size and the computational complexity is linear with the number $N$ of observations. From a statistical standpoint, when the presence of the pattern is determined using a hypothesis test, sliding window techniques lead to the multiple testing problem~\cite{window1, window2}: due to the overlap of the data in the subsequent windows, it becomes difficult to control the false positive rate. Despite these limitations, sliding window techniques are straightforward to implement and remain largely employed. They underpin state-of-the art RFI detection frameworks such as AOFlagger. As AOFlagger slides a window whose width increases exponentially at each iteration, the number of hypothesis tests scales as $\mathcal{O}(N\log_2 N)$.

\hspace{0pt}\\Pysegments is a search algorithm which can be used to mitigate the deficiencies of sliding window techniques. Given a real interval $I$ and a binary function $\chi:P(I)\to\{\mathsf{True},\mathsf{False}\}$ on the set $P(I) = \{J~|~J \subseteq I\}$ of subintervals of $I$, Pysegments is an algorithm that identifies the set $\mathcal{S}$ of disjoint intervals in $P(I)$ of maximum length for which the function returns $\mathsf{True}$. Given access to an oracle capable of telling for any $J\in P(I)$, whether a pattern is present or not, Pysegments has an efficient strategy to determine on which interval to query the oracle next, to eventually obtain $\mathcal{S}$.

\hspace{0pt}\\The efficiency of Pysegments relies on the concept of dyadic intervals, which are real intervals with endpoints $j/2^n$ and $(j+1)/2^n$, with $j,n\in\mathbb{Z}$. Any dyadic interval (say at level $n$), can be uniquely written as the union of two disjoint dyadic intervals (at level $n+1$). Iterating this splitting procedure, any initial dyadic interval can be represented as a union of dyadic intervals at finer and finer resolutions. Consider a minimum resolution $n_\text{signal}$. First, Pysegments searches through this hierarchy of dyadic intervals for the first largest one where the function returns $\mathsf{True}$ (that is, conducts a breath-first search). Second, Pysegments attempts to enlarge it to a (non-dyadic) interval such that the function still returns $\mathsf{True}$ on the extended interval. Once the maximum extension has been identified, this two-step procedure is repeated on the complement. 

\hspace{0pt}\\The complexity of this algorithm is $\mathcal{O}(K)$ where $K$ is the number of disjoint intervals in $P(I)$ of size larger than $1/2^{n_{\text{signal}}}$ where the function is $\mathsf{True}$. This means that the best case complexity is $\mathcal{O}(1)$ obtained when $\chi(I)=\mathsf{True}$. Although the number of subintervals of $I$ is infinite, the number of queries is controlled by $K$. Although the intervals visited during the search may overlap with each other, the fact that complexity is controlled by $K$ significantly improves upon sliding windows.

\section{Method}\label{sec:method}
In this section, we provide a comprehensive description of the score derivation process, along with a clear definition of how outliers are detected and localized.

\subsection{RFI-score} 
We describe how we construct a scoring function that takes in input a set of visibility functions and returns a novelty score, which reflects the degree of being contaminated with RFI. Without loss of generality we consider the visibilities as functions of time. For a fixed frequency $\nu\in \Omega$ we write 
\begin{align*}
  \gamma_{i,j,\nu}:I\to \mathbb{C}, ~t\mapsto \gamma_{i,j}(t,\nu).
\end{align*}
We note that the same analysis can be carried out for the visibilities viewed as functions of the frequency variable.

\hspace{0pt}\\\textbf{Vectorization}  As explained in \Cref{ssec:signature}, a natural feature map for paths, such as baseline visibility signals, is provided by the signature $x_{i,j,\nu} = \mathrm{sig}(\gamma_{i,j,\nu})$.
We propose to consider as input data the antenna visibilities, that is $N_A$ complex-valued signals associated to an antenna-frequency pair $(i,\nu)$, and vectorize this data through the expected signature \cite{ni2012expected, chevyrev2016characteristic, chevyrev2018signature}
            \begin{align}
                x_{i,\nu} = \frac{1}{N_A}\sum_{j=1}^{N_A}\mathrm{sig}(\gamma_{i,j,\nu}) 
            \end{align}
\hspace{0pt}\\\textbf{Novelty score} Now that we have obtained a convenient vectorial representation of the antenna visibility data, we compute the nearest neighbor Mahalanobis distance as explained in \Cref{ssec:conformance}. 

\hspace{0pt}\\We assume that we have access to some visibility data which are not corrupted by RFI. We refer to this set of RFI-free data as a \textit{corpus} and denote it by $\glossentrysymbol{corpus}=\{x_1, \ldots, x_n\}$.
Note that to simplify the notation, we switch to a single integer to index the baseline-frequency $(i,j,\nu)$ now enumerated by $i=1, \ldots, n$. 
Given a corpus we compute two types of statistics, namely the \textit{sample mean} and the \textit{sample covariance matrix}, under the transform described above
\begin{align*}
 \bar{x}=\frac{1}{n}\sum_{i=1}^{n}x_i \quad\text{and}\quad \glossentrysymbol{covn}=\frac{1}{n-1}\sum_{i=1}^{n}(x_i-\bar{x})(x_i-\bar{x})^\top.
\end{align*}
  The Mahalanobis distance between an input $x$ and an element $x_i$ of the corpus, is then defined as 
 \begin{align*}
d_{\glossentrysymbol{corpus}}(x,x_i)&=\sqrt{(x-x_i)^\top \glossentrysymbol{covn}^{-1}(x-x_i)}.
\end{align*}
Given an input $x$, we compute its Mahalanobis distance to every element of the corpus, and use this collection of distances to construct a score $\alpha(x;\glossentrysymbol{corpus})$. For example, a popular choice, is to define the score as the nearest-neighbor (NN) distance, namely
\begin{align*}
    \alpha(x;\glossentrysymbol{corpus})=\min_{i=1,\ldots,n} d_{\glossentrysymbol{corpus}}(x,x_i). 
\end{align*}

\subsection{From RFI-scores to RFI detection}\label{ssec:Distfit}
\begin{figure}[!t]
    \centering
    \includegraphics[width=3in]{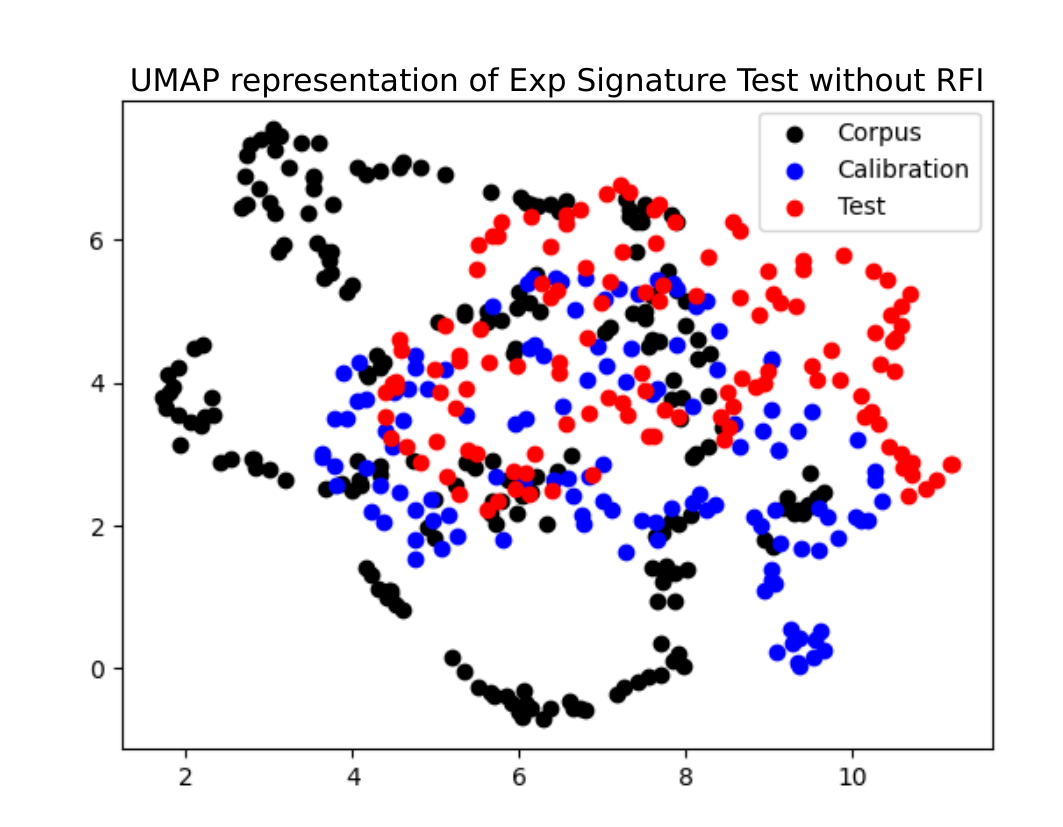}\\
    \includegraphics[width=3in]{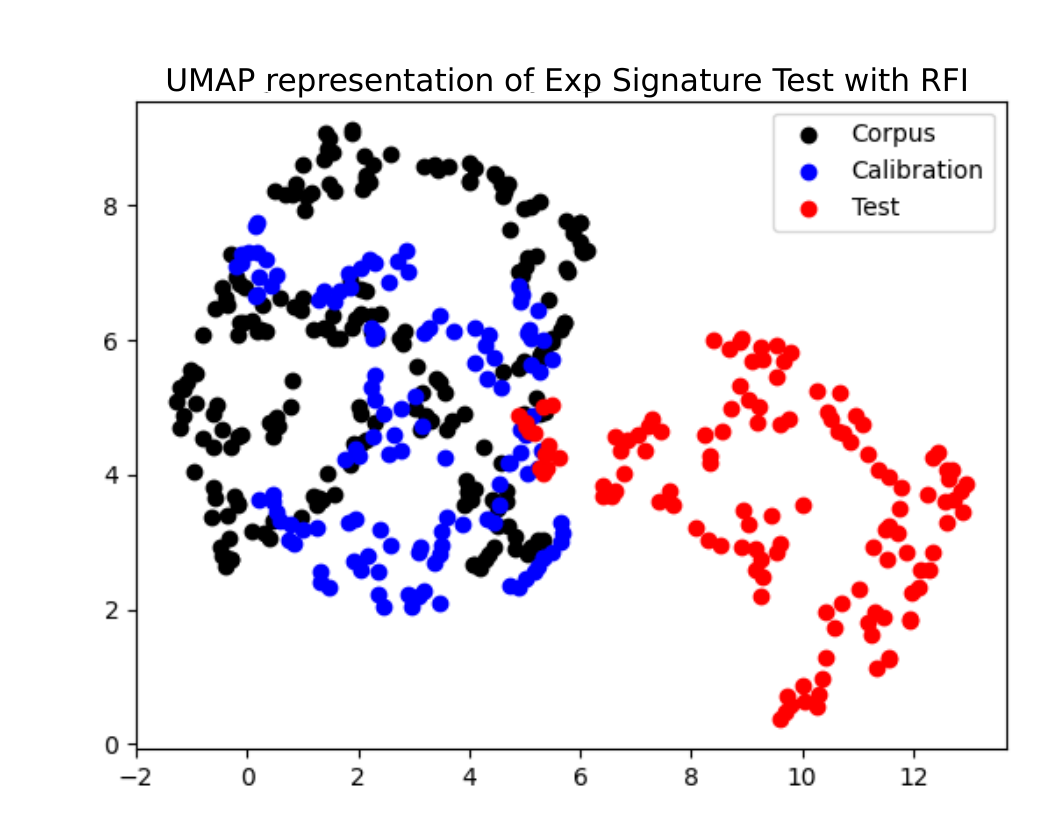}
    \caption{A UMAP~\cite{mcinnes2018uniform} representation of the expected signature for each antenna in the corpus, calibration, and test sets. The dataset dimensions are denoted as (N\_Ant, Sig), with an example size for the corpus being (213, 62) due to truncation of the signature at level 5. UMAP projects this into a lower dimension of (213, 2). The top plot illustrates a test set without RFI, while the bottom one depicts a test set with high RFI contamination. This example uses simulated data, with intentionally high RFI for explanatory purposes.}
    \label{fig:UMap}
\end{figure}

Now, we explain how we build a one-class classifier for RFI detection, in other words, how given a score we decide whether the signal is contaminated with RFI or not. To this aim, we use another set of RFI-free data, which we refer to as the \textit{calibration set} and denote (similarly to the corpus) by
\begin{align*}
    \widetilde{\mathcal{D}}_m = \{\tilde{x}_1, \ldots, \tilde{x}_m\}.
\end{align*}
We precompute the scores of each element of the calibration set $\alpha(\tilde{x}_1;\glossentrysymbol{corpus}),~\ldots,~ \alpha(\tilde{x}_m;\glossentrysymbol{corpus})$ and use the $\mathsf{distfit}$ Python package \cite{Taskesen_distfit_is_a_2020} to fit a generalized extreme value (GEV) distribution \cite{pickands1975statistical} to these scores \cite{roberts2000extreme,siffer2017anomaly,vignotto2020extreme}. We then read the quantile $\alpha_\epsilon$ at level $\epsilon$, that is, the smallest value such that $\mathbb{P}[\alpha\geq \alpha_\epsilon]<\epsilon$. \Cref{fig:UMap} shows a projection of how the expected signatures of the \textit{corpus, calibration} and \textit{test} set might look like.

\hspace{0pt}\\\textbf{RFI detection for one antenna} For any new expected signature $x^*$, the RFI detector is defined by
\begin{align*} \text{R}^\epsilon(x^*;\glossentrysymbol{corpus}) = \left\{
    \begin{array}{ll}
        \text{RFI-free} & \mbox{if } \alpha(x^*;\glossentrysymbol{corpus}) \leq \alpha_\epsilon \\
        \text{RFI} & \mbox{otherwise}.
    \end{array}
\right.
\end{align*}

\hspace{0pt}\\Next, we propose a way to obtain a single flagging mask for the whole array of antennas. In arrays that contain several antennas, we believe it might be useful for the radioastronomer to get a rough first diagnostic at the level of the entire array, in the form of a single flagging mask. 

\hspace{0pt}\\\textbf{RFI detection for a group of antennas} For flagging a collection of antennas, we use
\begin{align*}
    \text{R}^\epsilon\left(\{x^*_i\}_{i=1}^{N_A};\glossentrysymbol{corpus}\right) = \left\{
    \begin{array}{ll}
        \text{RFI-free} & \mbox{if }  \frac{1}{N_A}\sum_{i=1}^{N_A}\alpha(x^*_{i};\glossentrysymbol{corpus}) \leq \alpha_\epsilon \\
        \text{RFI} & \mbox{otherwise}.
    \end{array}
\right.
\end{align*}

\begin{remark}
    As highlighted in \cite{wilensky2019absolving}, clean visibility signals have a frequency dependence. For this reason, we calibrate an RFI detector for each frequency channel. In other words, we construct a corpus $\mathcal{D}^{(\nu)}_{n}$ and a calibration set $\widetilde{\mathcal{D}}^{(\nu)}_m$ for each frequency channel $\nu$. 
\end{remark} 

\noindent We note that the two proposed approaches for RFI detection (for a single antenna or a group of antennas) coincide when $N_A=1$.

\subsection{Localizing the RFI}
   Given a real interval $I$ and a binary function $\chi:P(I)\to\{\mathsf{True},\mathsf{False}\}$ on the set $P(I) = \{J~|~J \subseteq I\}$ of subintervals of $I$, Pysegments is a search algorithm that identifies the set of disjoint intervals in $P(I)$ of maximum length for which the function returns $\mathsf{True}$. We propose to use this algorithm for RFI detection. In this context, the input is a visibility signal indexed on a time (or frequency) interval and the binary function is an anomaly detector. The latter takes in input the visibility signals restricted to a subinterval of time (or frequency) and returns $\mathsf{True}$ if it is RFI-free. The output of the segmentation algorithm is a union of disjoint RFI-free time intervals. 
    
    \hspace{0pt}\\For each set of antenna visibilities $\{\gamma_{i,j,\nu}\}_{j=1}^{N_A}$ in a frequency channel $\nu$, observed over the time domain $I$, we run the segmentation algorithm with the binary function defined by
    \begin{align*}
        \chi_\gamma(J)=R^{\epsilon}\left(\big\{\frac{1}{N_A}\sum_{j=1}^{N_A}\mathrm{sig}(\gamma_{i,j,\nu}|_J)\big\}_{i=1}^{N_A};\glossentrysymbol{corpus}\right)
    \end{align*}

\subsection{Complexity analysis}
We discuss the complexities of the operations that are performed during the flagging a new dataset.

\hspace{0pt}\\\textbf{Anomaly score} For each vector $x\in\mathbb{R}^D$ we need to compute its anomaly score, that is conduct the nearest neighbor search $\min_{i=1,\ldots,n} d_{\glossentrysymbol{corpus}}(x,x_i)$ where each distance can be computed in $\mathcal{O}(D^3)$ time. Therefore, the complexity of brute force nearest neighbor search is $\mathcal{O}(nD^3)$. This complexity can be improved by leveraging approximate nearest neighbor search algorithms such as the NN-Descent algorithm \cite{dong2011efficient} or the \textsc{Faiss} library \cite{johnson2019billion}. Furthermore, the inverse of the covariance matrix is only computed once, hence the cubic cost is amortized, and the Mahalanobis distances can be computed in $\mathcal{O}(D^2)$. 
   
\hspace{0pt}\\\textbf{Segmentation} Let $I$ be an interval and $A$ be a set of $K$ disjoint subintervals $J_1,\ldots, J_K$ of $I$. Consider the characteristic function $\chi_A(J)=\mathsf{True}$ if $J\subseteq J_i$ for some $J_i\in A$ and $\chi_A(J)=\mathsf{False}$ otherwise. Pysegments is an algorithm which "approximates" $A$ by evaluating $\chi_A$ on $\mathcal{O}(K')$ subintervals where $K'=\#\{J\in A~|~|J|\geq 1/2^{n_s}\}$ following a simple and efficient strategy. In the context of RFI detection, $A$ is the collection of disjoint RFI-free intervals. We do not have access to $\chi_A$, but to a proxy given by the anomaly detector. Suppose that the anomaly detector is perfect, in the sense that it identical to $\chi_A$. Given a tolerance $n_{s}\in\mathbb{Z}$, the complexity of the algorithm is determined by the number $K'$ of disjoint RFI-free intervals whose length is larger than $1/2^{n_{s}}$. In particular, if the whole sequence of size $N$ is RFI-free, the complexity is $\mathcal{O}(1)$. This is much more efficient than the \textsc{SumThreshold} or \textsc{SSINS} algorithms which scale at least as $\mathcal{O}(N\log_2 N)$ and at best $\mathcal{O}(N)$.

\vspace{1pt}
     \begin{algorithm}[H]
    \caption{\acrshort{name}}
    \label{alg:full_algo}
    \begin{algorithmic}[1]
    \STATE \textbf{Input:} Frequency-antenna pair $(\nu,i)$, interval $[t_L,t_U]$ and threshold $\epsilon$
    \STATE \textbf{Output:} Set of integration times contaminated with RFI 
    \STATE Initialize set of clean intervals $\mathcal{C} = \{\}$
    \STATE First interval to query $[s,t]=[t_L, t_U]$ 
    \WHILE{$[s,t]$ is not \textsf{None}}
    \STATE $\text{RFI}=R^\epsilon\left(\{x^{(i,\nu)}_{j}\}_{j=1}^{N_A}; \mathcal{D}^{(\nu)}_n\right)_{s,t}$ 
    \STATE If RFI is \textsf{False}, add $[s,t]$ to $\mathcal{C}$
    \STATE Determine next interval $[s,t]$ to query 
    \ENDWHILE
    \RETURN $[t_L, t_U]\backslash \mathcal{C}$
    \end{algorithmic}
\end{algorithm}

\section{Results}\label{sec:results}
In this section, we present the performance of \acrshort{name} in comparison to the \textsc{ssins} and \textsc{AOFlagger} frameworks using both simulated and real data. We utilized the latest version of \textsc{AOFlagger} (3.1.0 as of June 2023) and an updated version of \textsc{ssins} for obtaining the results. We primarily present the results of \textsc{AOFlagger} using the ``integrate all baselines" modality, displaying the average values. Additional analyses using the baseline-per-baseline modality are provided in the supplementary material.

\hspace{0pt}\\The parameters for \acrshort{name} are chosen as follows: the signature truncation level is determined through optimization studies, and we found that level 5 provides excellent results without requiring further transformations. As for the segmentation algorithm, it depends on the data's time steps and is a function of the interval length it will loop over.

\subsection{Simulated Data}\label{ssec:SimulData}
We generated synthetic data using CASA 6.2.1.7 software~\cite{TheCASATeam2022} with 64 frequency channels, 50 integration times, and one circular polarization RR, while also including the presence of thermal noise. For the \textit{corpus} data used in \acrshort{name}, we employed the ngVLA configuration file, setting it up to 214 antennas and applying the appropriate scaling factor for thermal noise~\cite{CASATuotrial}. The \textit{calibration} set was also simulated using the same configuration file but with 110 antennas and a different scaling factor. Subsequently, for the \textit{testing} data, we simulated 127 antennas and introduced different types of noise while adjusting their intensities across the various frequency channels. We manually added RFI to the simulated dataset as follows: the first and last 5 frequency channels with high and constant RFI over time (multiplying by 30 the thermal noise), frequency channels 20 to 25 with a mid-high, constant RFI over time (10 times greater than the thermal noise), and frequency channels 40 to 50 with an increasing RFI over time. 

\begin{figure*}[!t]
    \centering
\includegraphics[scale=0.22]{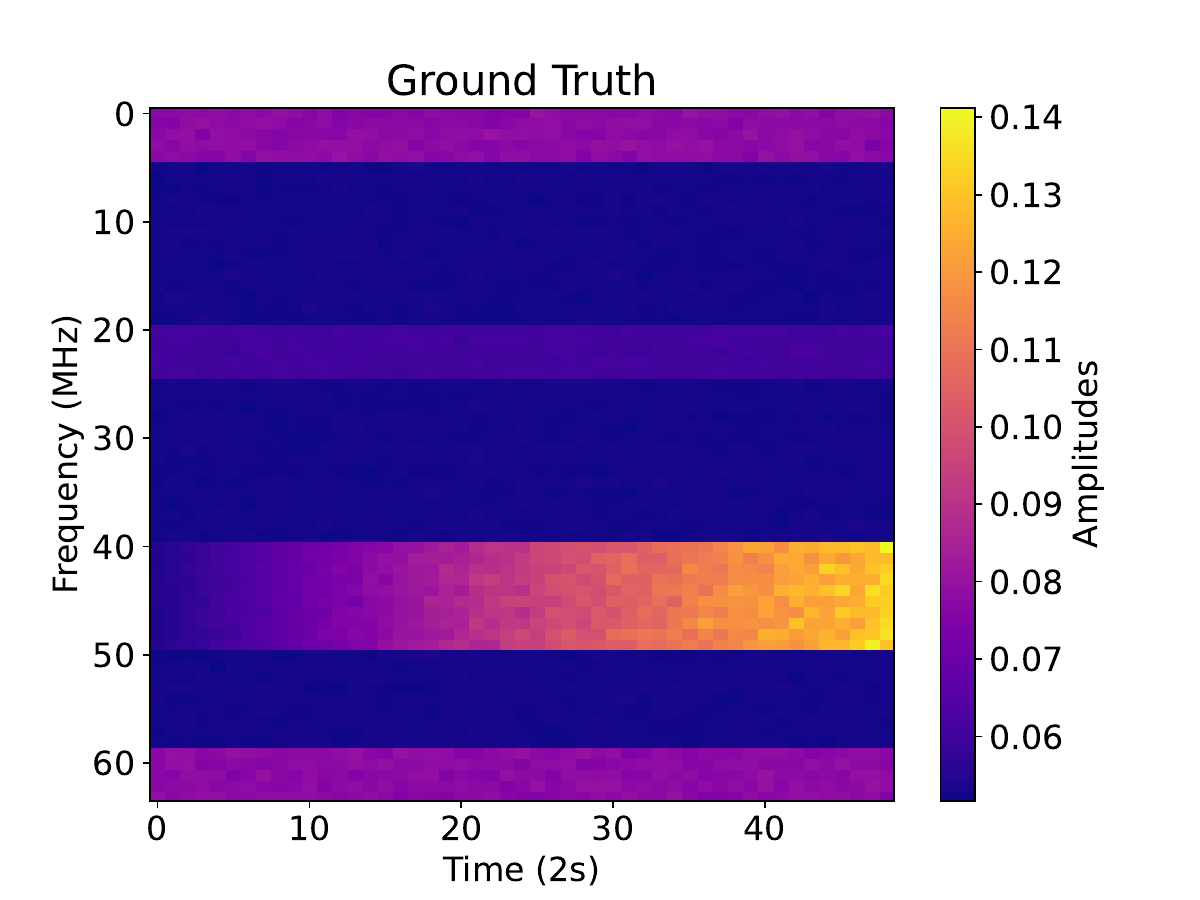}\includegraphics[scale=0.22]{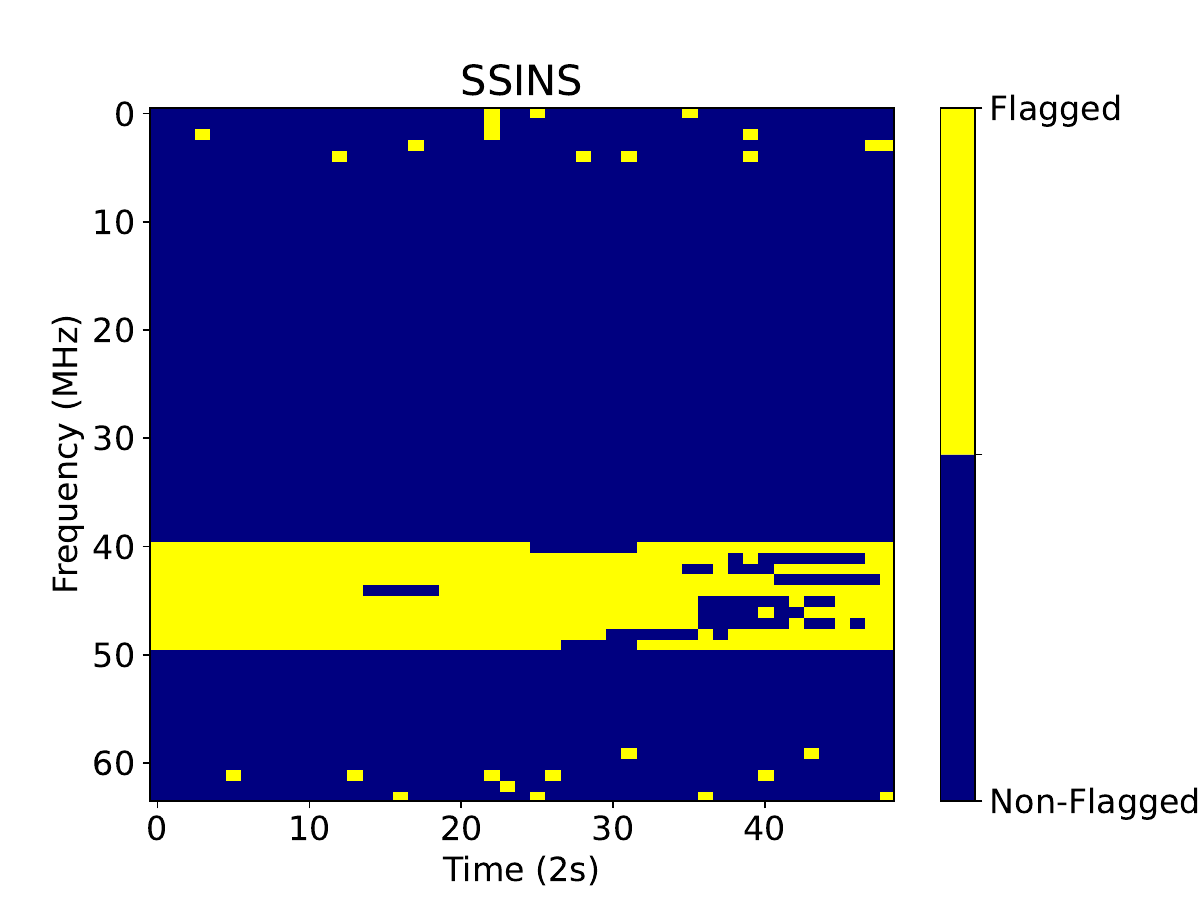}\includegraphics[scale=0.22]{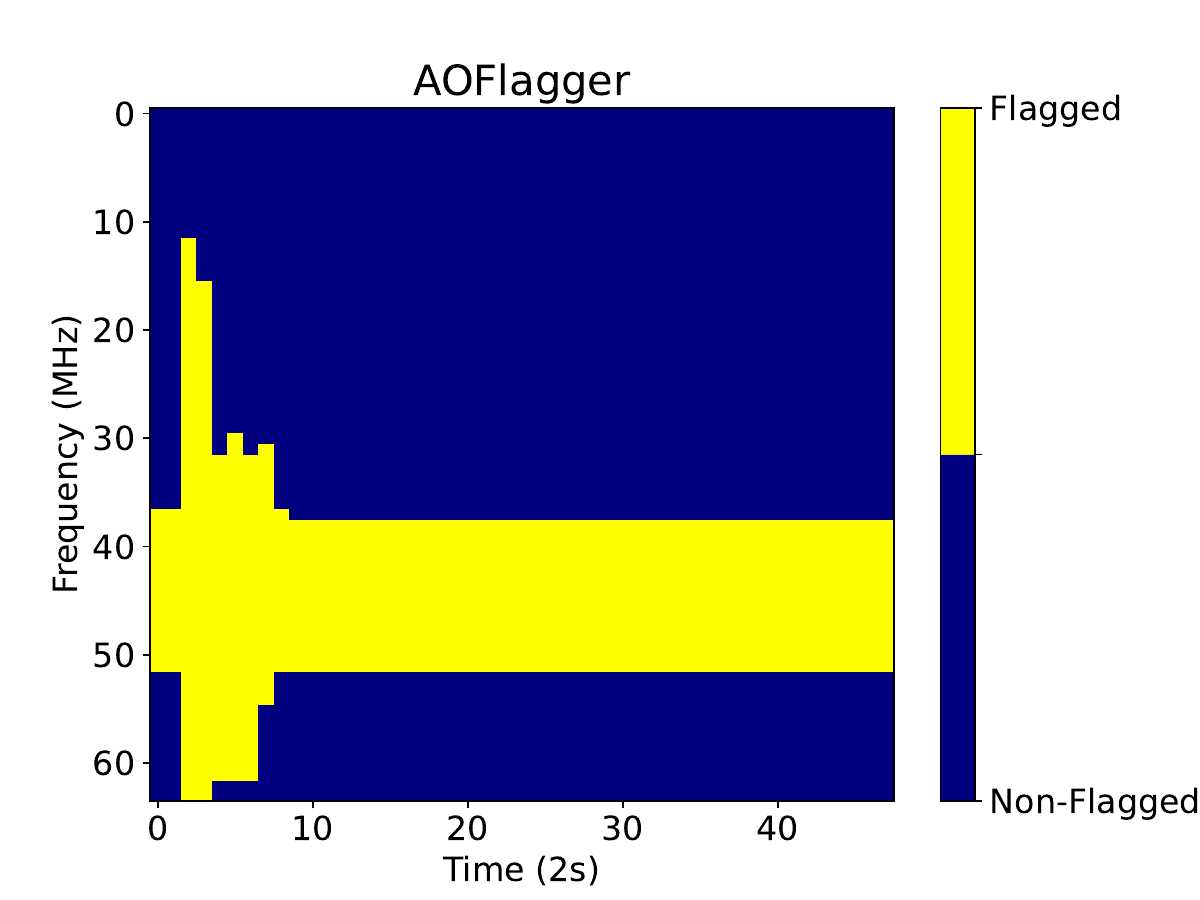}\includegraphics[scale=0.22]{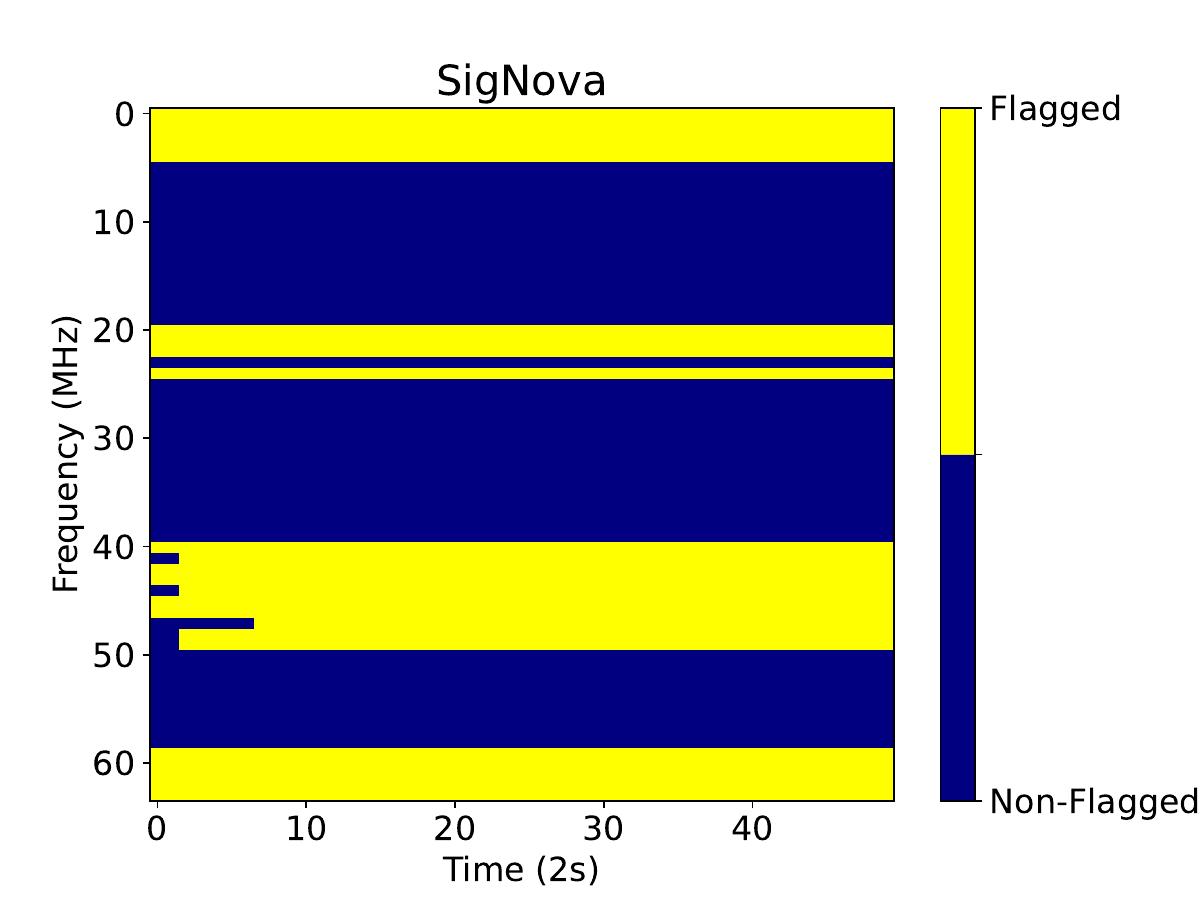}
    \caption{CASA simulations with different types of RFI contaminating only antenna 1.  The ground truth, illustrating the amplitude difference, is depicted in the plot on the right. The subsequent plots feature \textsc{ssins}, \textsc{AOFlagger}, and \acrshort{name}, respectively.}
    \label{fig:RFI_Simulation1}
\end{figure*}

\hspace{0pt}\\The results presented in \Cref{fig:RFI_Simulation1} show the contamination previously explained only applied to antenna 1. The leftmost plot in the figure displays the average amplitudes across all baselines, serving as the ground truth for comparison. The second plot shows the results obtained from \textsc{ssins}, the third plot displays the results from \textsc{AOFlagger}, and the fourth plot represents the results from \acrshort{name}. Upon initial observation, it is evident that both \textsc{ssins} and \textsc{AOFlagger} encounter challenges when flagging constant RFI, even when it is of high intensity. Only the varying RFI is effectively flagged, with a notable concentration occurring early in the time domain in the \textsc{AOFlagger} plot. Further studies with different thresholds were maid with no improvements observed. In contrast, \acrshort{name} clearly identifies and localizes the three different types of RFI in the respective frequency channels.

\begin{figure*}[!t]
    \centering
\includegraphics[scale=0.22]{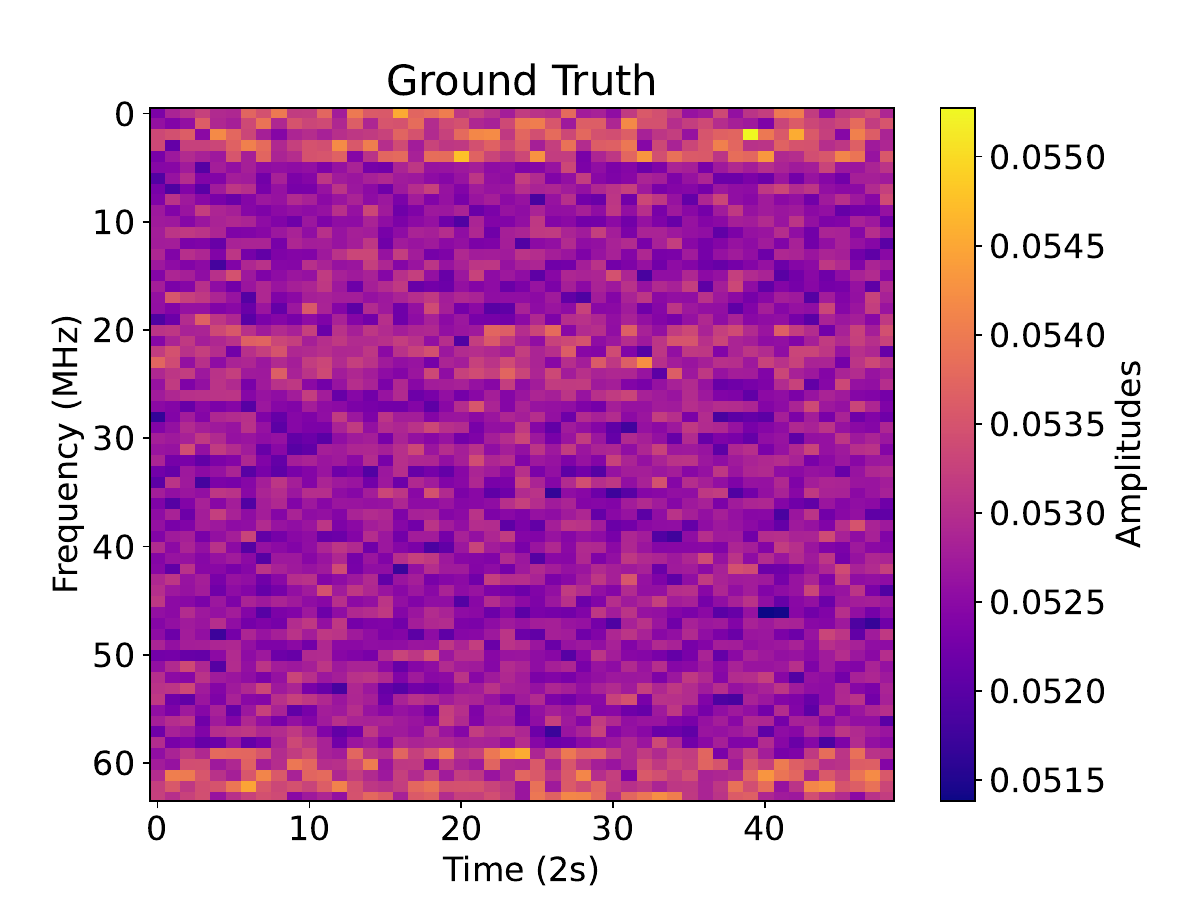}\includegraphics[scale=0.22]{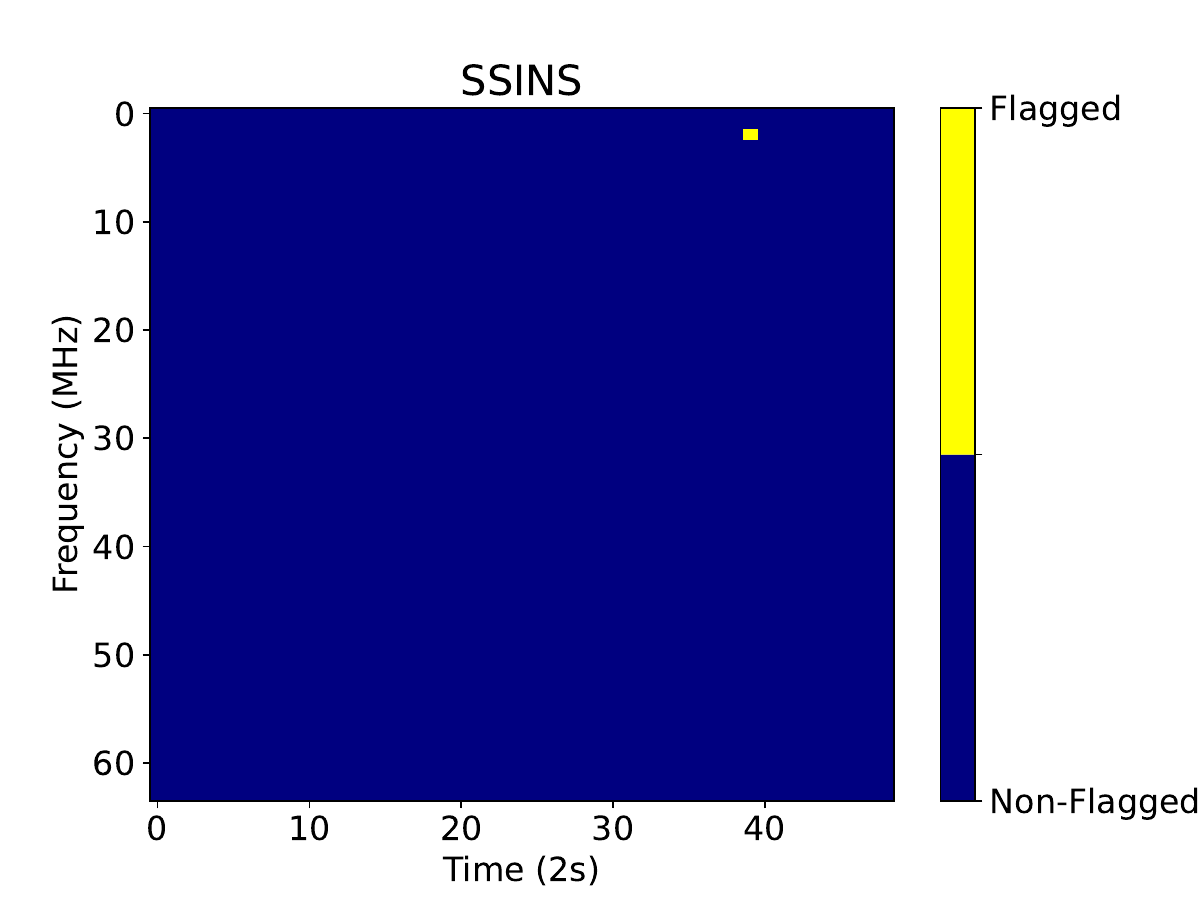}\includegraphics[scale=0.22]{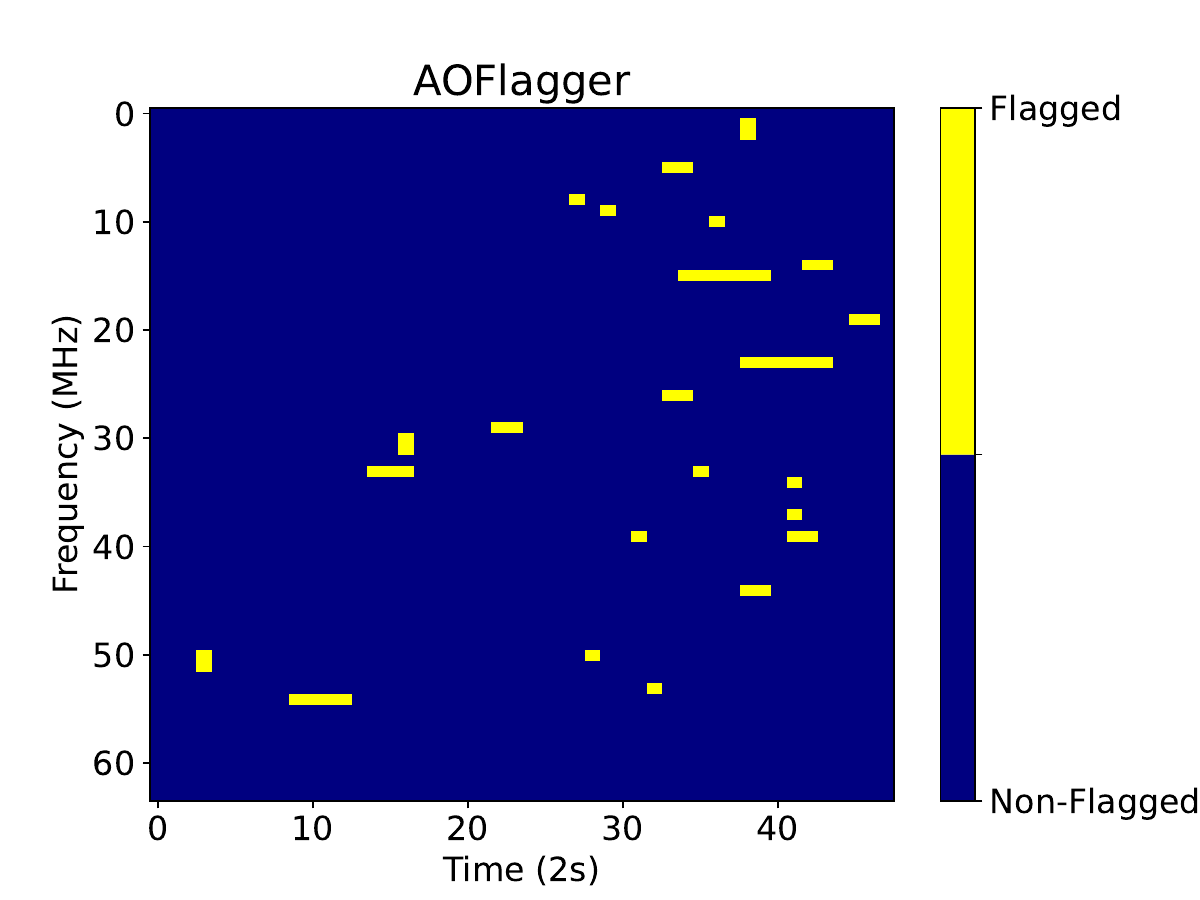}\includegraphics[scale=0.22]{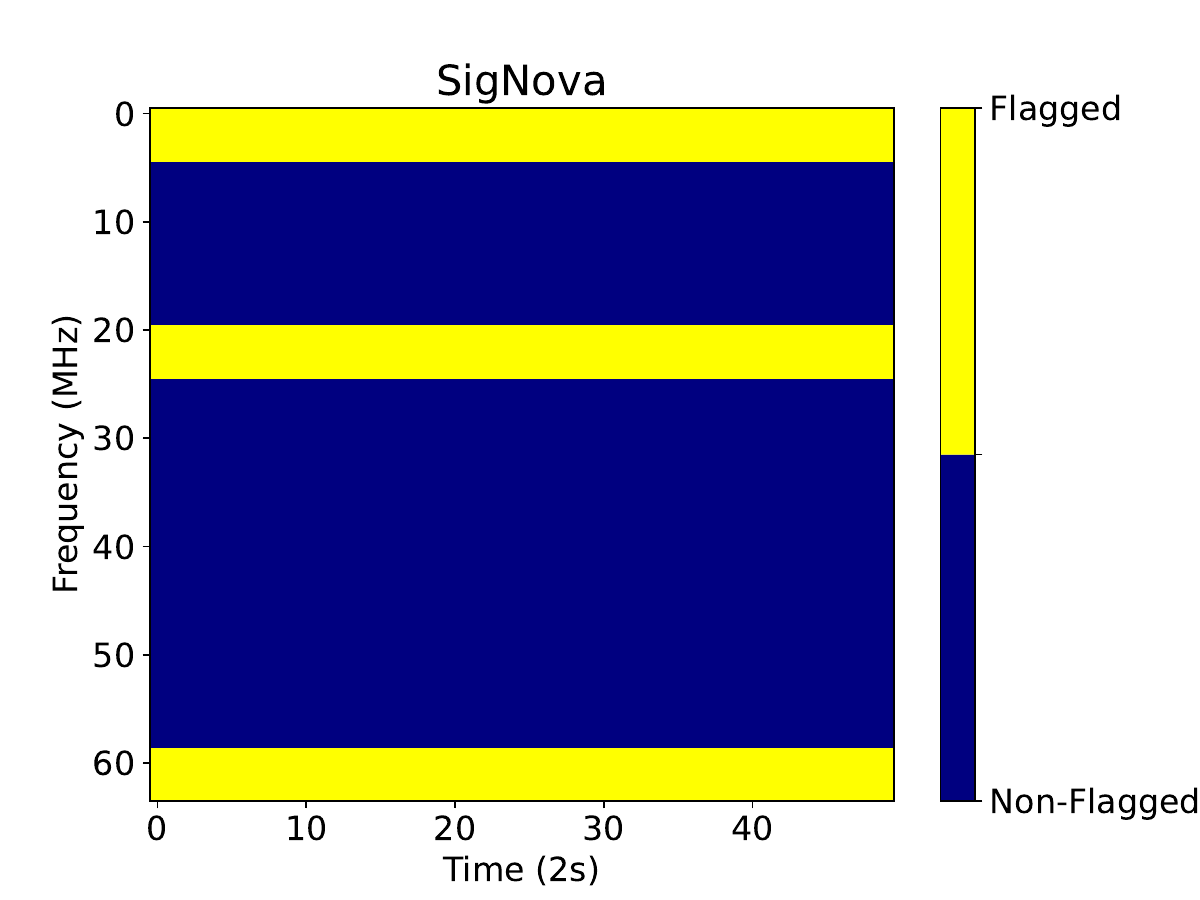}
    \caption{CASA simulations with different types of RFI contaminating only baseline 1 (antenna 1 and antenna 2).  The ground truth, illustrating the amplitude difference, is depicted in the plot on the right. The subsequent plots feature \textsc{ssins}, \textsc{AOFlagger}, and \acrshort{name}, respectively.}
    \label{fig:RFI_Simulation2}
\end{figure*}

\hspace{0pt}\\In order to assess the ability to identify RFI in a single baseline, we applied the same manual RFI contamination to the simulated datasets, but now specifically targeting the first baseline between antenna 1 and antenna 2. In this example, the contamination appears less prominent, with only slight differences in amplitudes visible in the ground truth plot in \Cref{fig:RFI_Simulation2}. Both \textsc{ssins} and \textsc{AOFlagger} encounter difficulties in detecting the RFI, while \acrshort{name} successfully identifies two types of RFI (with the first and last channels exhibiting the same type). However, the RFI in channels 40 to 50, which varies over time, does not appear to be detected by \acrshort{name}, it is not even visible in the ground truth. Nonetheless, \acrshort{name} performs correctly in identifying RFI in a single baseline scenario. It is worth noting that \textsc{AOFlagger} provides the option to examine results on a baseline-by-baseline basis, and in this mode, it manages to detect some RFI, albeit in incorrect frequency channels.

\subsection{Real Data}\label{ssec:real_data}
To verify the authenticity of the clean real data used in \acrshort{name}, we cross-referenced the downloaded data specifications from the webpage to ensure there were no references to RFI. Furthermore, we performed rigorous testing on the ``clean" dataset using \acrshort{name}, \textsc{ssins}, and \textsc{AOFlagger} to validate its cleanness. Only when no indications of RFI were detected, we selected the dataset as a reliable \textit{corpus} for \acrshort{name}. This same procedure was followed during the \textit{calibration} step. To facilitate further studies, we randomly subsampled the real ``clean" data, which was initially selected as the \textit{corpus}, in multiple iterations as outlined in the supplementary material. This approach ensures the integrity of the \textit{corpus} data, as we observed that across different iterations, the maximum score obtained was minimal and comparable to the thermal noise expectation.

\subsubsection{Narrowband Interference}\label{sssec:Narrowband}
We downloaded MWA data via the All-Sky Virtual Observatory (ASVO) web page~\cite{ASVO}. The data has a continuum spectrum at 167 MHz with 384 frequency channels, almost  2 minutes of observation with integration times of 2 seconds.

\hspace{0pt}\\To ensure a sufficiently large \textit{corpus}, we combined two RFI-free datasets (IDs 1065280704 and 1068809832) to provide 254 instances for every frequency channel. For the \textit{calibration} set, we used a separate RFI-free dataset (ID 1065280824) with 127 points in each frequency channel. We selected a signature truncated at level 5 for all results, and set the tolerance for the segmentation algorithm to $-3$, consistent with the 50 integration times in these datasets. To establish the decision threshold, we fit a curve using $\mathsf{distfit}$, as explained in more detail in~\Cref{ssec:Distfit}, at a 0.05 confidence intersection interval (CII).

\begin{figure*}[!t]
    \centering

\includegraphics[scale=.27]{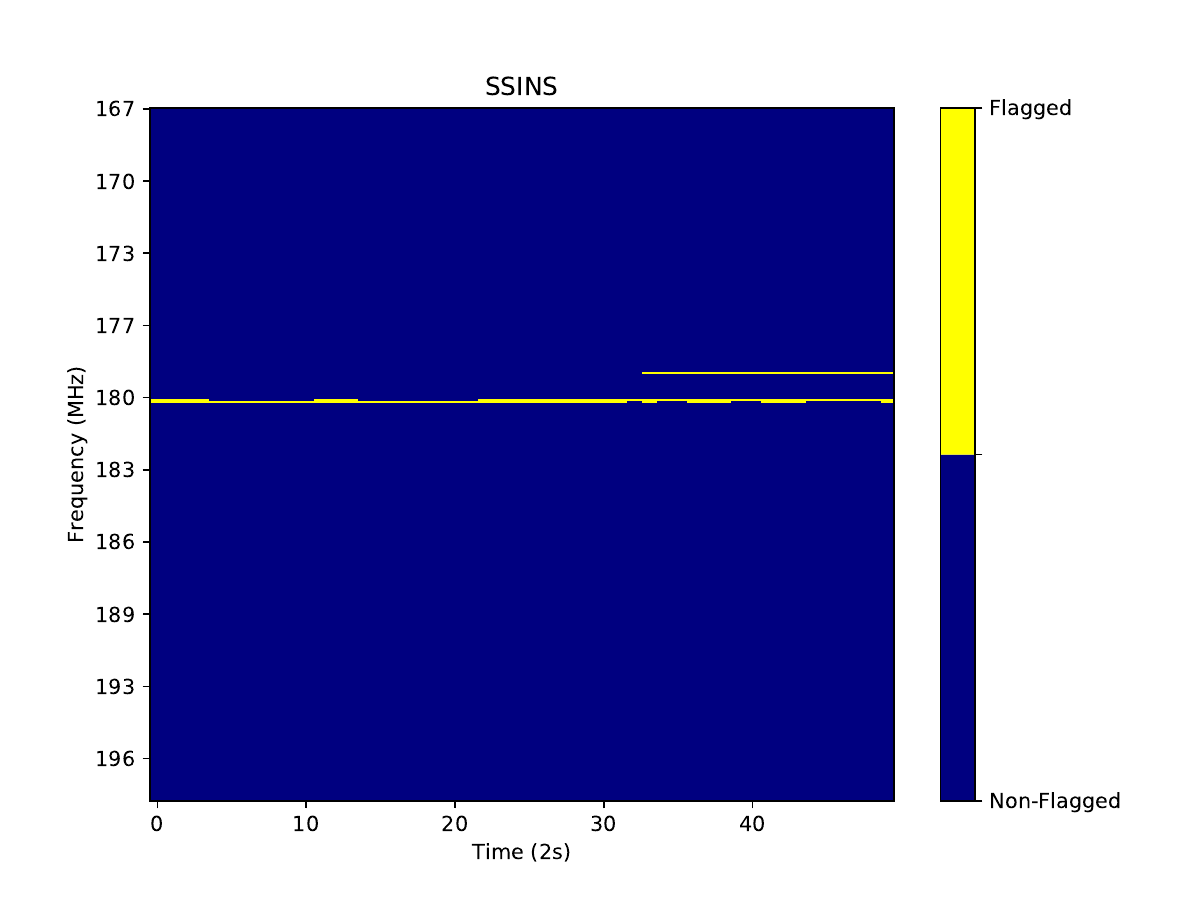}\includegraphics[scale=.27]{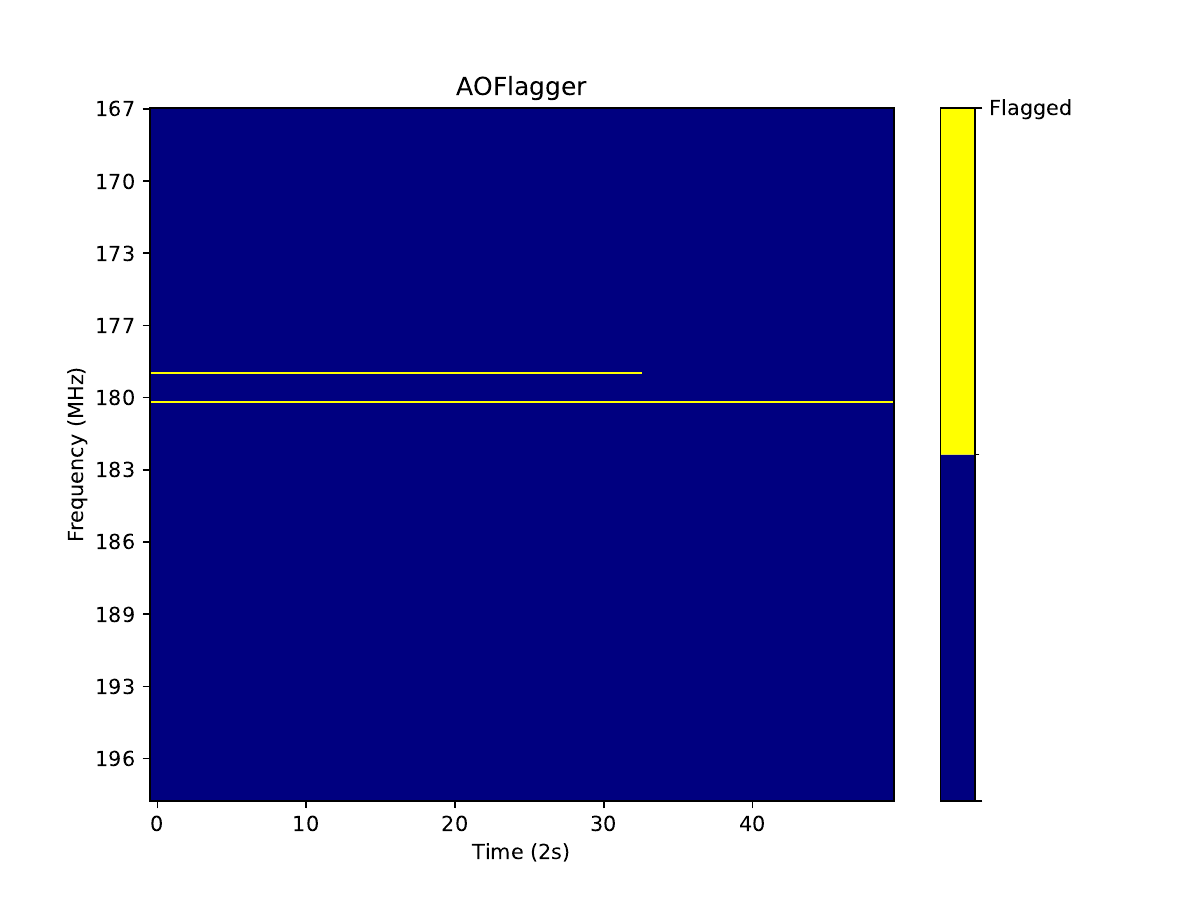}\includegraphics[scale=.27]{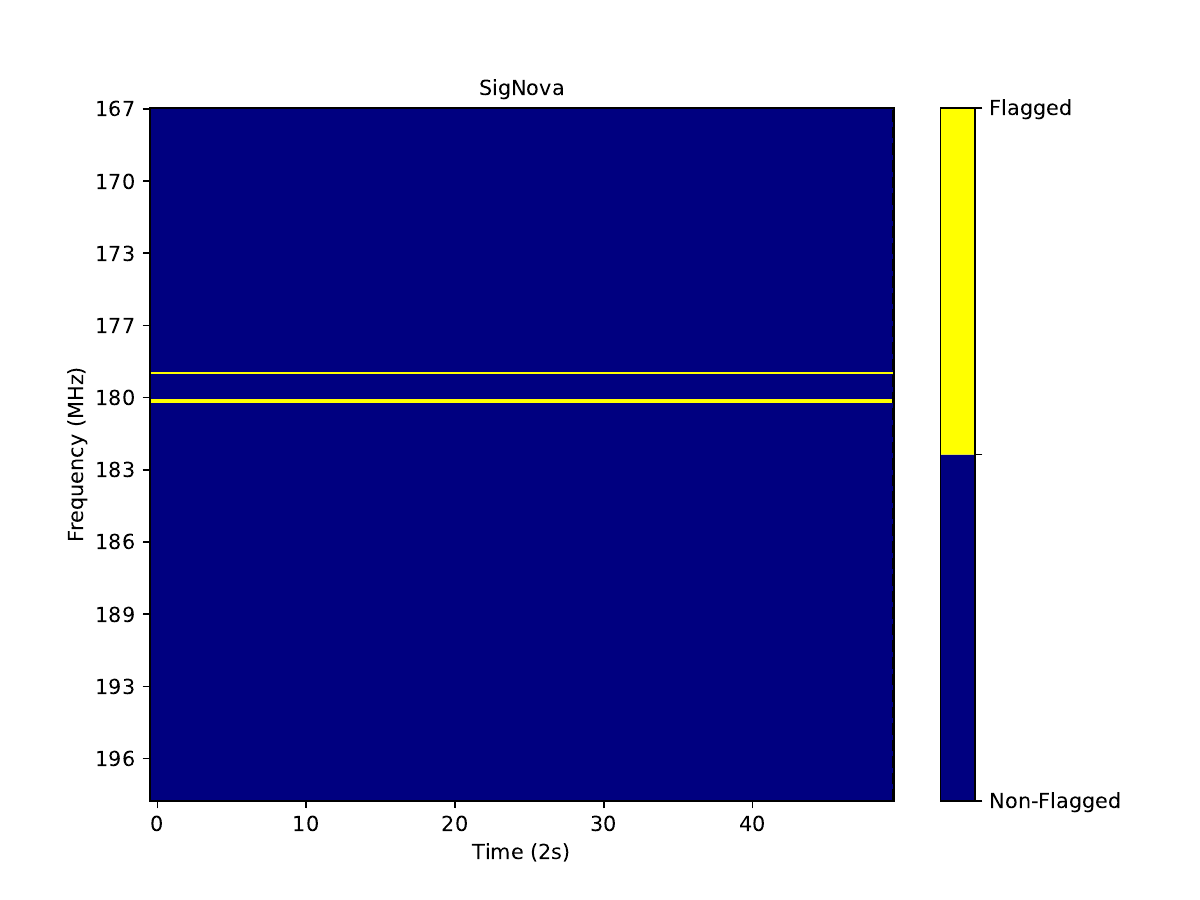}

    \caption{Narrowband MWA data example. The \textsc{ssins} results are shown on the left, the \textsc{AOFlagger} one in the center, and \acrshort{name} on the right.}
    \label{fig:RFI_Narrowband}
\end{figure*}

\hspace{0pt}\\\Cref{fig:RFI_Narrowband} displays the results of one polarization (RR) obtained from \textsc{ssins}, \textsc{AOFlagger}, and \acrshort{name} using the MWA dataset ID 1061318984 as an example of narrow-band RFI (refer to Figure 7 in paper~\cite{mclachlan1999mahalanobis}). This representation aims to directly compare the outcomes. The highlighted yellow areas represent the regions flagged as RFI. Notably, one frequency channel exhibits significant RFI impact, leading to contamination in its neighboring channels. For \textsc{ssins}, the authors' recommendation was followed, setting the threshold to 5. Additional \textsc{ssins} studies (supplementary material) were conducted with varying thresholds, and even when the thresholds were relaxed, the faint RFI frequency channel remained inadequately flagged, allowing noise to show. \textsc{AOFlagger} required a higher base threshold of 6.2 to effectively detect RFI. Without this adjustment, a substantial number of frequency channels would be flagged. \acrshort{name}'s rotated waterfall plot illustrates the segmentation algorithm's output, which is further discussed in \Cref{ssec:pysegments}. \acrshort{name} shows its capability to identify RFI across the entire main frequency spectrum, a more challenging task for~\textsc{ssins} and \textsc{AOFlagger}. Furthermore, our framework excels in accurately localizing faint RFI over time.

\begin{figure*}[!t]
    \centering
    \includegraphics[scale=.27]{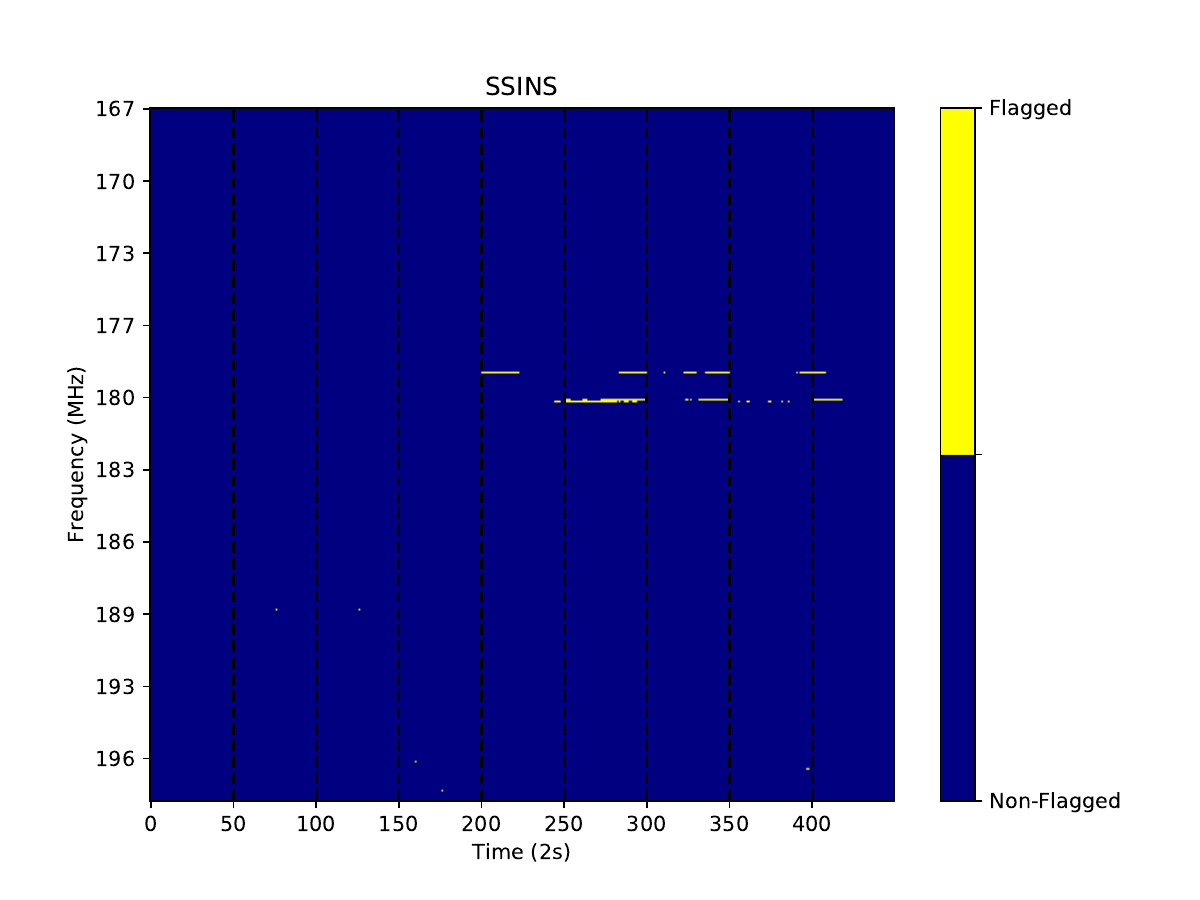}\includegraphics[scale=.27]{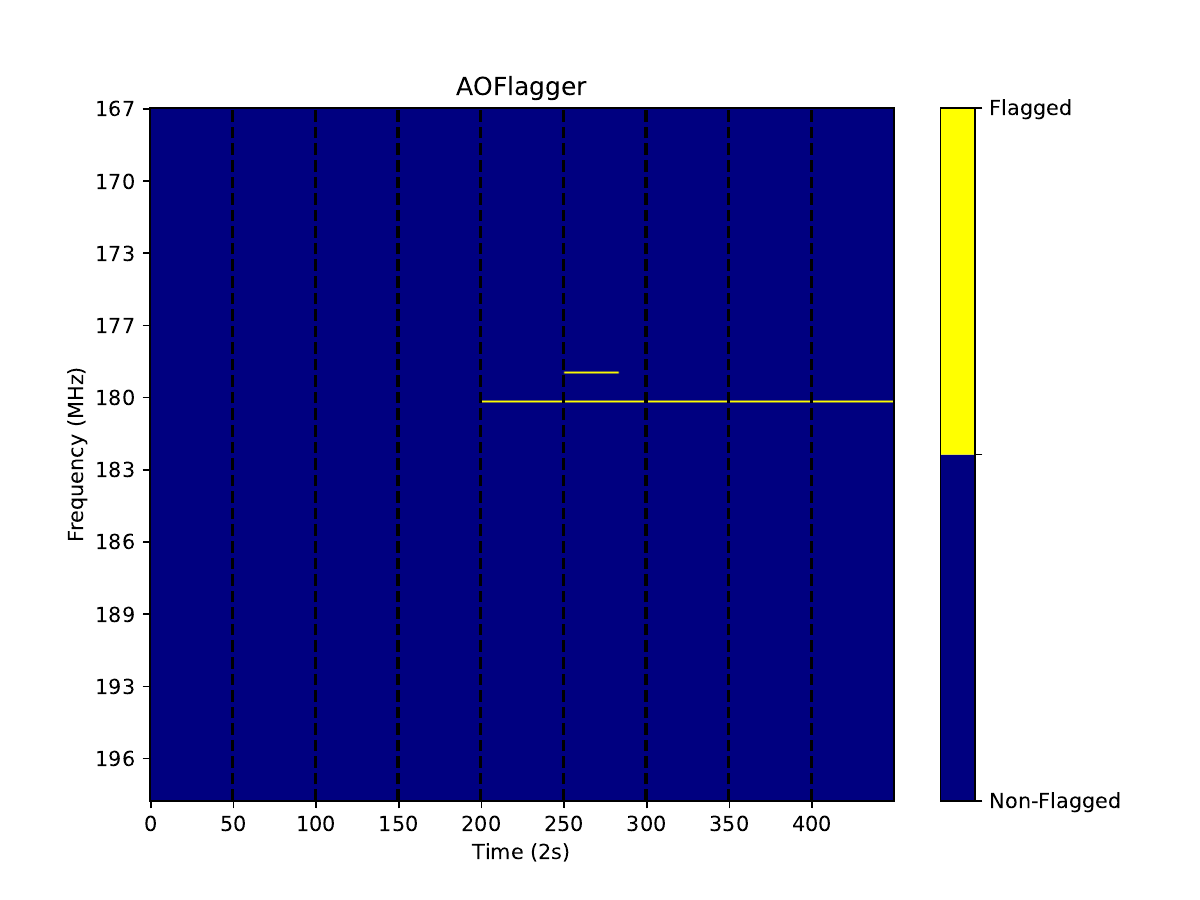}\includegraphics[scale=.27]{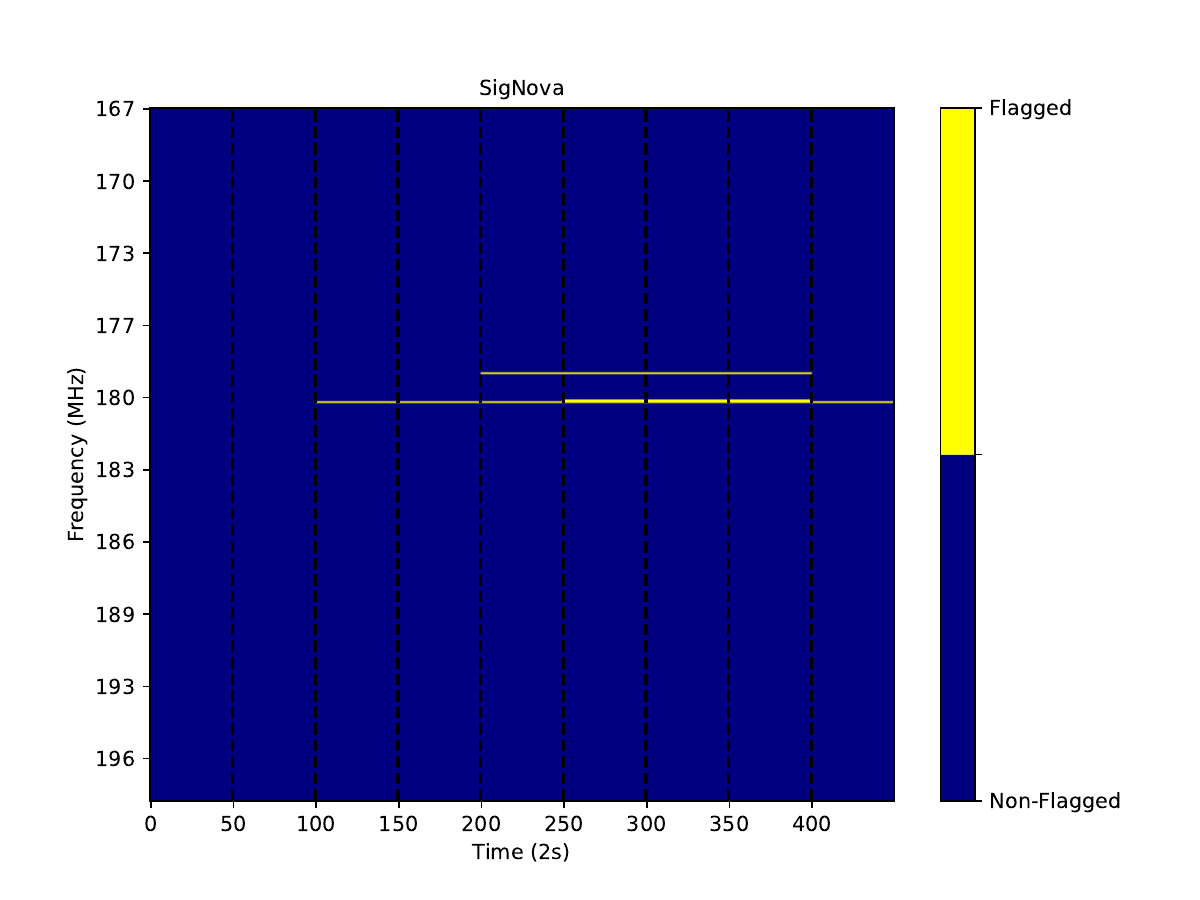}
    \caption{Extension of 9 consecutive datasets including the one shown in Figure~\ref{fig:RFI_Narrowband} from time 250 to 300. The \textsc{ssins} results are shown on the left, the \textsc{AOFlagger} one in the center, and \acrshort{name} on the right.}
    \label{fig:RFI_FullDatasets}
\end{figure*}

\hspace{0pt}\\To gain a comprehensive understanding of the onset of RFI in \Cref{fig:RFI_Narrowband}, we acquired eight adjacent and consecutive datasets and analyzed them separately. Our analysis employed the same corpus and calibration scores in each result. \Cref{fig:RFI_FullDatasets} displays the concatenation of the nine datasets, arranged chronologically, with the \textsc{ssins} result presented on the left, \textsc{AOFlagger} in the center, and \acrshort{name} on the right. Notably, the highly contaminated frequency channel identified in \Cref{fig:RFI_Narrowband} is still evident across different datasets. Examining the time-axis of the plots, we observe that \acrshort{name} detects the RFI at 110 integration times, while \textsc{ssins} detects it at 250, and \textsc{AOFlagger} at 200. This suggests that \acrshort{name} detected the incoming RFI approximately 4.6 minutes before \textsc{ssins}  and 3.8 minutes before \textsc{AOFlagger}. These results show that the RFI started as a faint signal and eluded detection by \textsc{ssins} and \textsc{AOFlagger}. We experimented with varying thresholds for \textsc{ssins} and \textsc{AOFlagger}, but none of them detected the contaminated frequency channel in the early datasets, as further studies demonstrated in the supplementary material.

\subsection{Real and Simulated Data}

We simulated clean data using $\mathsf{hera\_sim}$ tool~\cite{hera_sim}, testing it against real data from the HERA observatory. The HERA data has potential RFI interference in the averaged closure phase data, prompting further investigation. The closure phase concept~\cite{10.1093/mnras/stad371}, summing three visibility phases within an antenna triangle, was applied to eliminate gain phases. HERA closure phase data represents an average of all antenna triads, condensed into a single time-frequency spectrogram for a triad, providing information on only one antenna rather than an array. Noteworthy, this characteristic did not pose any issues for SigNova.

\begin{figure*}[t!]
    \centering
    \includegraphics[scale=0.27]{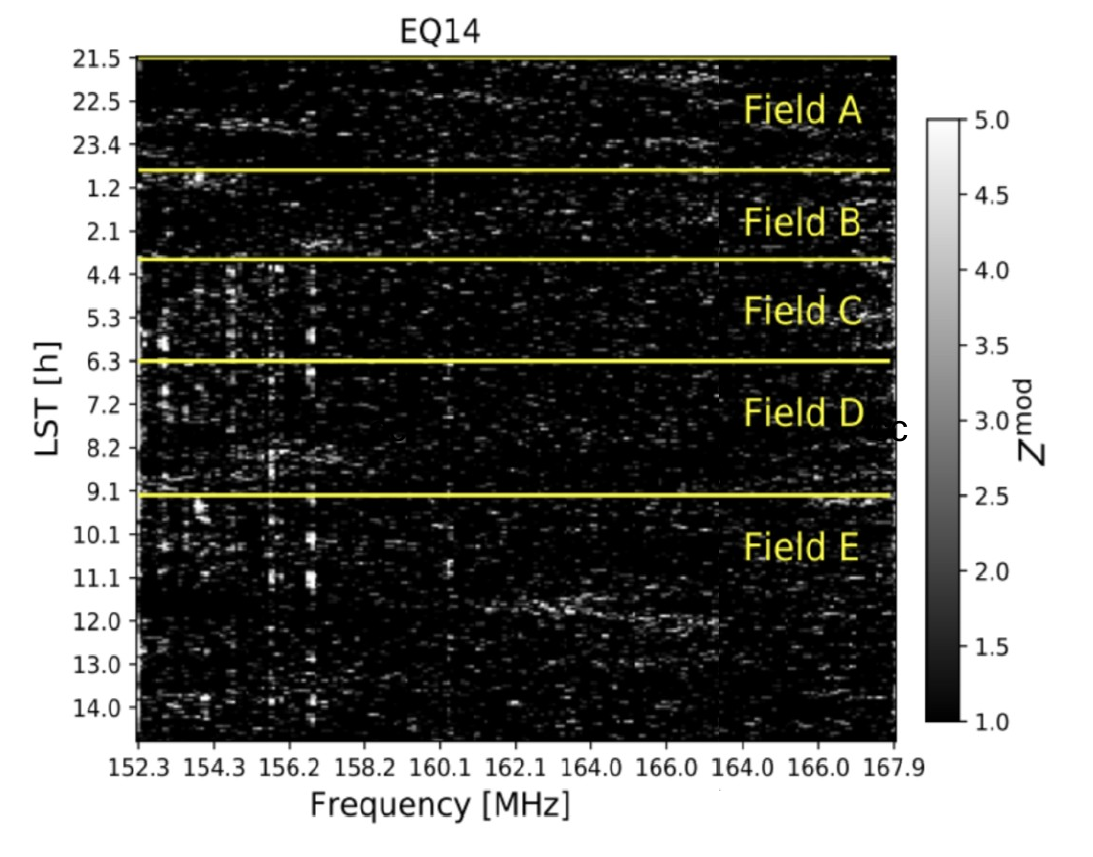}\includegraphics[scale=0.27]{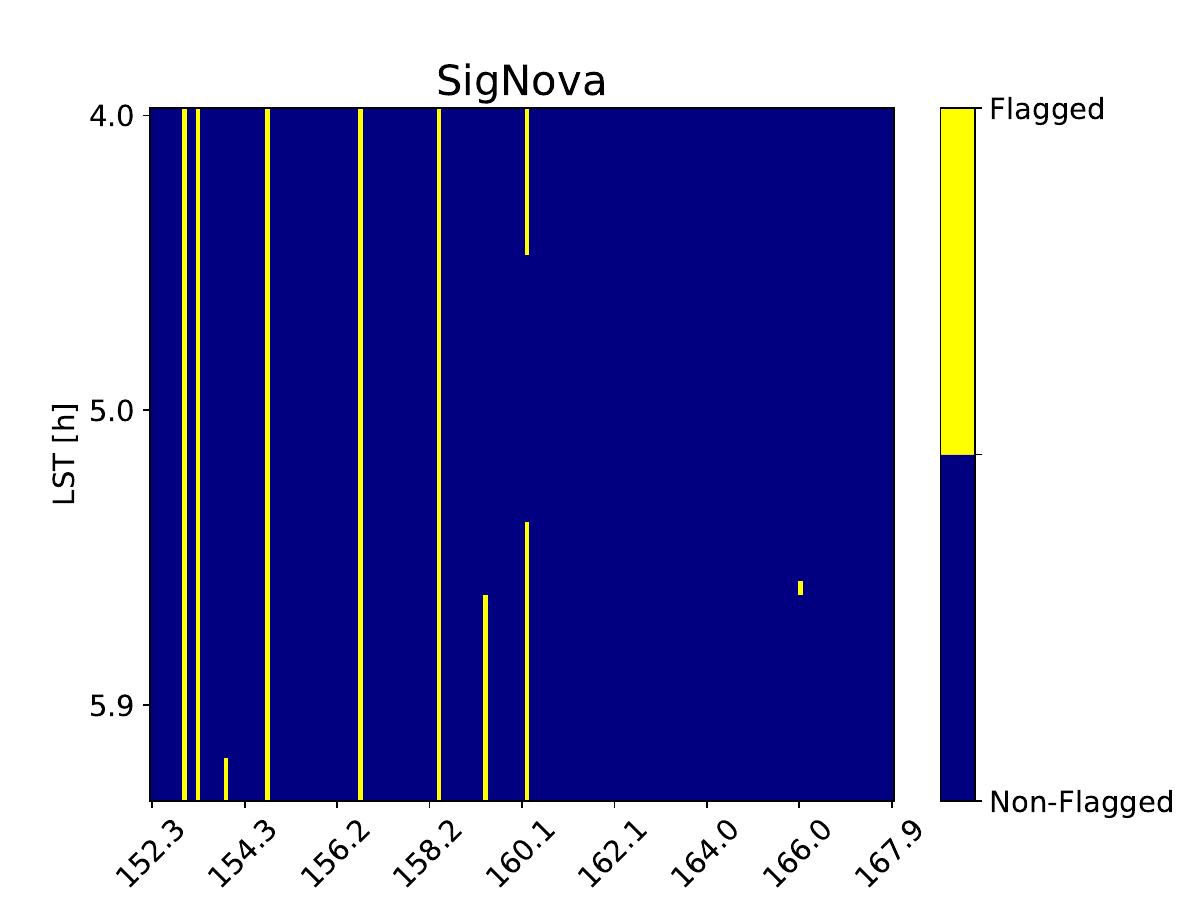}
    \caption{The left plot displays the HERA analyzed observations of five designated fields: A, B, C, D, and E. These observations span the frequency band from 152.25 to 167.97 MHz (showed with permission from Pascal Keller). On the right, SigNova's results are presented, showing the flagged RFI within field C.}
    \label{fig:HERA_Results}
\end{figure*}

\hspace{0pt}\\SigNova, without prior foreground emission studies, efficiently replicated HERA's RFI results using $\mathsf{hera\_sim}$ simulations~\cite{hera_sim}. This approach demonstrated the synergy between simulations and real data analysis, underscoring SigNova's effectiveness in identifying RFI. Notably, SigNova identified RFI in the expected frequency channels using real HERA data, shown in~\Cref{fig:HERA_Results}, but also in a very difficult area to flag around 160 MHz.

\hspace{0pt}\\To simulate HERA data, we used the $\mathsf{hera\_sim}$ tool~\cite{hera_sim}, generating a dataset featuring 61 antennas and 161 frequency channels, with the number of antennas being randomly selected. Subsequently, we tested this simulated data against real HERA data, comprising one antenna with 161 frequency channels.

\section{Discussion}\label{sec:discussion}
We present examples using both simulated and real data, demonstrating \acrshort{name}'s capability in detecting constant RFI (low and high) as well as faint RFI. Our evaluation encompassed both large and small datasets, with no issues encountered in either case. We are aware of \textsc{AOFlagger's} baseline-by-baseline approach and acknowledge that it may be preferable in certain situations. \acrshort{name} also offers this modality if the user chooses to utilize it.

\hspace{0pt}\\\textsc{AOFlagger} is highly sensitive to variations in data between neighboring channels, enabling it to effectively identify a significant amount of RFI-contaminated data. However, this sensitivity can pose a challenge when flagging MWA. In further studies, we noticed that \textsc{AOFlagger} incorrectly identifies the coarse channels as RFI. A comprehensive analysis of RFI presence requires careful consideration of the flagging that occurs every 16 frequency channels. If genuine RFI happens to occur within one of these flagged channels, \textsc{AOFlagger} may struggle to effectively detect it.

\hspace{0pt}\\This methodology exhibits versatile applications, extending to telescopes like HERA and others. Specifically, the technique is well-suited for data formats akin to visibilities~\cite{10.1093/mnras/stad371}. 


\subsection{RFI Identification Speed}
\Cref{tab:time}~shows the computation times of \acrshort{name} for training/testing one frequency channel with 50 integration times. Our framework's computational time is influenced by the number of antennas, frequency channels, and time. In the comparative analysis for generating the final plots of the previously presented examples, we assessed the runtime performance of \acrshort{name}, \textsc{AOFlagger}, and \textsc{ssins}. While \textsc{AOFlagger} exhibited slightly faster performance than \acrshort{name}, it is noteworthy that \textsc{AOFlagger} limited us to use exclusivly measurement set (MS) files. Additionally, the large size of these files posed challenges for storage, leading to multiple processing, particularly for the 9 datasets of real data results in~\Cref{fig:RFI_FullDatasets}. The additional steps involved in file management, coupled with the subsequent concatenation of results on the remote machine, significantly prolonged the time required to obtain the final results.

\hspace{0pt}\\One of \acrshort{name}'s advantages is its use of a Python pickled data format, which greatly enhances flexibility and accessibility for reading and sharing, especially when compared to the uvfits or ms data formats. Important to note that \acrshort{name} remains suitable for real-time processing and subsequent scientific studies. Once the parameters are chosen, \acrshort{name} provides reliable results with a complete plot obtained within the indicated times, requiring no further processing.

\begin{table*}[!t]
    \begin{center}
    \caption{Table with \acrshort{name} computation time for one frequency channel with 50 integration times with simulated data from~\Cref{ssec:SimulData}.}
    \label{tab:time}
    \begin{tabular}{c|c}
        \hline
       One frequency channel& \acrshort{name} \\
        \hline
        Training & 6 sec\\
        Test - With RFI & 22 sec\\
        Test- Without RFI & 6.5 sec\\        
        \hline
    \end{tabular}
    \end{center}    
\end{table*}

\section{Data Availability}\label{sec:data_availability}
The data used in this research from MWA is accessible through the Australian Square Kilometre Array (SKA) Regional Centre (ASVO) webpage. The datasets used are openly available for research purposes, and their access adheres to the principles of the UK Research and Innovation (UKRI) guidelines. To access the data, researchers can visit the ASVO webpage (\href{https://asvo.org.au}{https://asvo.org.au}) and follow the provided guidelines for data retrieval. The ID of the datasets are given in~\Cref{ssec:real_data}. It is crucial to comply with the terms and conditions outlined by the MWA and ASVO for the responsible and ethical use of the data. HERA data will soon be on the public domain.


\section{Conclusion}\label{sec:conclusion}
We presented \acrshort{name}, our anomaly detection framework that has demonstrated remarkable efficacy in identifying RFI in both real and simulated data. The modularity of our approach offers flexibility, facilitating the integration of other semi-supervised anomaly detectors once the data has been vectorized using the signature. These outcomes highlight the adaptability of our framework, representing a valuable addition to existing detection methods, and contributing to the refinement of anomaly detection in diverse datasets.

\hspace{0pt}\\We introduced a robust and versatile anomaly detection algorithm designed initially for detecting faint RFI, but its adaptability extends beyond this domain. This framework can be adapted across diverse datasets, it excels in discerning outliers by learning the characteristics of ''clean" data.



\subsection*{Acknowledgments}
We thank Sam Morley for his help with the implementation and follow-up of pysegments. For the purpose of open access, the authors have applied a Creative Commons Attribution (CC BY) license to any Accepted Manuscript version arising.

\bibliographystyle{IEEEtran}
\bibliography{references}

\appendix



\appendices



\section{Pysegments}

\begin{figure}[h]
    \centering
\includegraphics[scale=0.12]{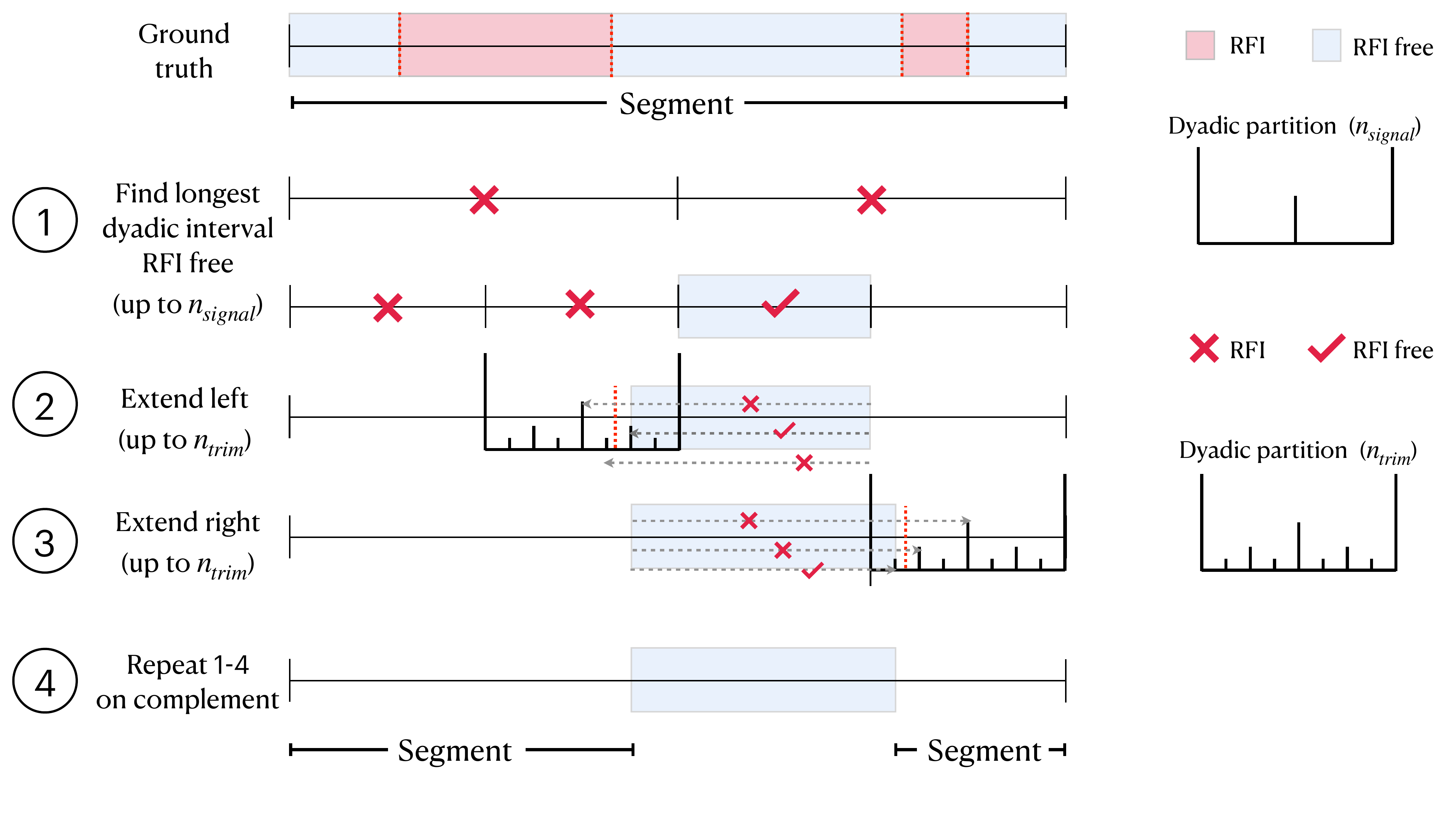}
    \caption{Schematic of the segmentation algorithm.}
    \label{fig:pysegment}
\end{figure}

\hspace{0pt}\\A schematic representation of the algorithm is shown on \Cref{fig:pysegment}.

~~~~~~~

\section{Simulated Data Different Thresholds}\label{sec:appendixSimulDataThresh}


\begin{figure*}[!t]
    \centering
    \includegraphics[scale=0.25]{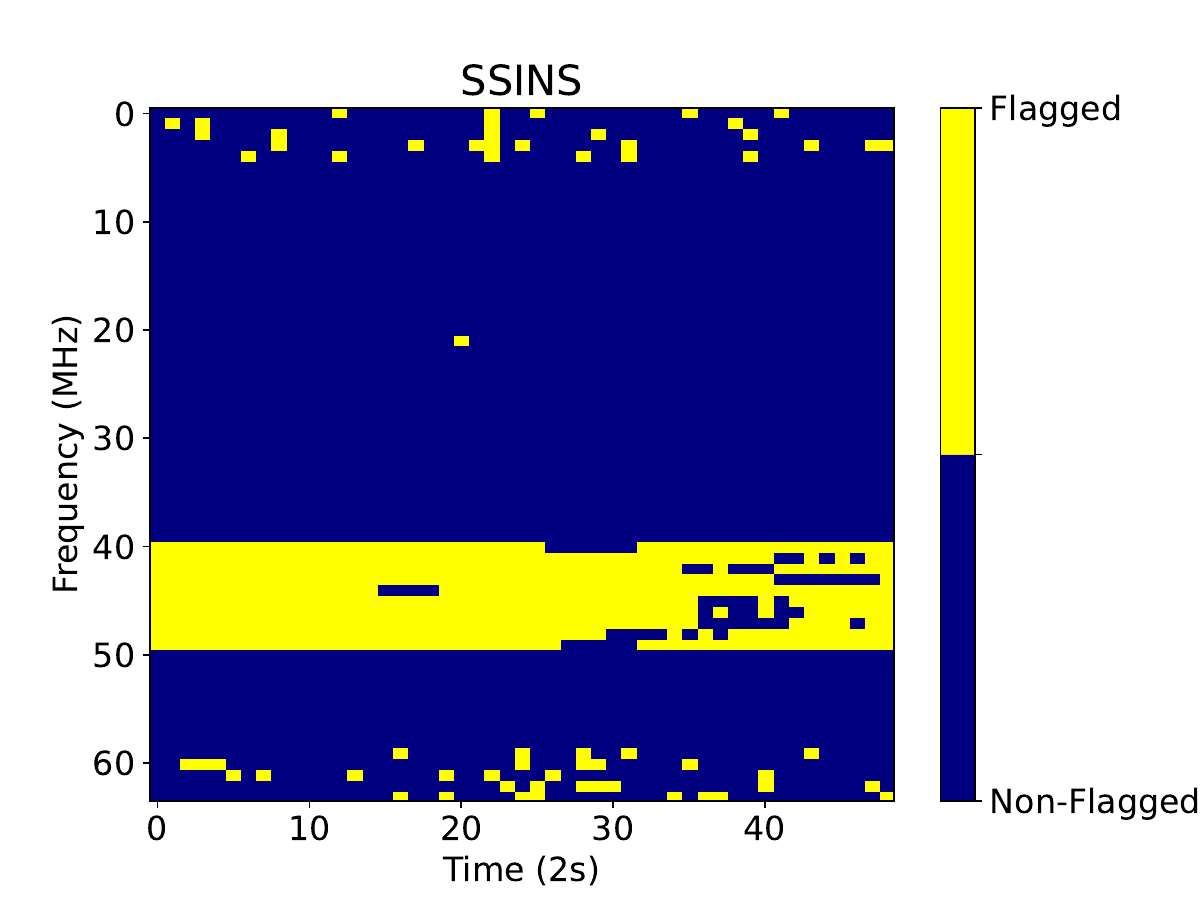}\includegraphics[scale=0.25]{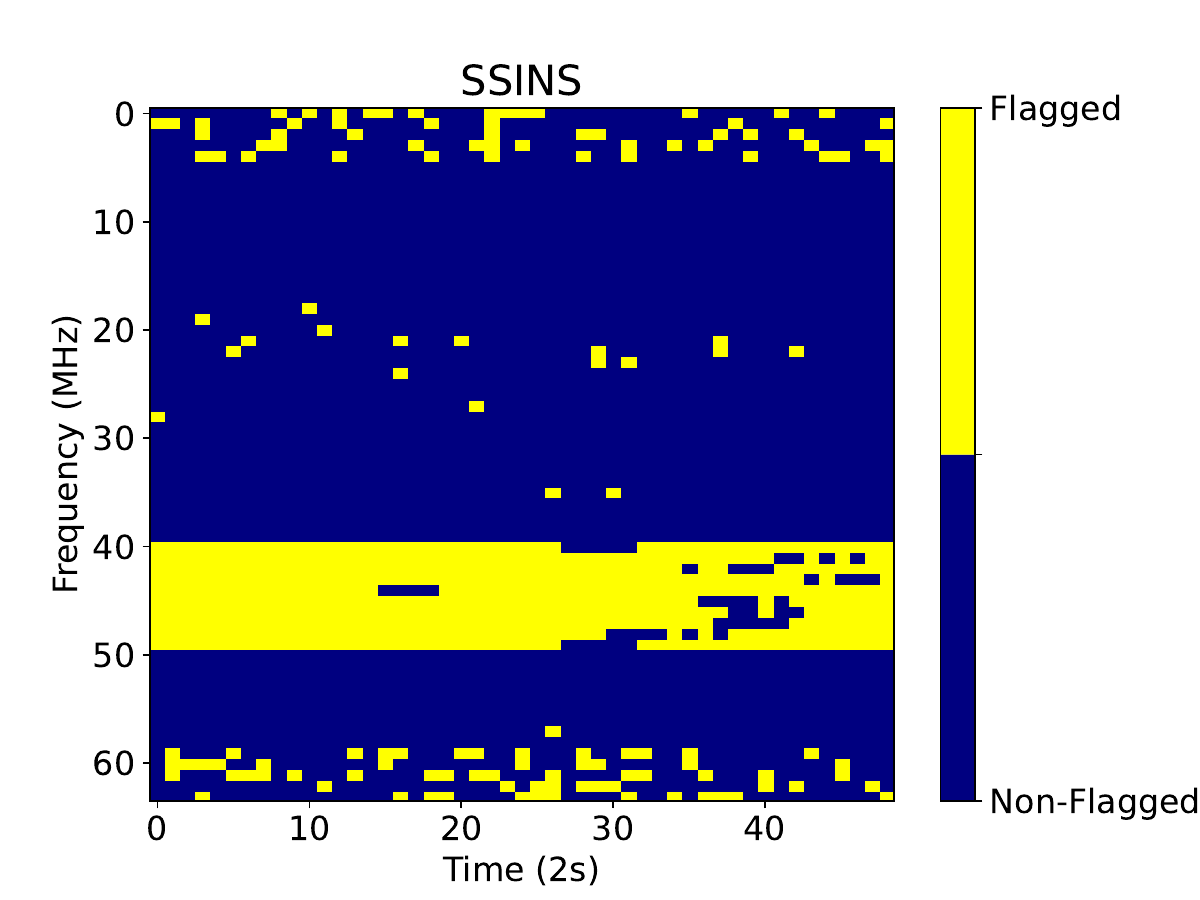}\\

    \includegraphics[scale=0.25]{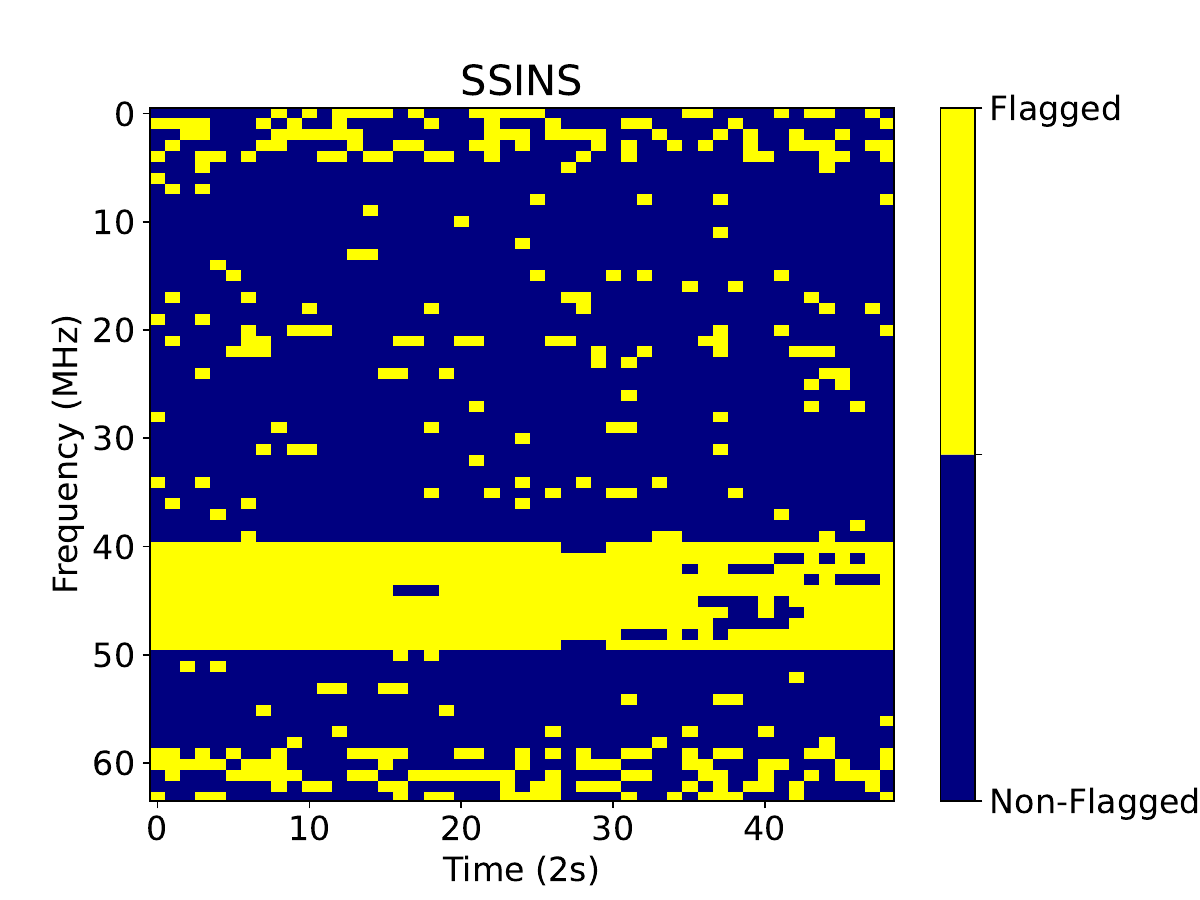}\includegraphics[scale=0.25]{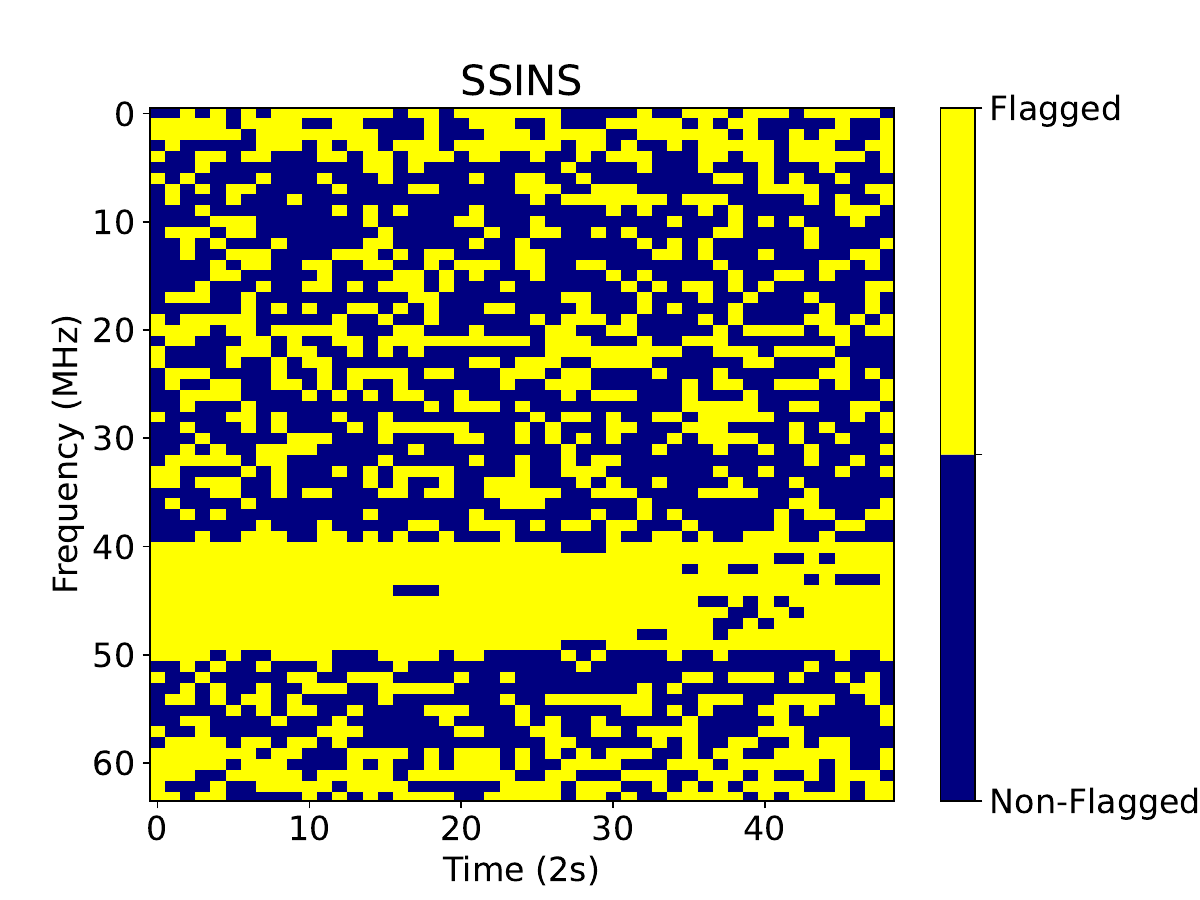}
    \caption{\textsc{ssins} output for different thresholds using simulated affecting only antenna 1 data from Figure 3 in section 4.2.1. The thresholds have been relaxed to 4 on the top left, 3 top right, 2 left bottom, and 1 right bottom. There are no clear traces of the RFI examples except for the varying over time.}
\label{fig:SSINSSimulThreshAnt1}
\end{figure*}

\begin{figure*}[!t]
    \centering
    \includegraphics[scale=0.25]{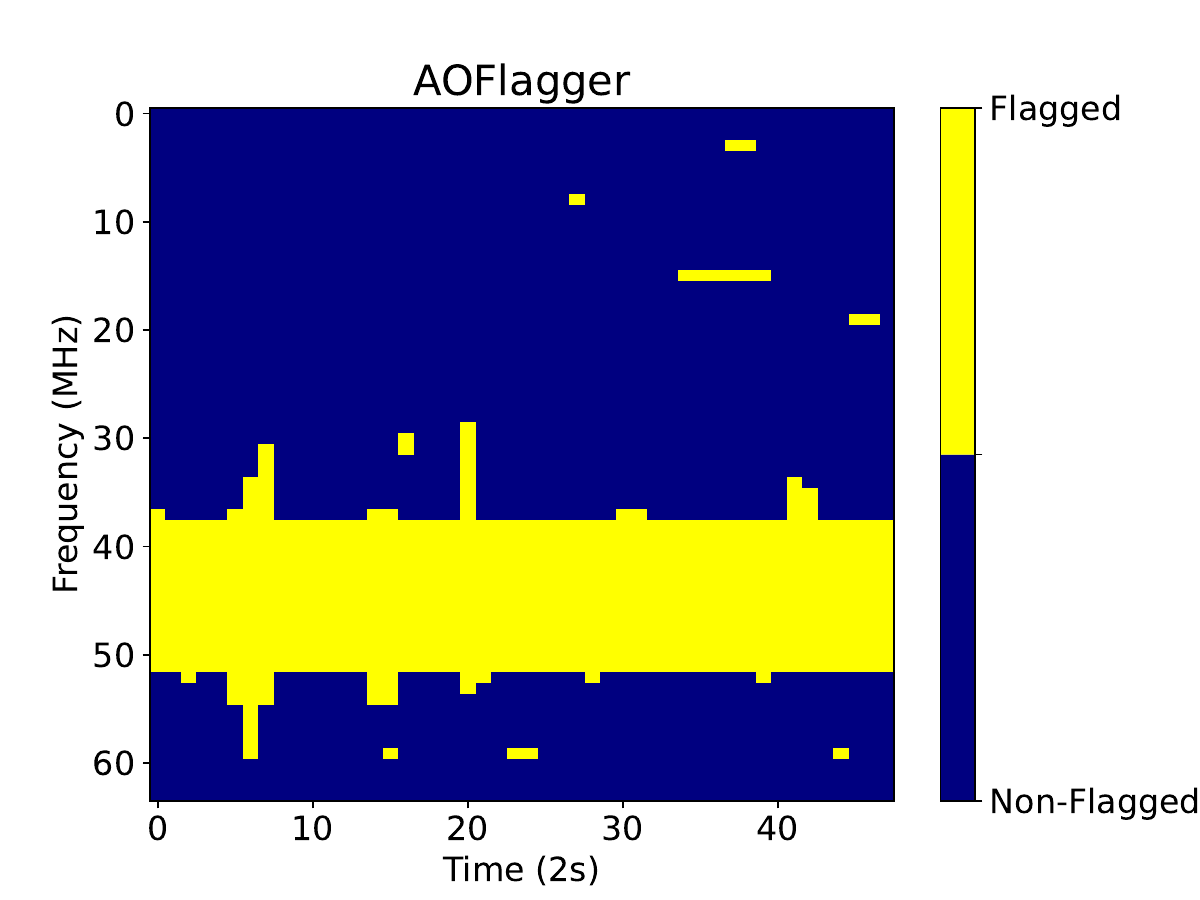}\includegraphics[scale=0.25]{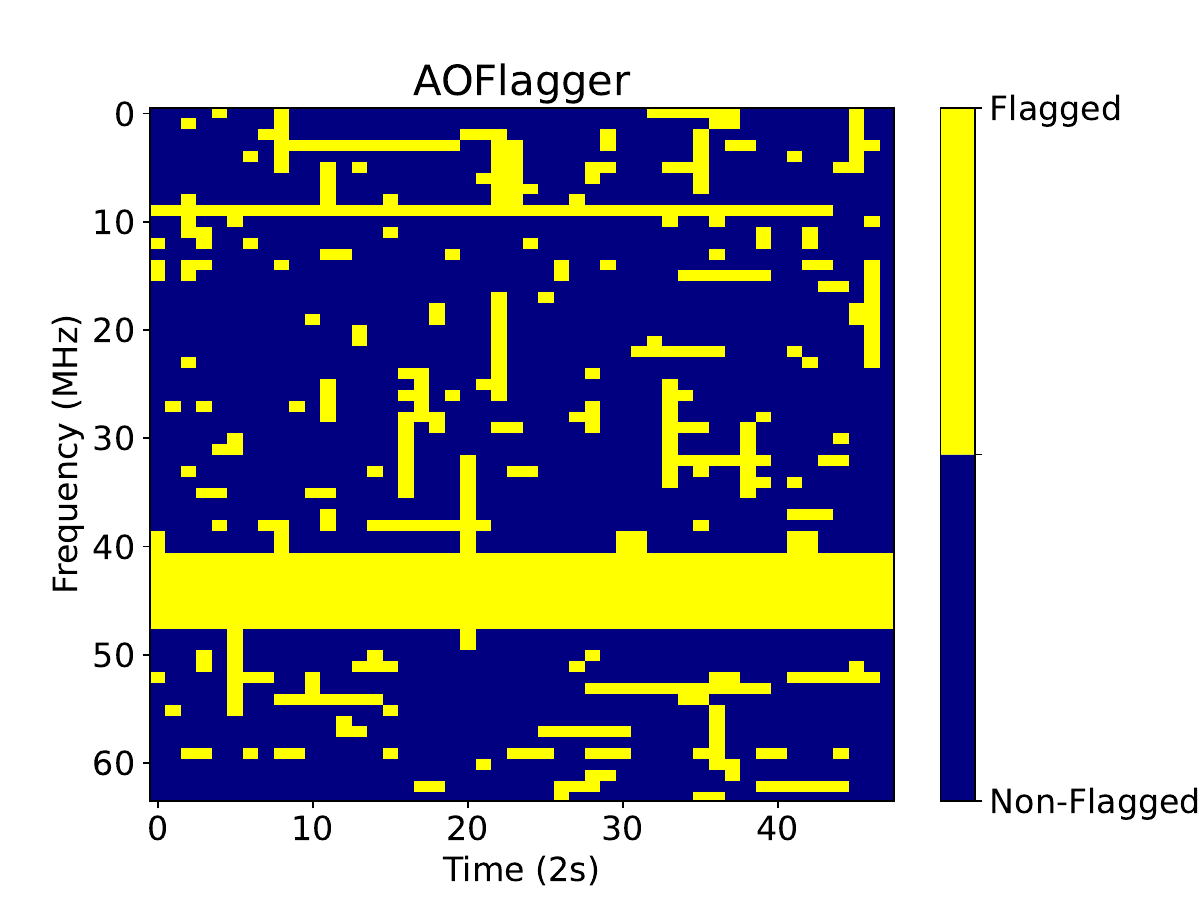}\\

    \includegraphics[scale=0.25]{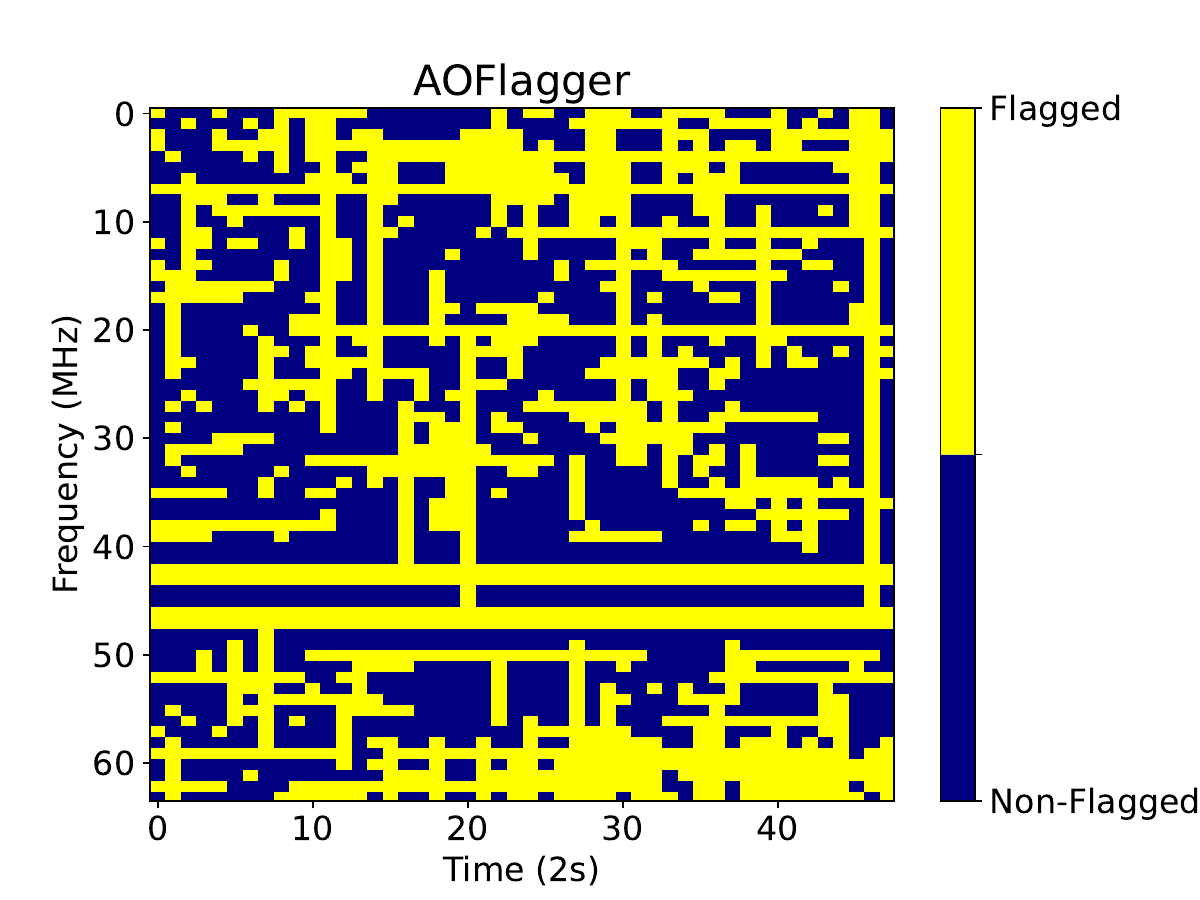}
    \caption{\textsc{AOFlagger} output for different thresholds using simulated affecting only antenna 1 data from Figure 3 in section 4.2.1. The thresholds have been relaxed to 0.8 on the top left, 0.5 top right, and 0.4 on the right bottom. There are no clear traces of the RFI examples except for the varying over time.}
\label{fig:AOFlaggerSimulThreshAnt1}
\end{figure*}


\begin{figure*}[!t]
    \centering
    \includegraphics[scale=0.25]{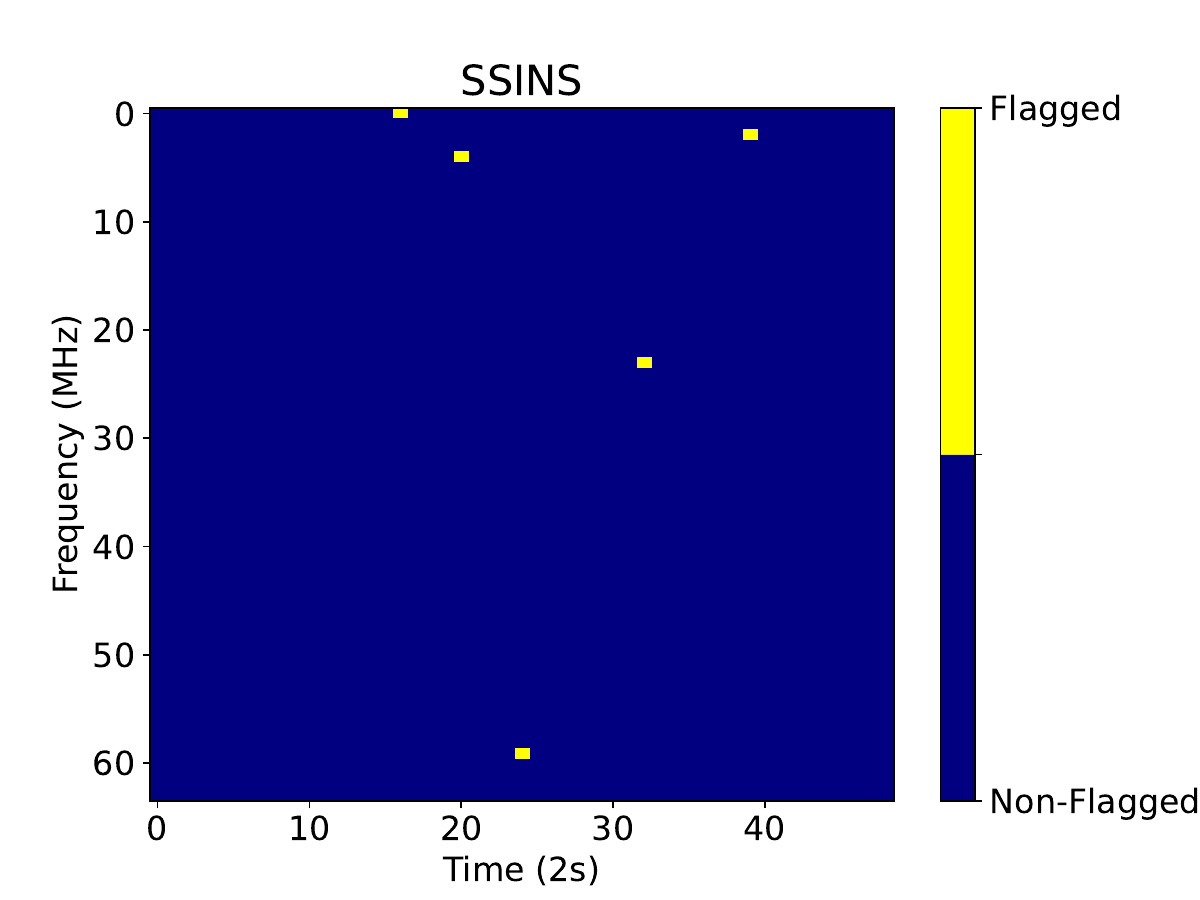}\includegraphics[scale=0.25]{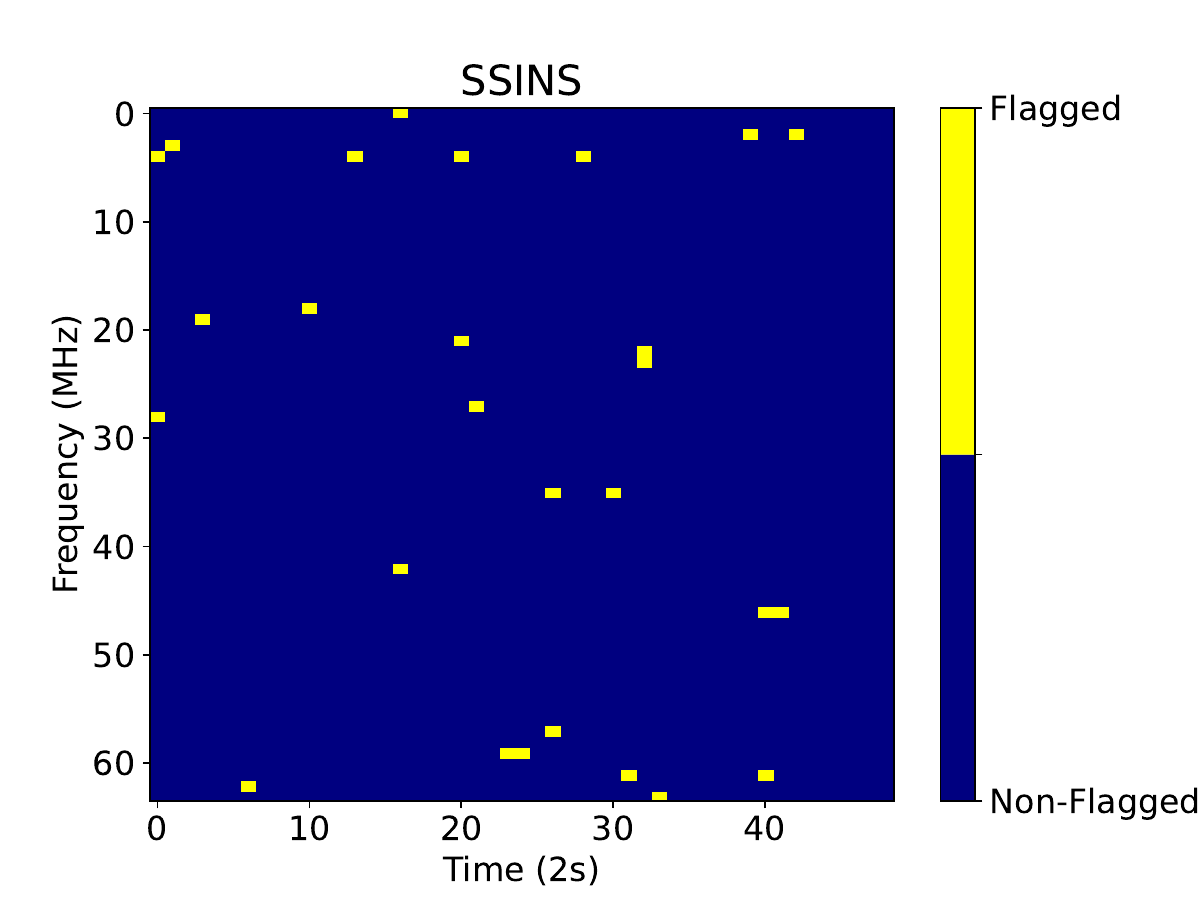}\\
    \includegraphics[scale=0.25]{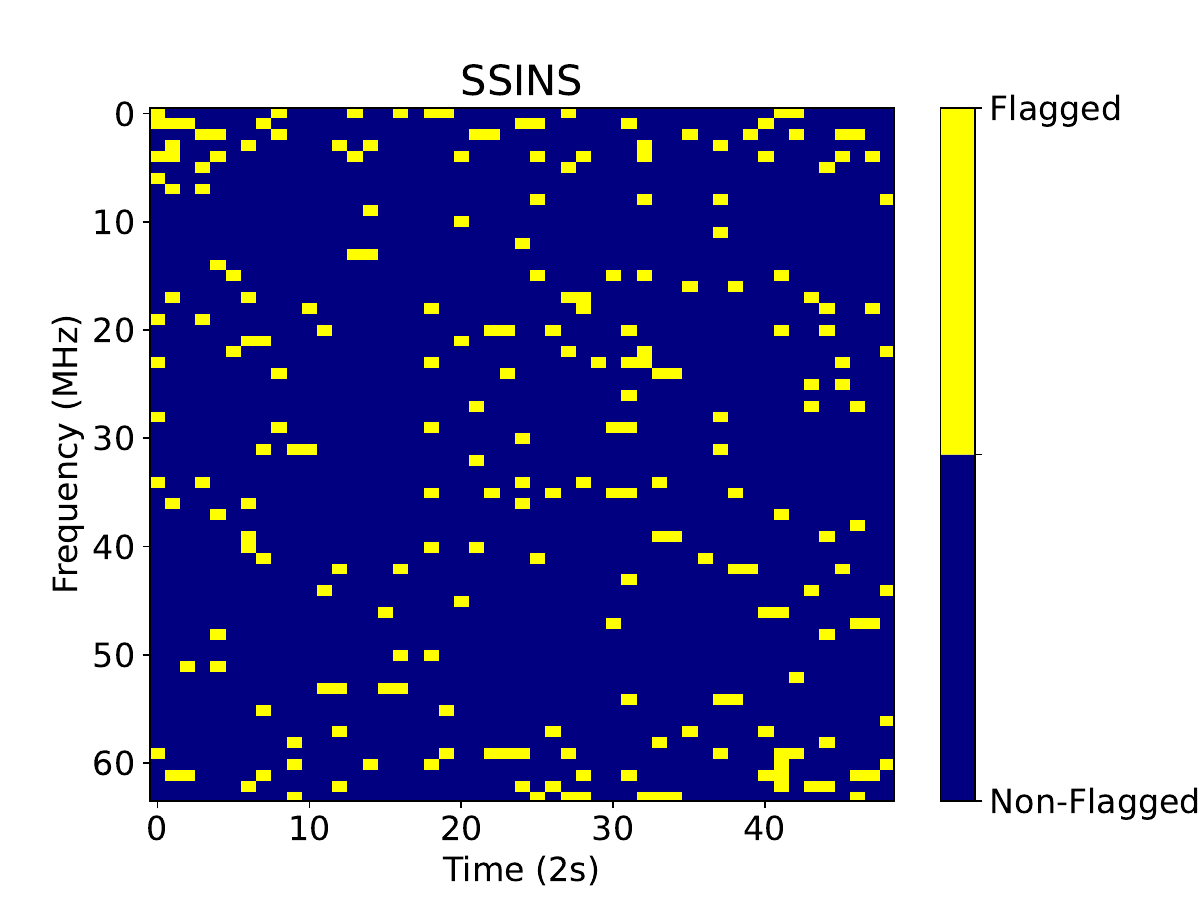}\includegraphics[scale=0.25]{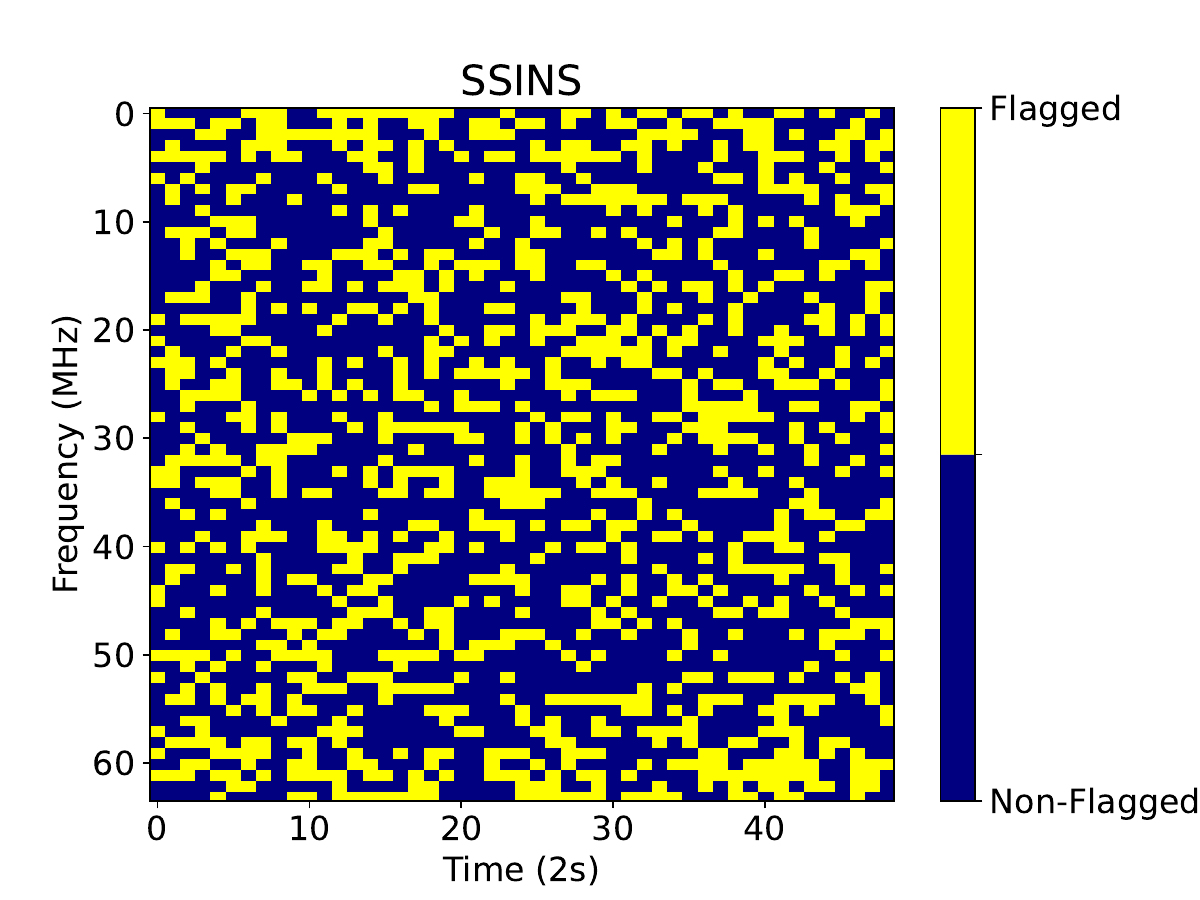}
    \caption{\textsc{ssins} output for different thresholds using simulated affecting only baseline 1 data from Figure 4 in section 4.2.1. The thresholds have been relaxed to 4 on the top left, 3 top right, 2 left bottom, and 1 right bottom. There are no clear traces of the RFI examples.}
\label{fig:SSINSSimulThreshBase1}
\end{figure*}

\begin{figure*}[!t]
    \centering
    \includegraphics[scale=0.25]{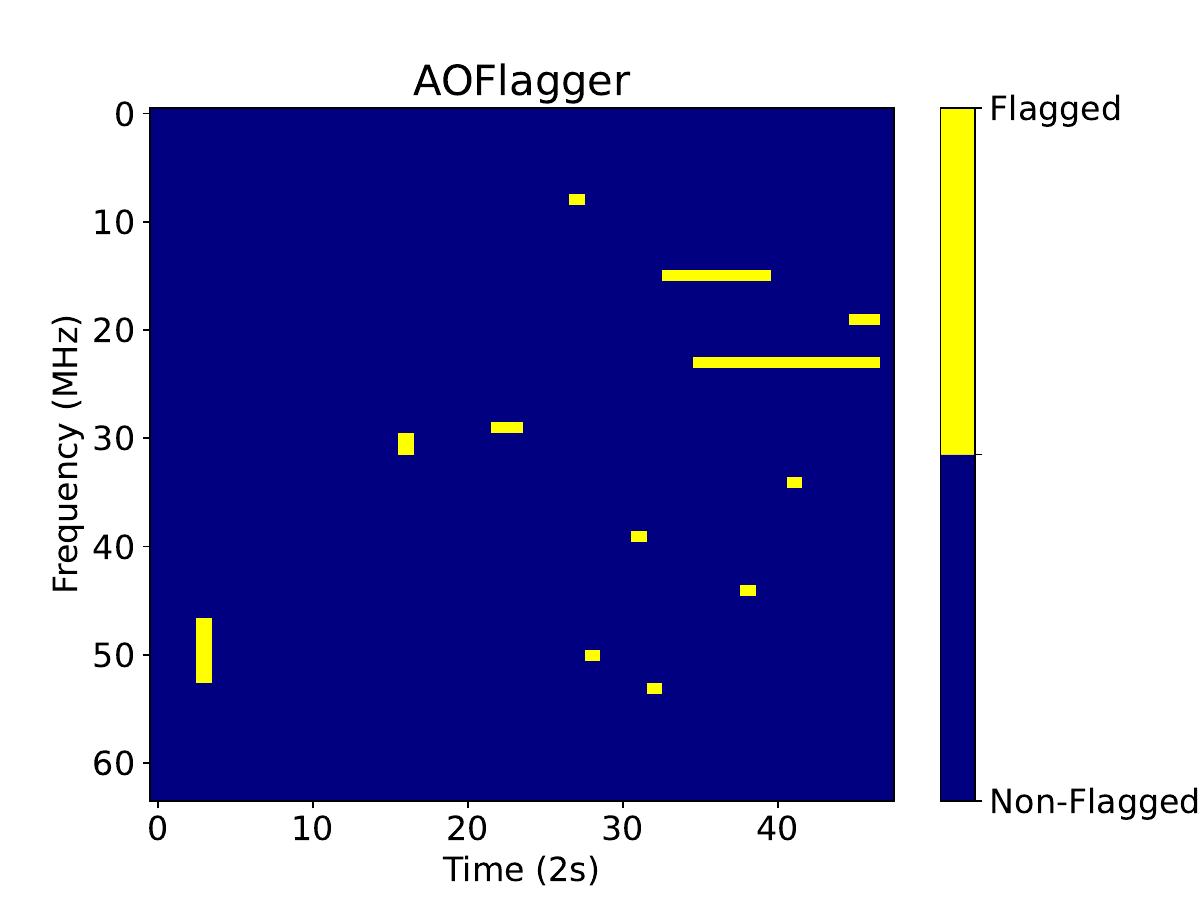}\includegraphics[scale=0.25]{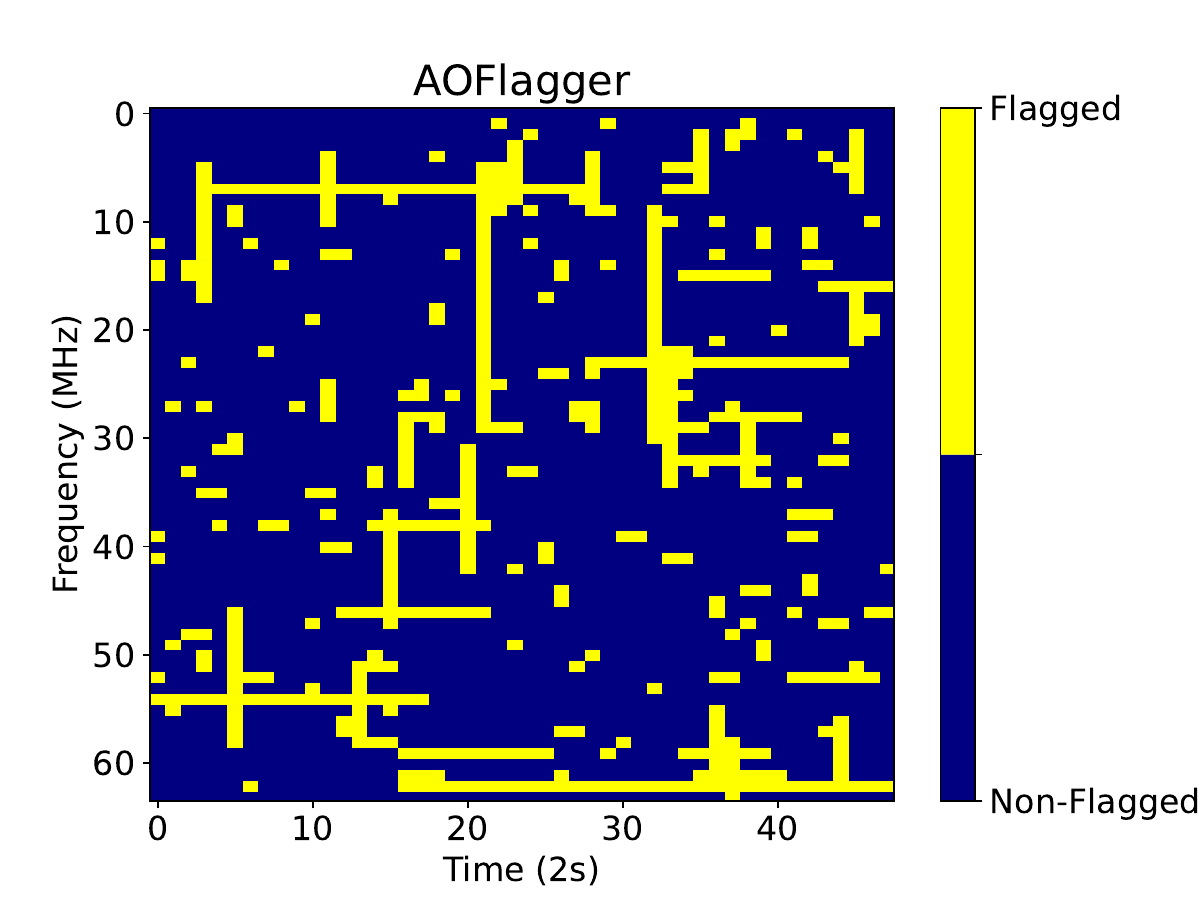}\\\includegraphics[scale=0.25]{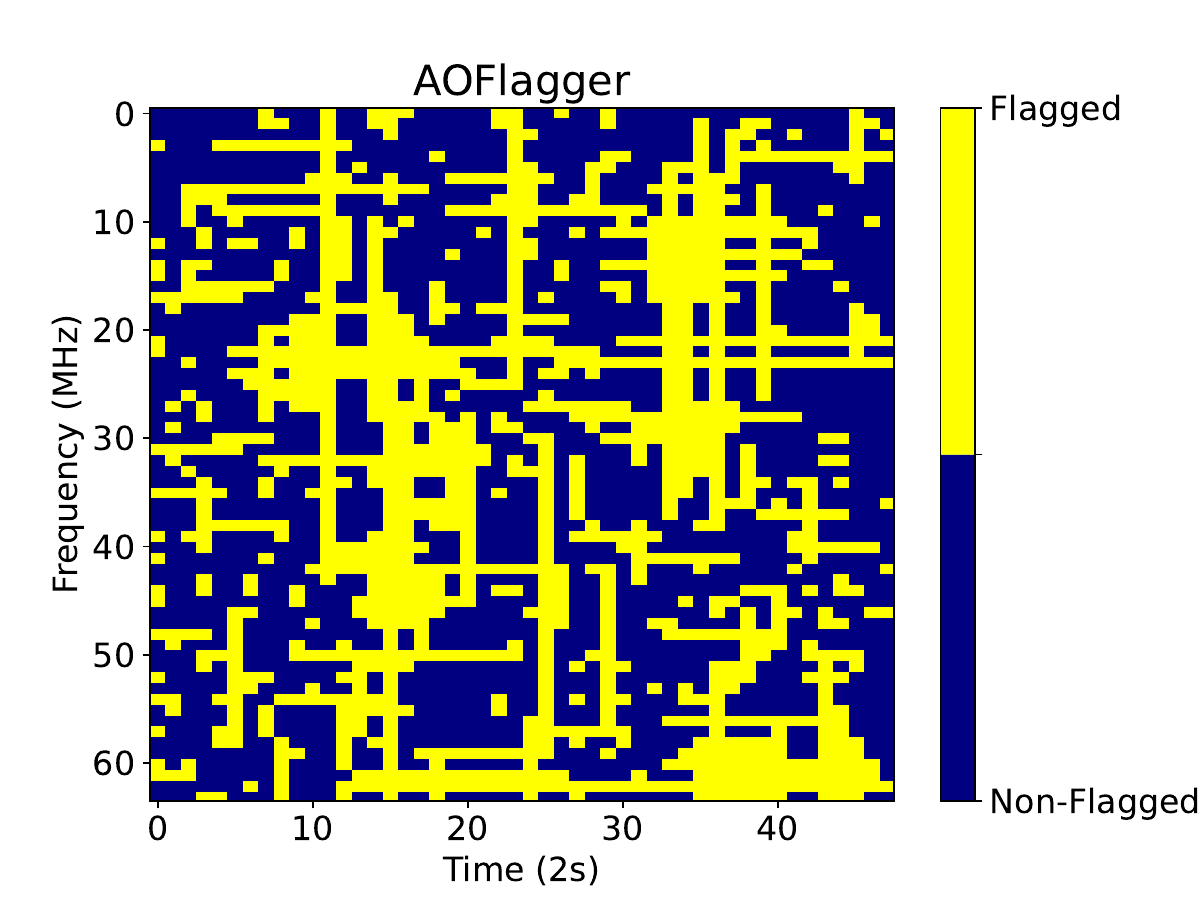}
\caption{\textsc{AOFlagger} output for different thresholds using simulated affecting only baseline 1 data from Figure 4 in section 4.2.1. The thresholds have been relaxed to 0.8 on the top left, 0.5 top right, and 0.4 on the left bottom. There are no clear traces of the RFI examples.}
\label{fig:AOFlaggerSimulThreshBase1}
\end{figure*}


\begin{figure*}[!t]
    \centering
    \includegraphics[scale=0.25]{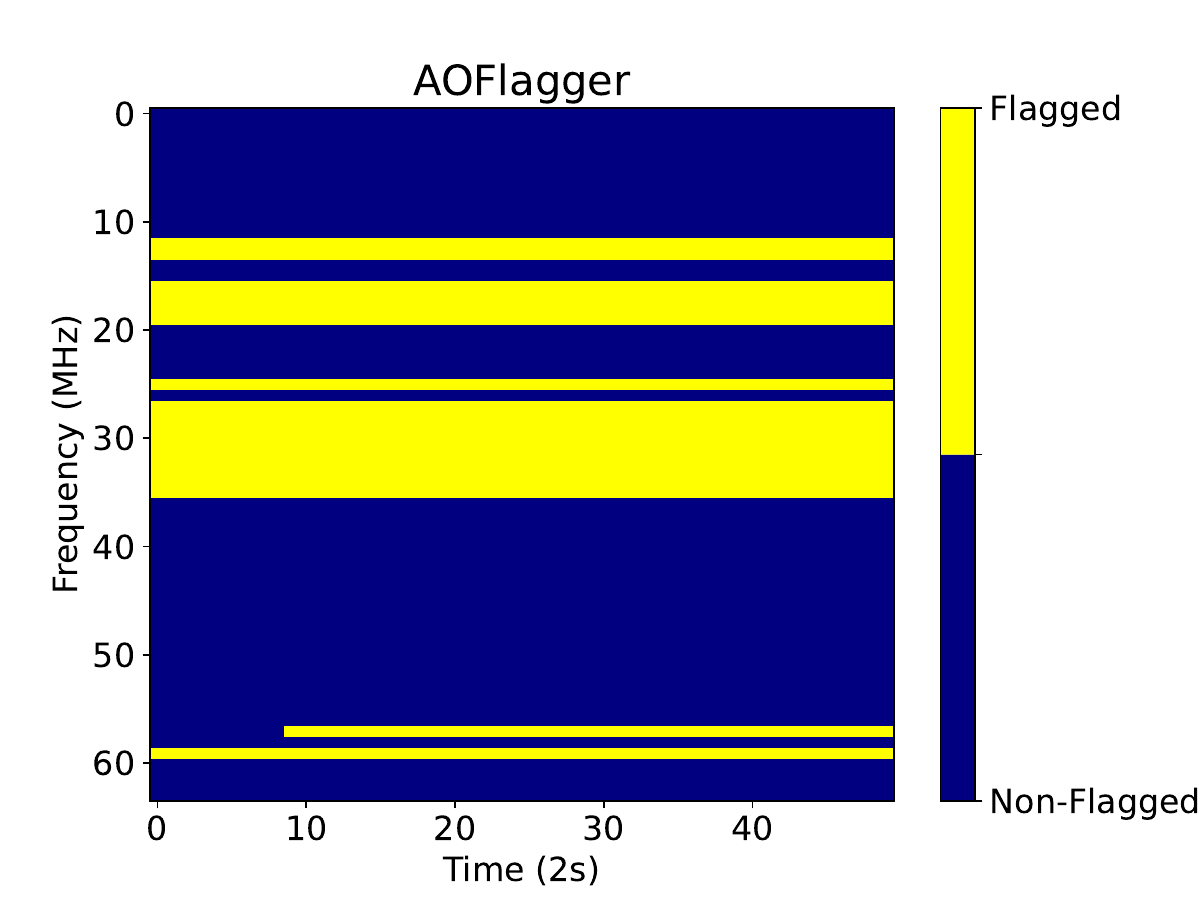}\includegraphics[scale=0.25]{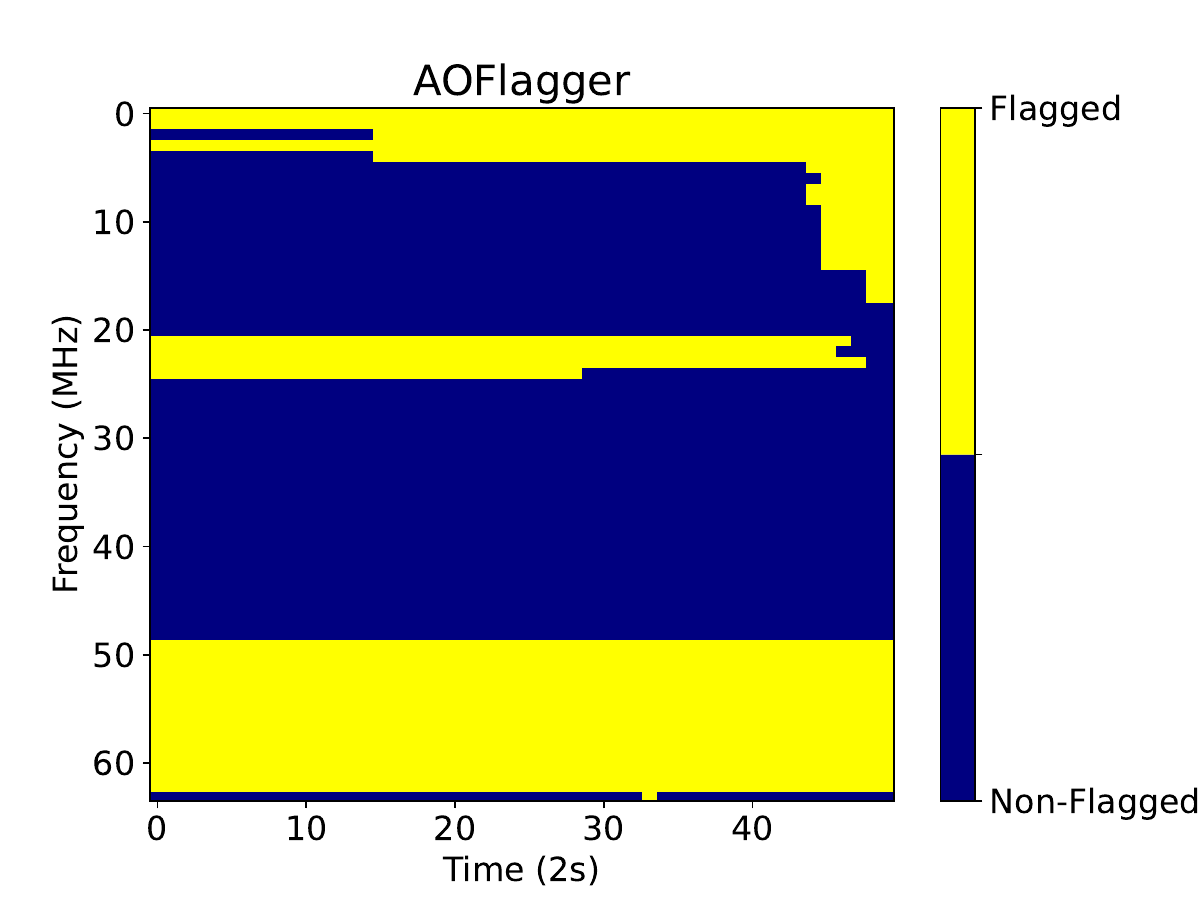}\\
    \includegraphics[scale=0.25]{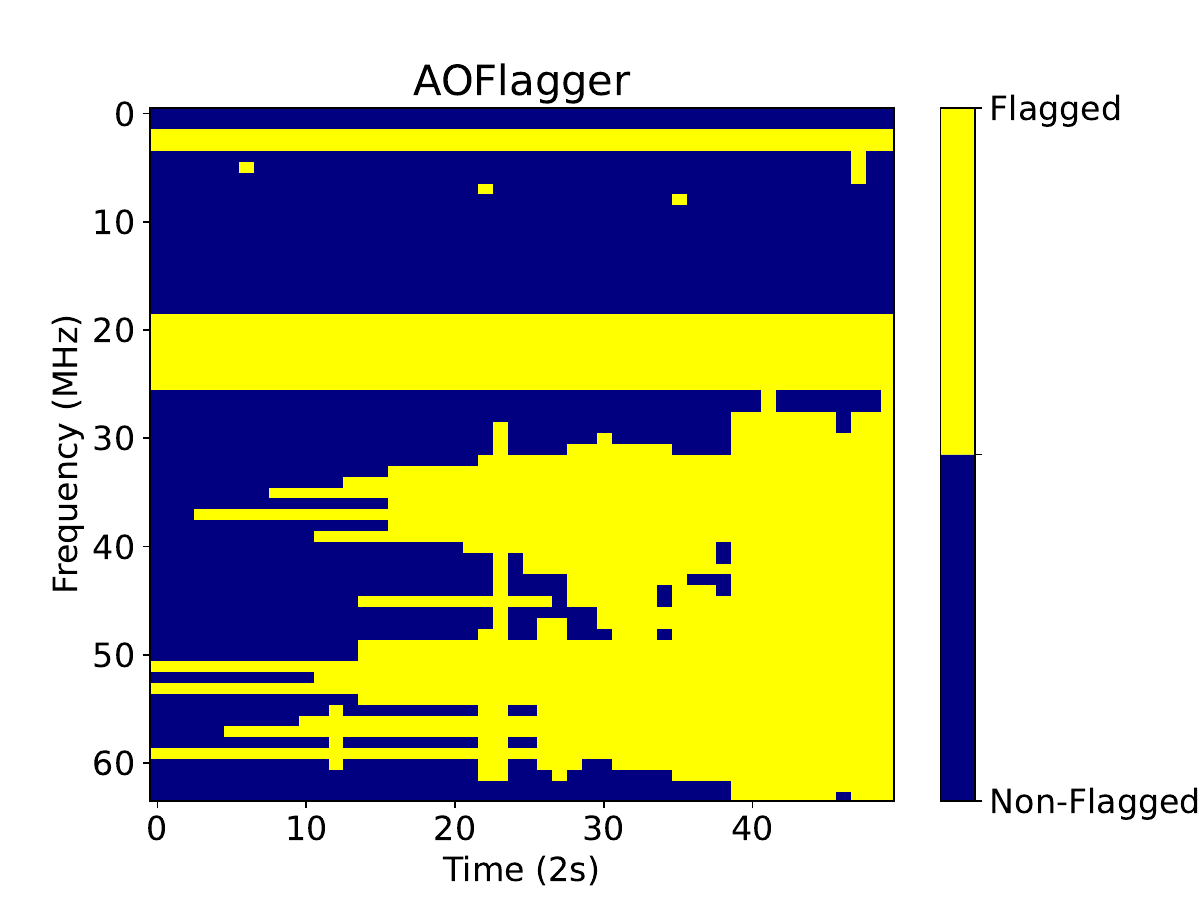}\includegraphics[scale=0.25]{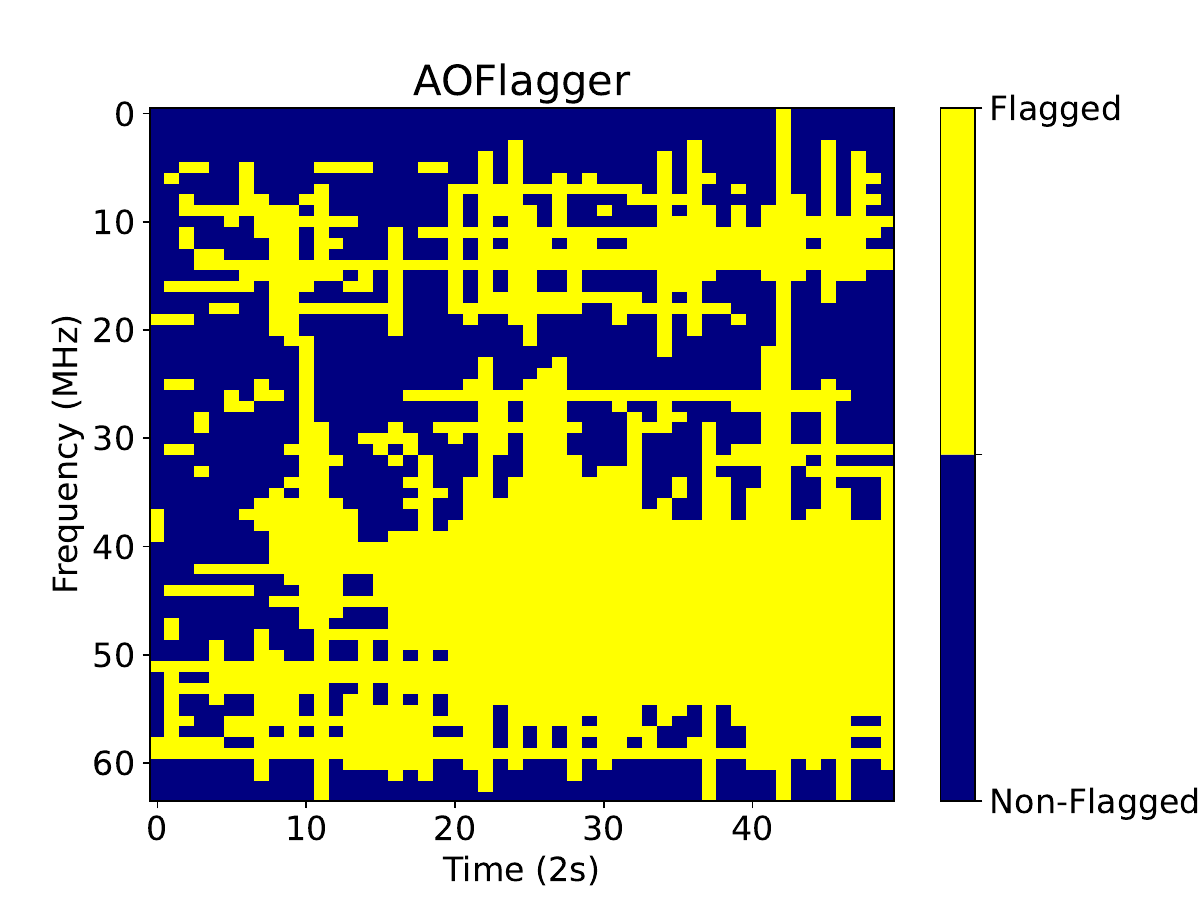}
    \caption{\textsc{AOFlagger} output for different thresholds using simulated affecting only baseline 1 data from Figure 4 in section 4.2.1. This figure shows \textsc{AOFlagger} baseline-per-baseline modality, showing the results only for baseline 1 (the manually contaminated one). The thresholds have been set to 1 on the top left, 0.9 top right, 0.5 on the left bottom, and 0.4 on the right bottom. The results wrongly flaggs the frequency channels containing the RFI.}
\label{fig:AOFlaggerSimulThreshBase1OnlyBase1}
\end{figure*}

\section{Clean Data Studies}\label{sec:CleanData}
\hspace{0pt}\\We present verified studies that the ``clean'' real data used in SIgNova made no reference to RFI. We randomly subsampled the real ``clean" data, which was initially selected as the \textit{corpus}, in multiple iterations as outlined in here. This approach ensures the integrity of the \textit{corpus} data, as we observed that across different iterations, the maximum score obtained was minimal and comparable to the thermal noise expectation.

\hspace{0pt}\\The dataset used for our studies consisted of 242 antennas with 380 frequency channels for the \textit{corpus} and 128 antennas for the \textit{calibration}. To ensure robustness in our analysis, we adopted a methodology where we fixed the \textit{corpus} by selecting a random 90\% of the clean data, reserving the remaining 10\% for testing purposes. The \textit{calibration} dataset remained fixed in every iteration.

\hspace{0pt}\\We set a distfit threshold at 0.005, equivalent to the 0.5th percentile, indicating that only 0.5\% of the data points fell below this threshold. In~\Cref{tab:cleanstudies1}, we present the results for a specific frequency channel (number 100), showing the number of test scores that exceeded the threshold established by the \textit{calibration} set. These results provide insights into the ``cleanness" of the data used in the real data section. During the course of this study, we consistently noticed that the limited ``noisy" data primarily originated from specific antennas, as evidenced by the highest test scores listed in~\Cref{tab:cleanstudies1}. Antenna 4, in particular, was a recurring presence in the majority of tests, accompanied by antennas 20, 110, and 240, albeit to a lesser extent. This phenomenon can be attributed to the fact that when partitioning the data, some very localized effects on a specific antenna data become stronger.

  \begin{table*}
    \centering  
    \resizebox{\textwidth}{!}{%
    \begin{tabular}{cc|cccccccccc|c}
    Datasets & Interval Length & Test1 & Test2 & Test3&Test4&Test5&Test6&Test7&Test8&Test9&Test10&Mean\\
       \hline
        C: 90\%, In: Fixed, Test: 10\%  & 32 & 17/225 & 15/230 & 20/232 & 18/227 & 21/227 & 16/234 & 20/229 & 19/227 & 22/228 & 11/232 & \\
       & & 7\% & 6\% & 8\% & 8\% & 9\% & 7\% & 9\% & 8\% & 10\% & 5\% & 7.7\%\\
       & Ant1 $>$ thresh &4,20&4,20&4,20&4&110,211,240&110&4,20,41&4&110,211,240&4,20&\\
       \hline

C: 70\%, In: Fixed, Test: 30\% & 32 & 3/242 & 1/244 & 1/243 & 2/241 & 6/242 & 2/244 & 4/243 &2/247 &2/246 &2/248 &\\
& & 1\% & 0.4\% & 0.4\% & 0.8\% & 2\% & 0.8 \% & 2\% & 0.8\% & 0.8\% & 0.8\% & 0.98\% \\
& Ant1 $>$ thresh & 4,20&4&20 &4 & 4,240 &4&4,20&4&4,240 &4,20&\\

    \end{tabular}
    }
    \caption{Clean data studies.}
    \label{tab:cleanstudies1}
\end{table*} 

\hspace{0pt}\\When we shift to using the some random percentage of the corpus, with the remaining fraction serving as the \textit{calibration} set while keeping the \textit{test} data fixed, we consistently observed clean results. In this context, `clean' denotes that none of the \textit{test} scores exceeded the \textit{inlier} threshold.

\hspace{0pt}\\In the final round of our studies, we adopted a different approach. We randomly selected 90\% of the clean data to form the \textit{corpus}. For the \textit{inlier} dataset, which we utilized in the real data section and confirmed to be clean, we divided it randomly into two equal parts: one served as the fixed \textit{inlier}, and the other as the fixed \textit{test} set. Notably, across all ten iterations, our results remained consistently ``clean". This outcome reaffirmed the cleanliness of our real data, free from any radio frequency interference (RFI).

\section{Real Data Different Thresholds}\label{sec:RealDataThresh}

\begin{figure*}[!t]
    \centering
    \includegraphics[scale=0.25]{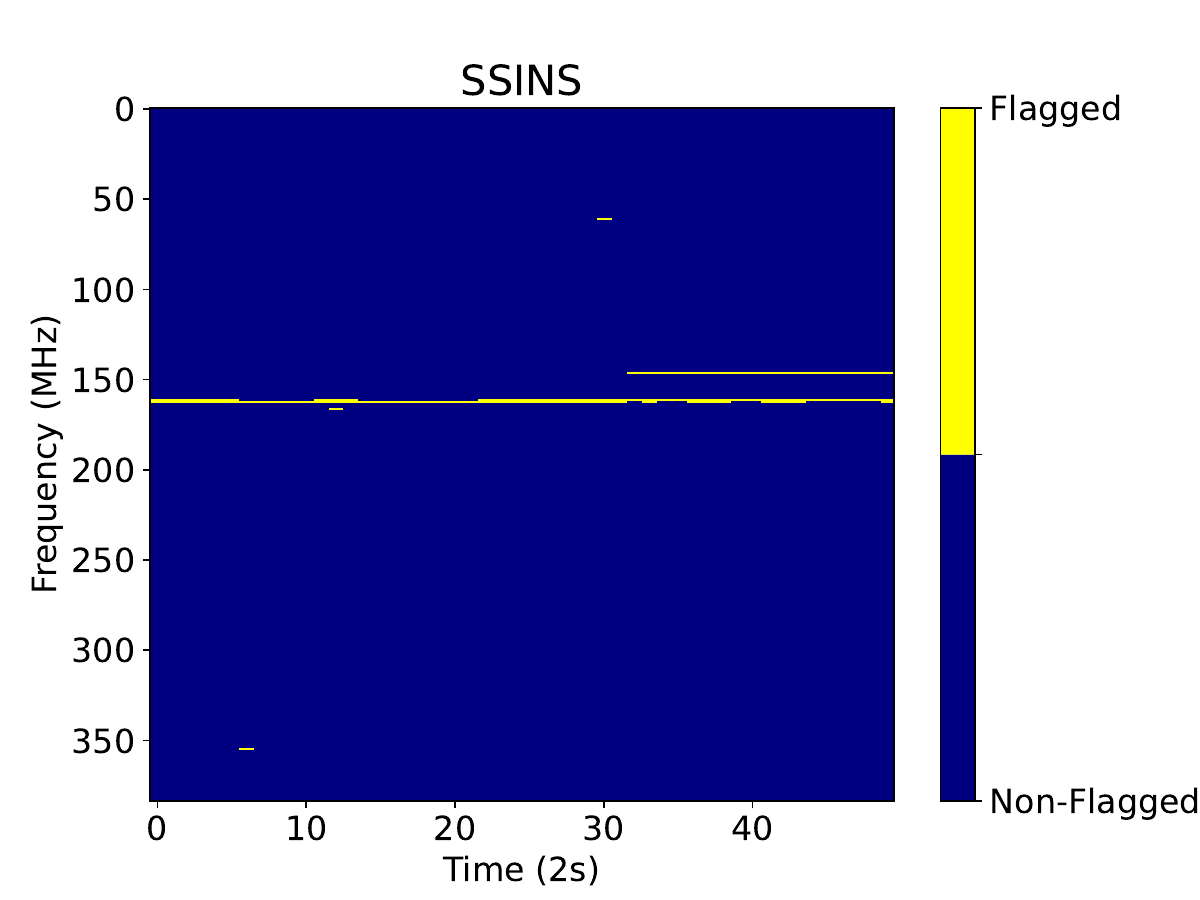}\includegraphics[scale=0.25]{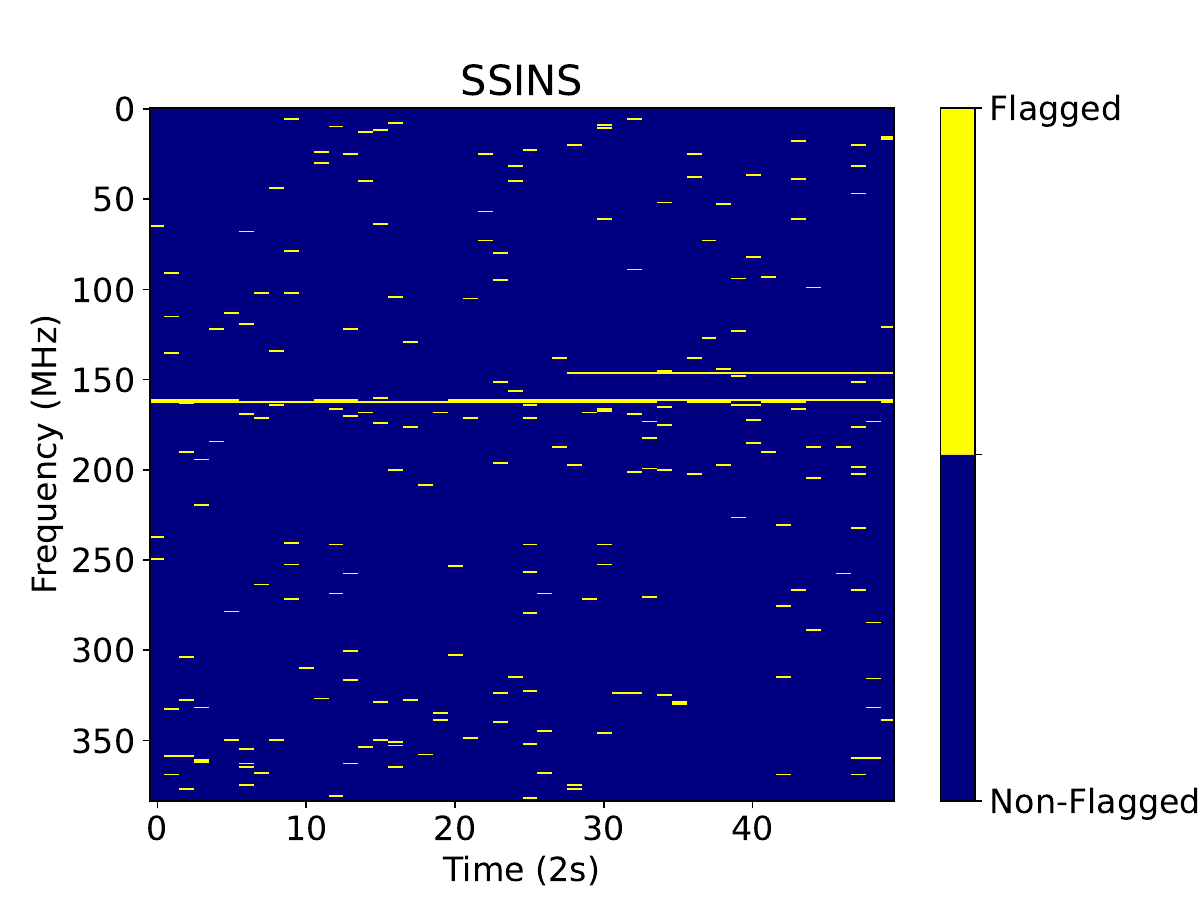}\\
    \includegraphics[scale=0.25]{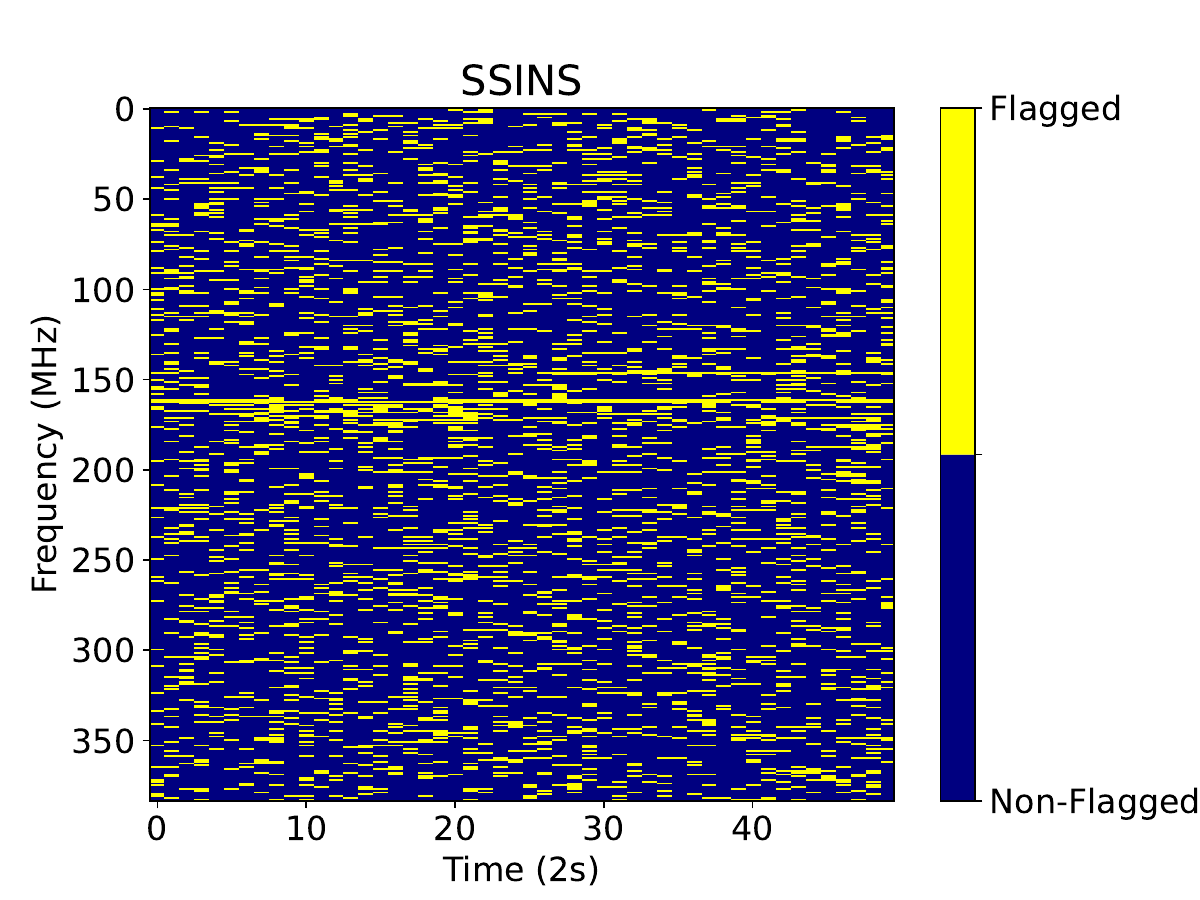}\includegraphics[scale=0.25]{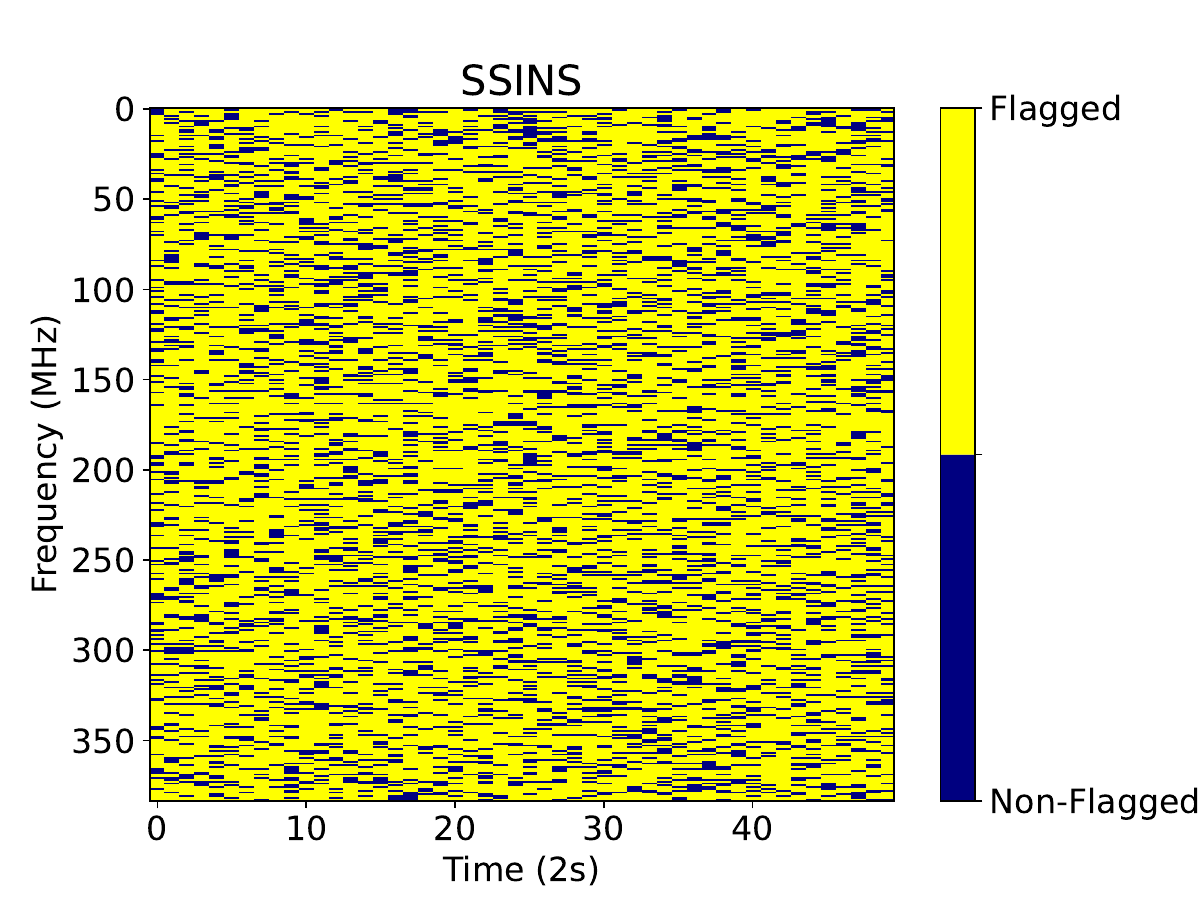}
\caption{\textsc{ssins} different thresholds for RFI narrowband MWA data example. Top left: threshold 4, top right: threshold 3, bottom left: threshold 2, bottom right: threshold 1. Even relaxing the thresholds the faint RFI frequency channel fails to get completely flagged and noise start to show.}
    \label{fig:SSINSNarrowbandThres}
\end{figure*}
\begin{figure*}[!t]
    \centering
    \includegraphics[scale=0.25]{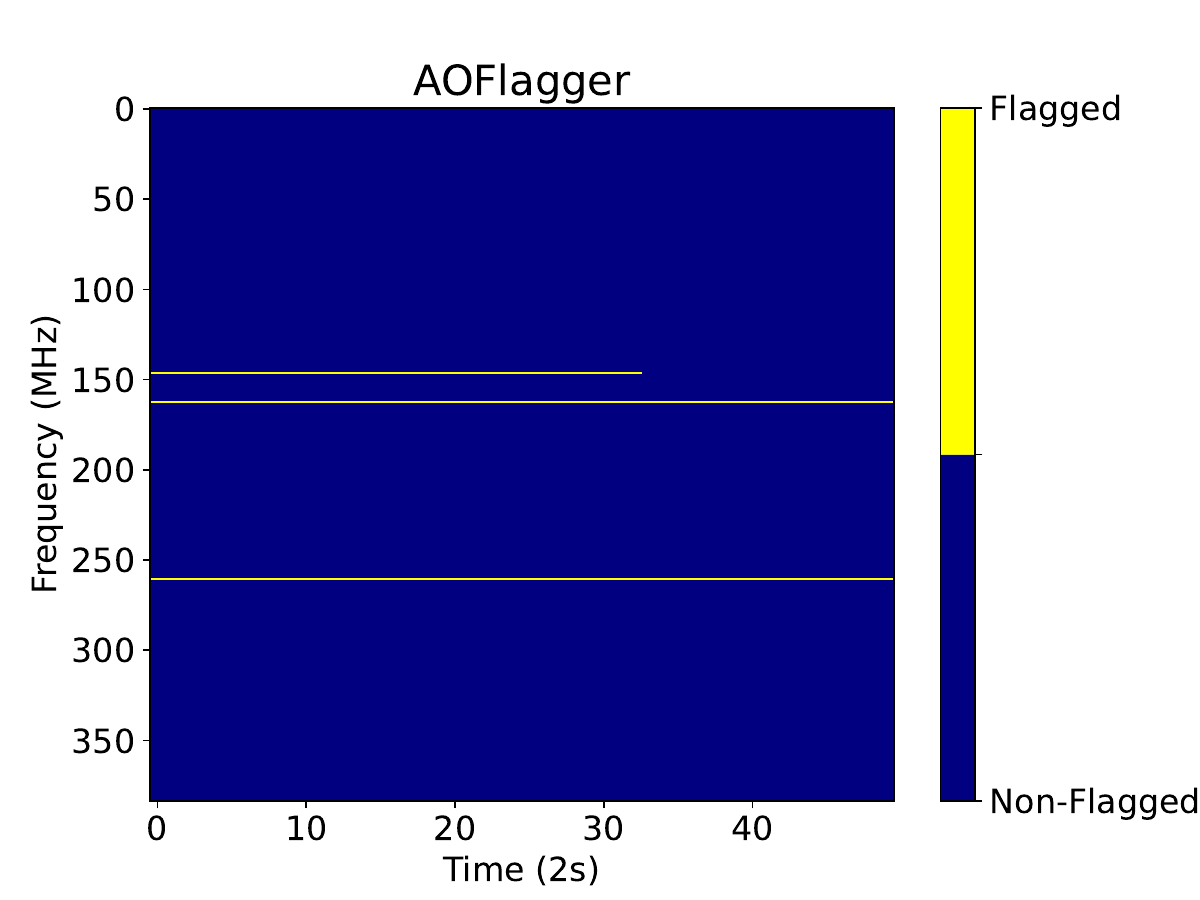}\includegraphics[scale=0.25]{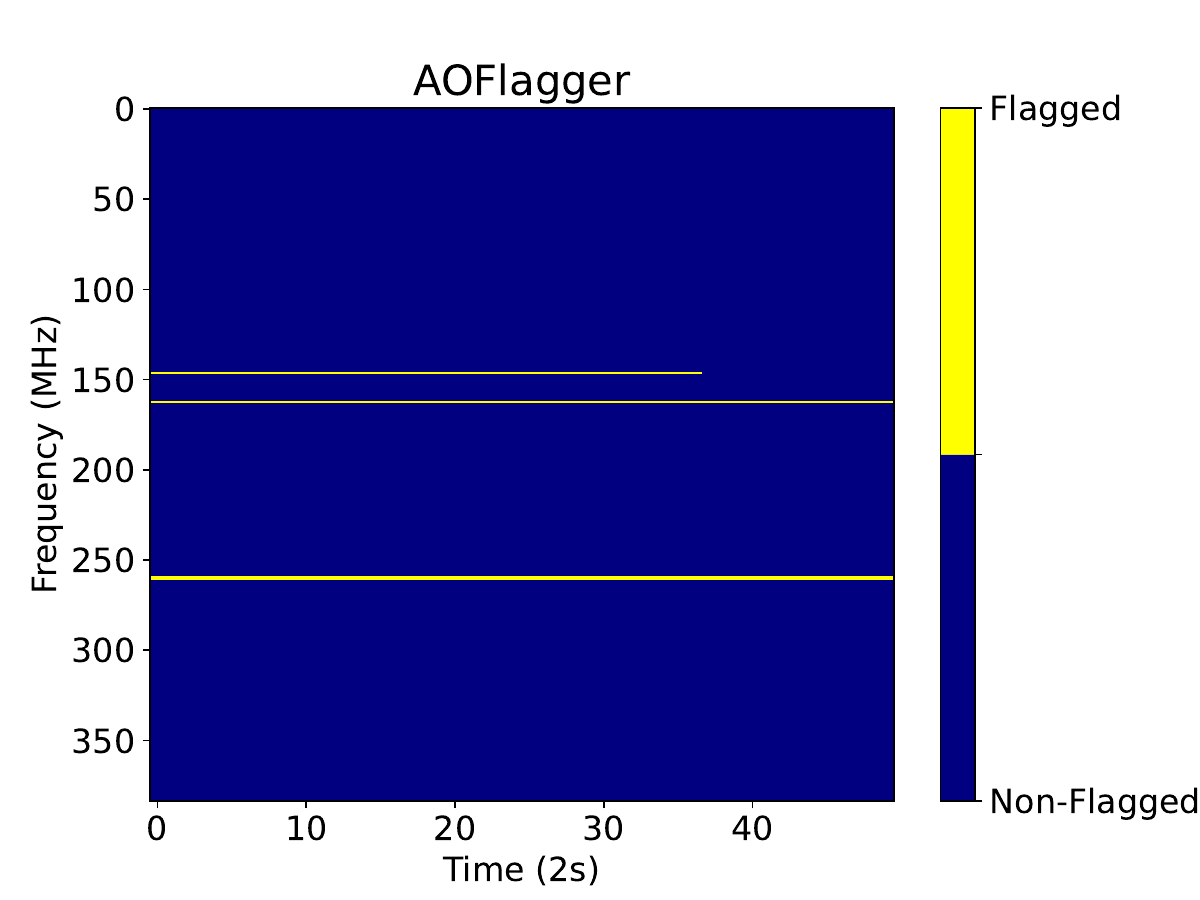}\\
    \includegraphics[scale=0.25]{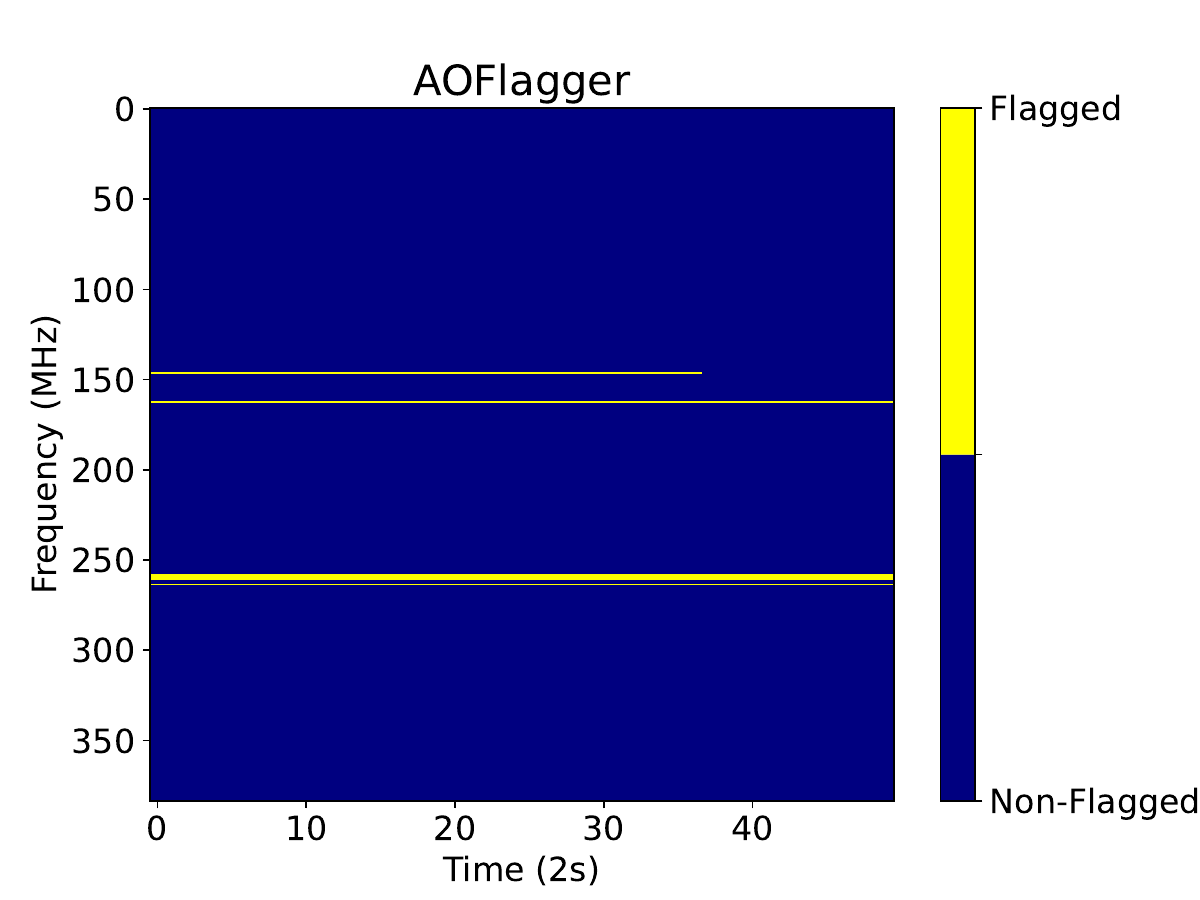}\includegraphics[scale=0.25]{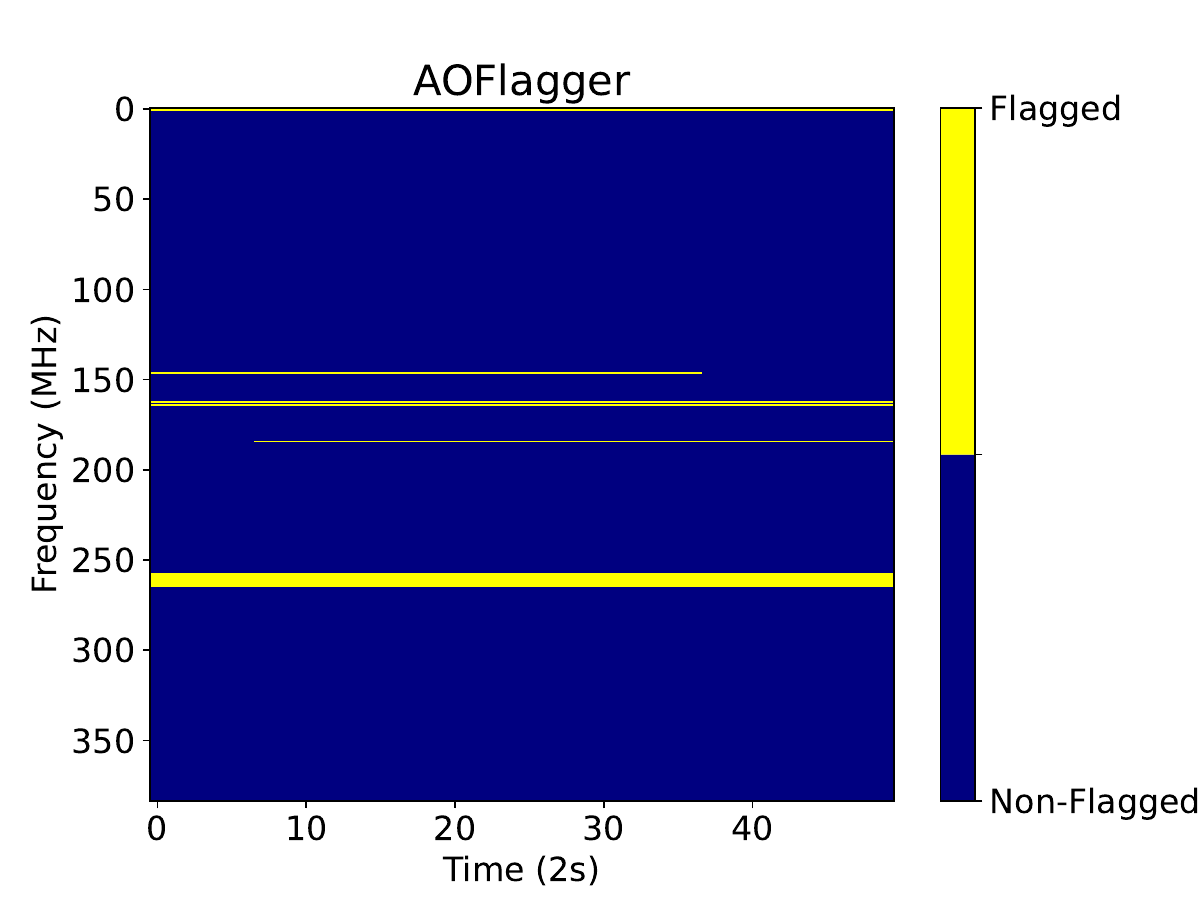}
\caption{\textsc{AOFlagger} different thresholds for RFI narrowband MWA data example. Top left: threshold 3.8, top right: threshold 3, bottom left: threshold 2.5, bottom right: threshold 2. Even relaxing the thresholds the faint RFI frequency channel fails to get completely flagged and noise start to show.}
    \label{fig:AOFlaggerNarrowbandThres}
\end{figure*}

\begin{figure*}[!t]
    \centering    \includegraphics[scale=0.25]{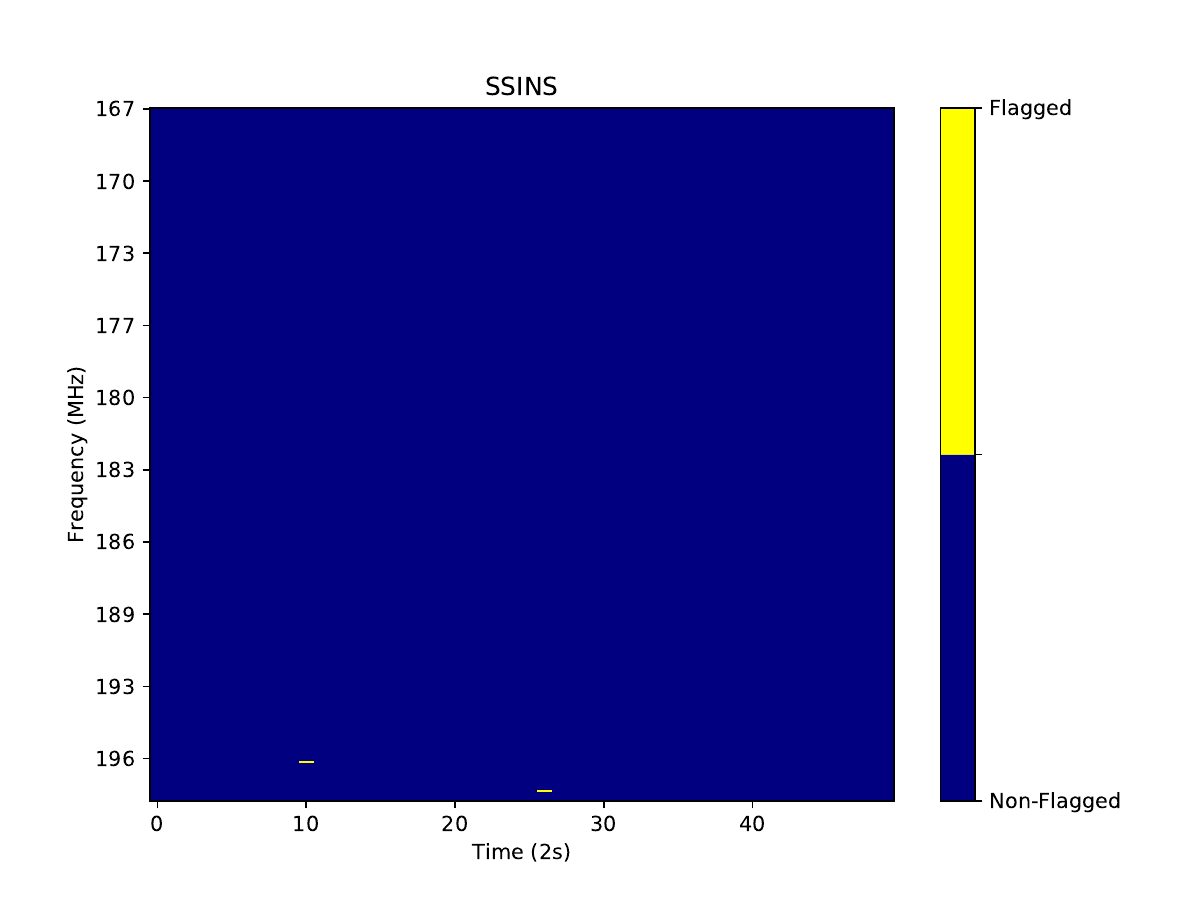}\includegraphics[scale=0.25]{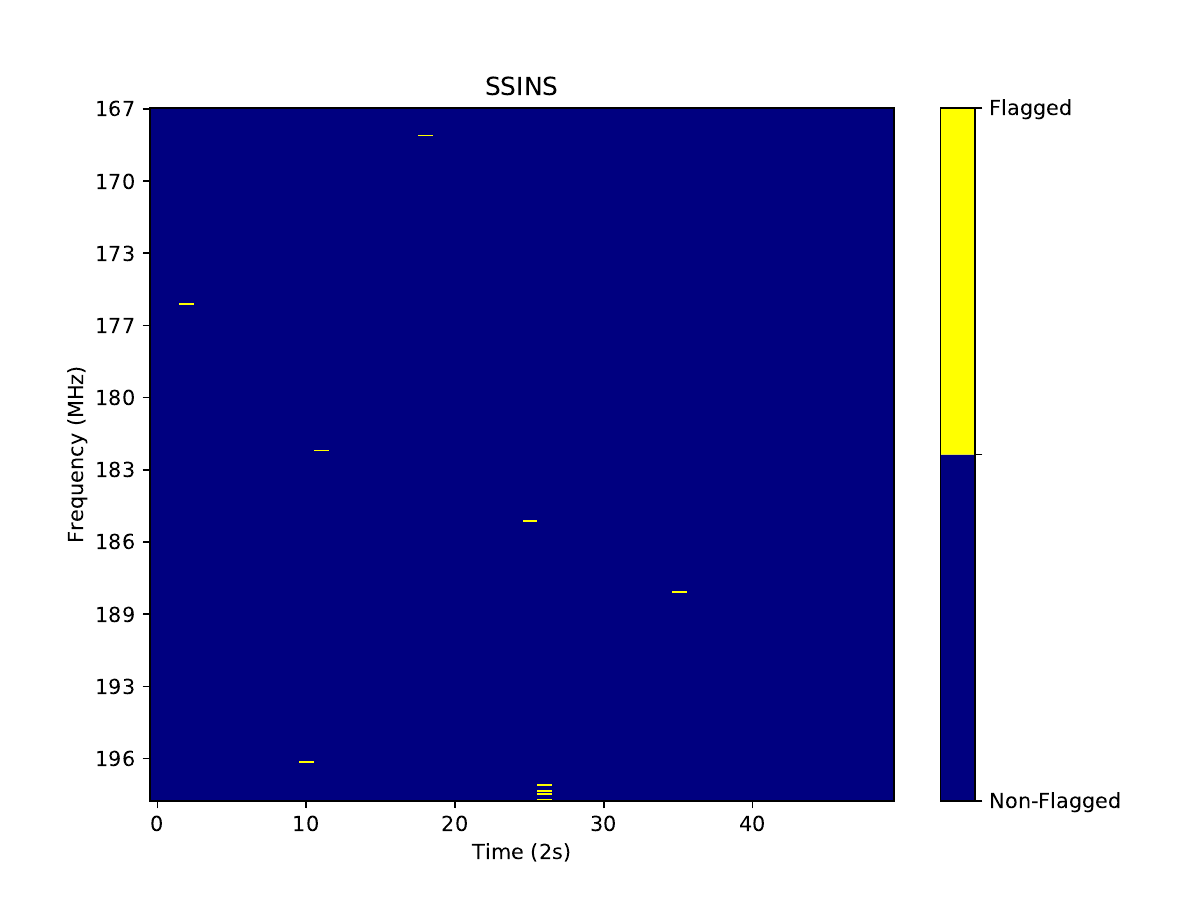}\\
    \includegraphics[scale=0.25]{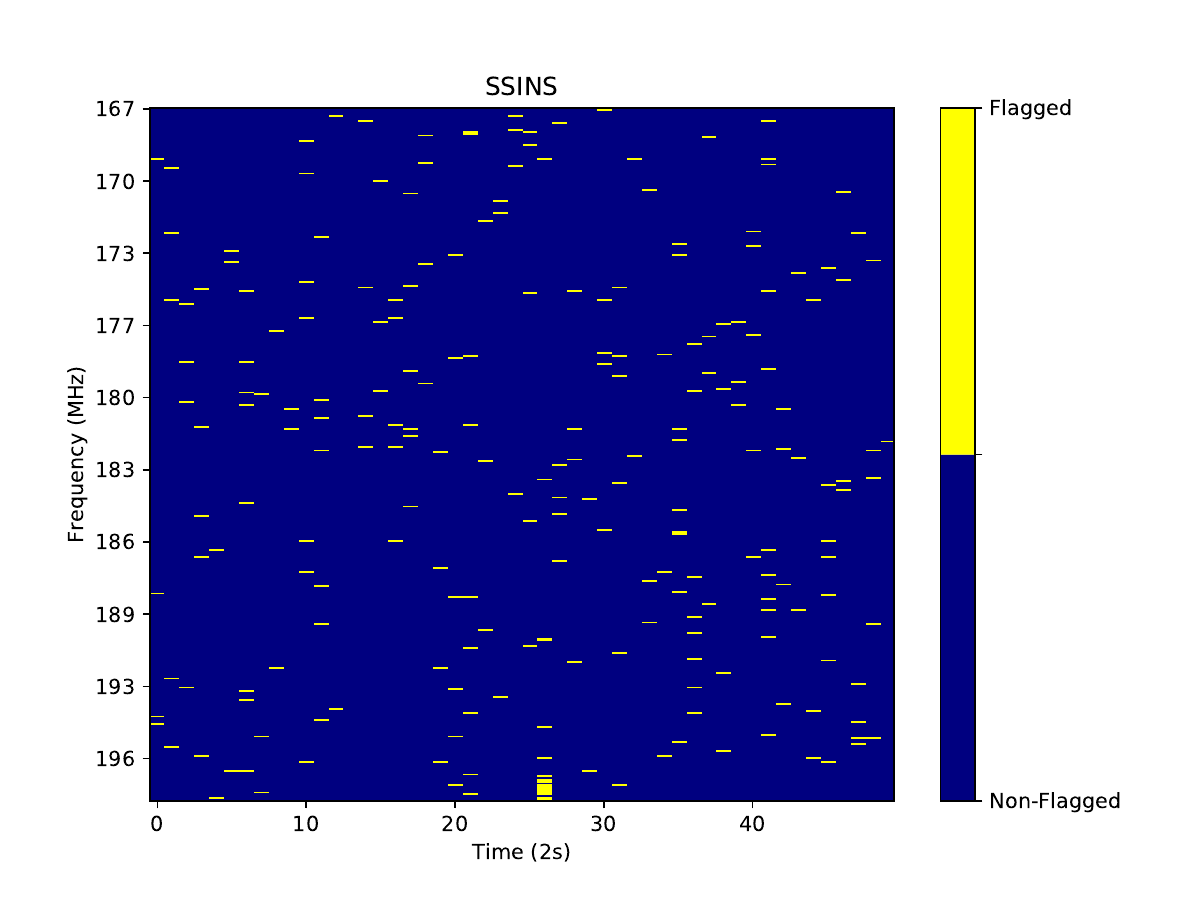}\includegraphics[scale=0.25]{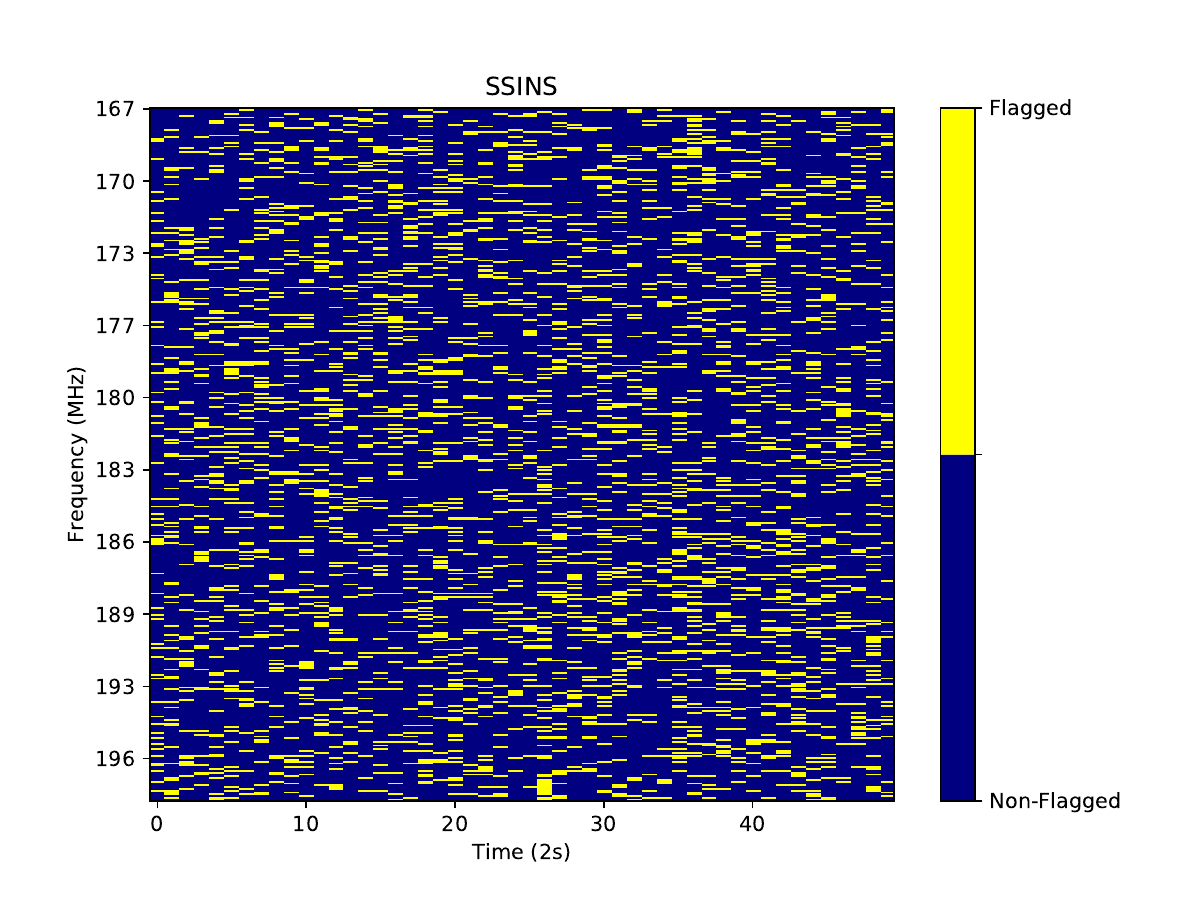}
    \label{fig:SSINSThresh}
    \caption{\textsc{ssins} different thresholds for RFI-faint contaminated dataset. Top left: threshold 5, top right: threshold 4, bottom left: threshold 3, bottom right: threshold 2.}
    \label{fig:SSINS2ObsBeforeThres}
\end{figure*}

In Figure~\ref{fig:SSINS2ObsBeforeThres}, we present various thresholds for SSINS using dataset ID 1061318736. This dataset includes faint RFI at 180.1 MHz that SSINS fails to detect. The Figure shows that even when the threshold is lowered, the contaminated channel remains undetected.

\begin{figure*}[!t]
\centering
\includegraphics[scale=0.25]{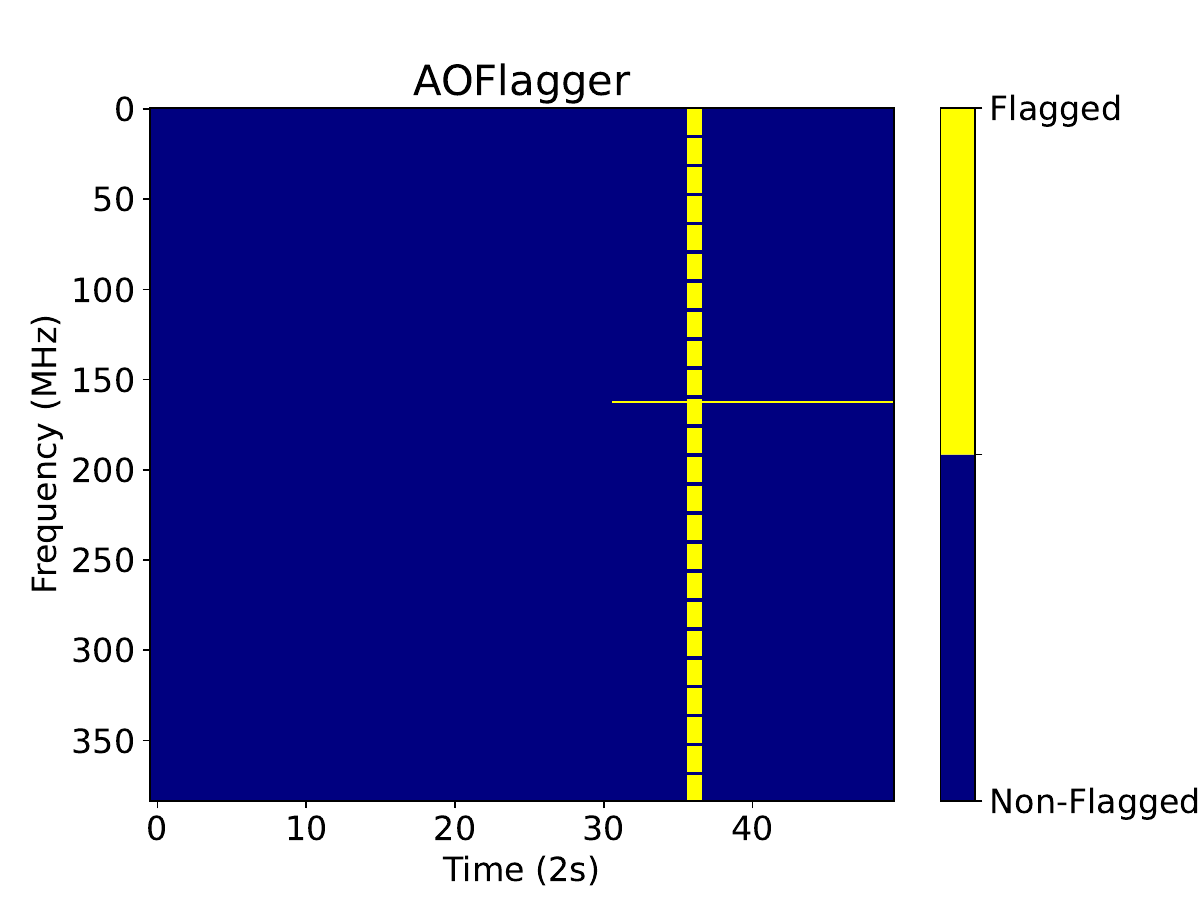}\includegraphics[scale=0.25]{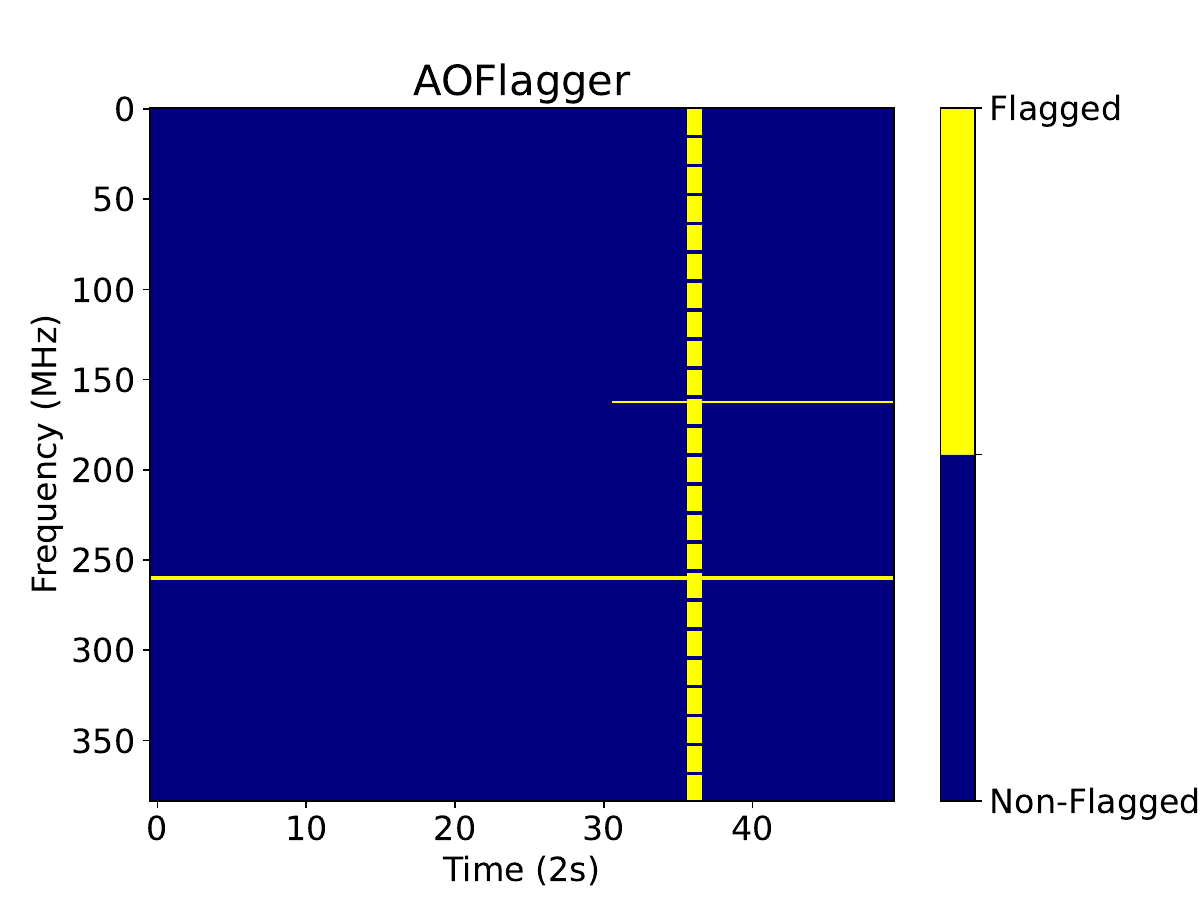}\\
\includegraphics[scale=0.25]{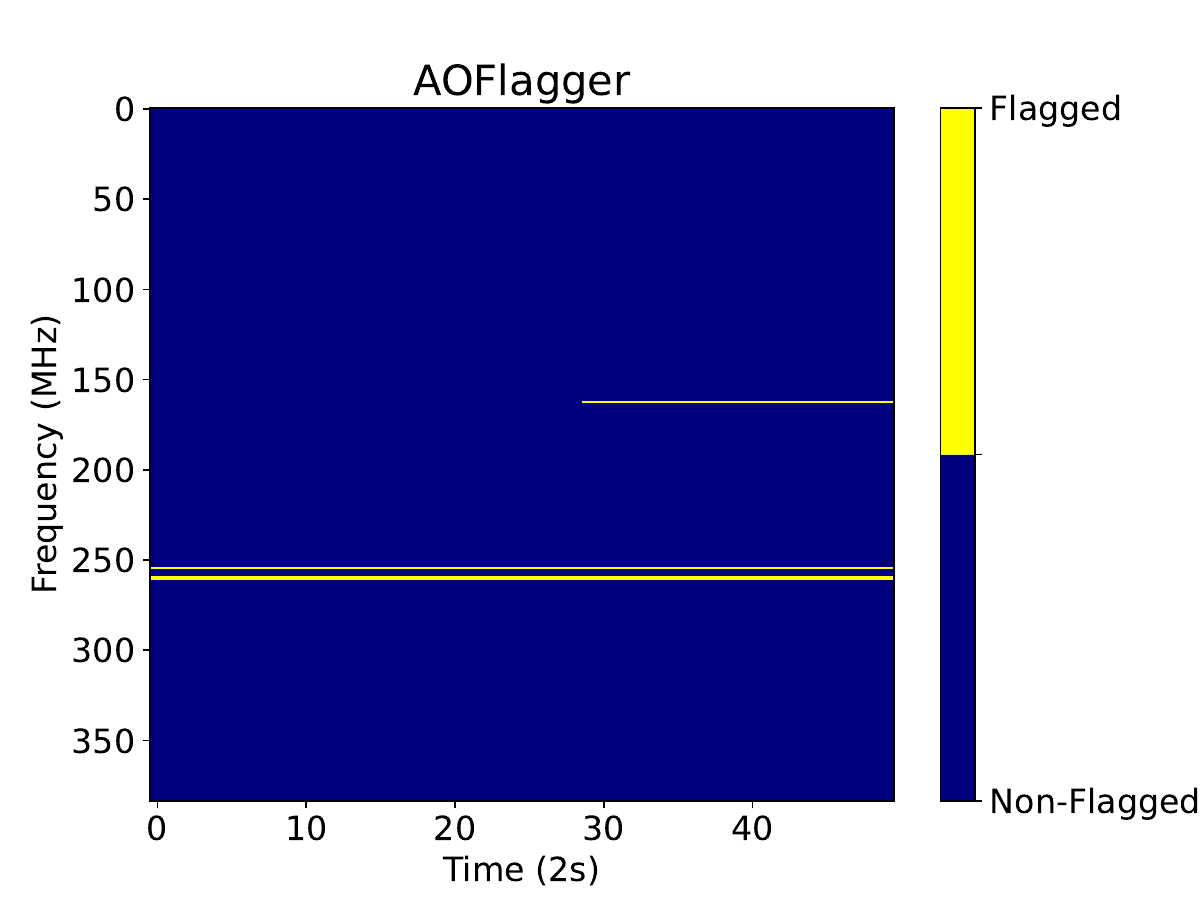}\includegraphics[scale=0.25]{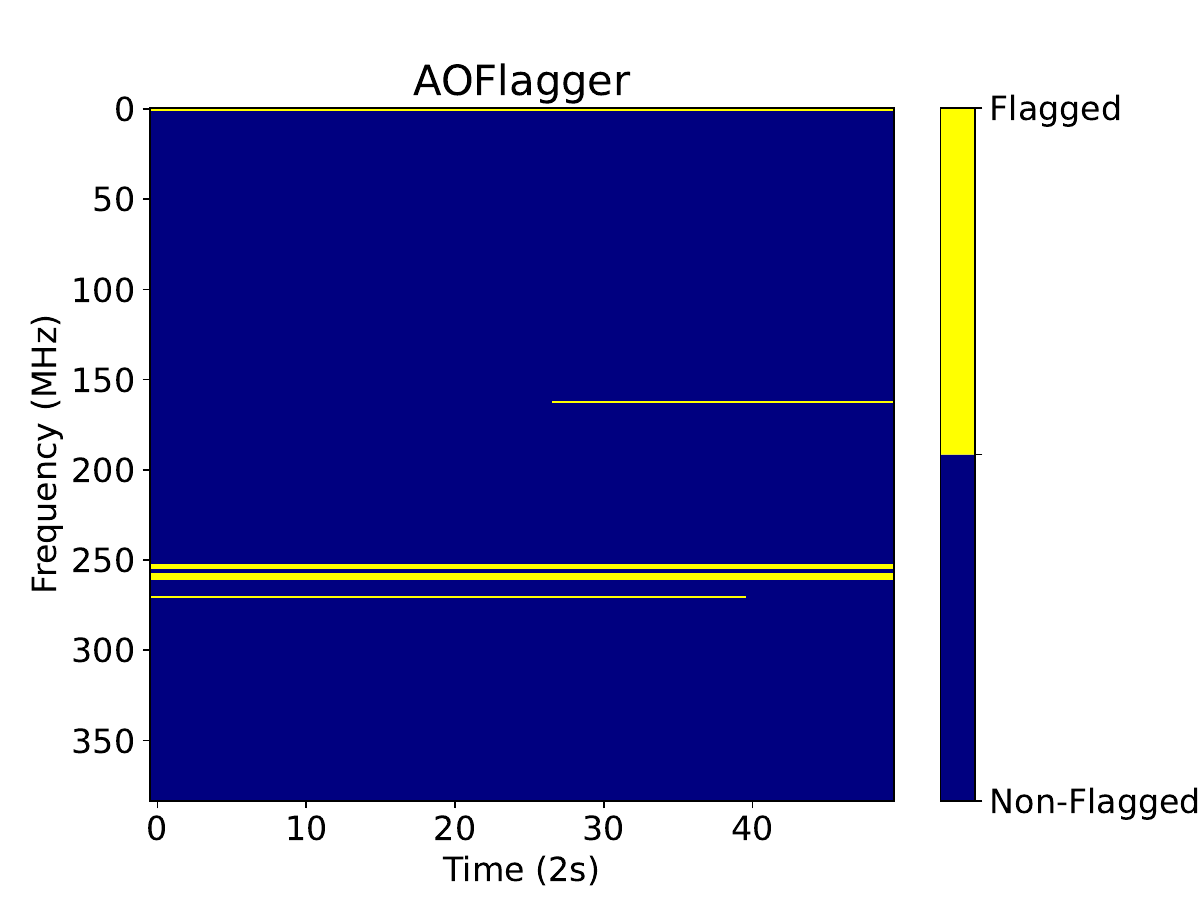}
\caption{\textsc{AOFlagger} different thresholds for RFI-faint contaminated dataset. Top left: threshold 3.8, top right: threshold 3, bottom left: threshold 2.5, bottom right: threshold 2. Even relaxing the thresholds the faint RFI frequency channel fails to get completely flagged and noise start to show.}
\label{fig:AOFlagger2ObsBeforeThresh}
\end{figure*}

\begin{figure*}[!t]
\centering
\includegraphics[scale=0.25]{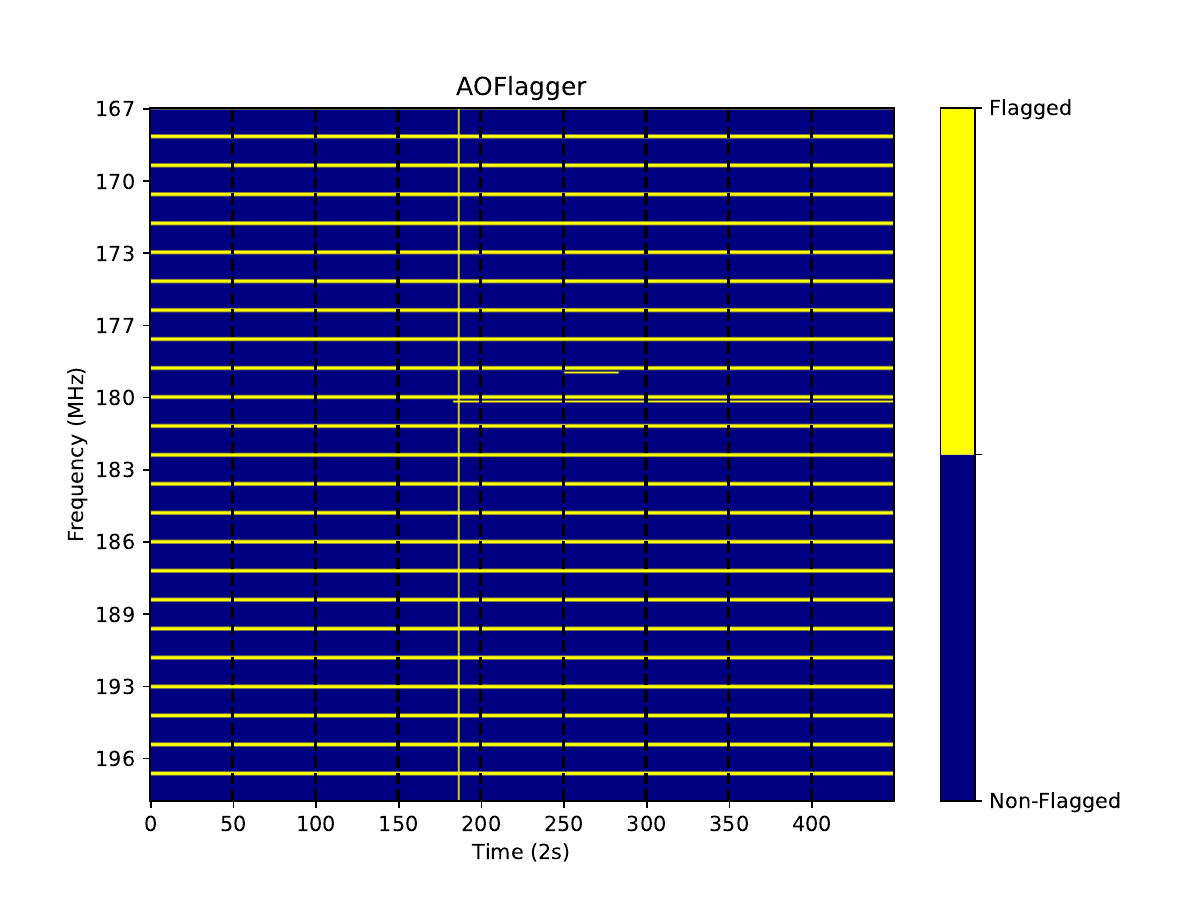}
    \includegraphics[scale=0.25]{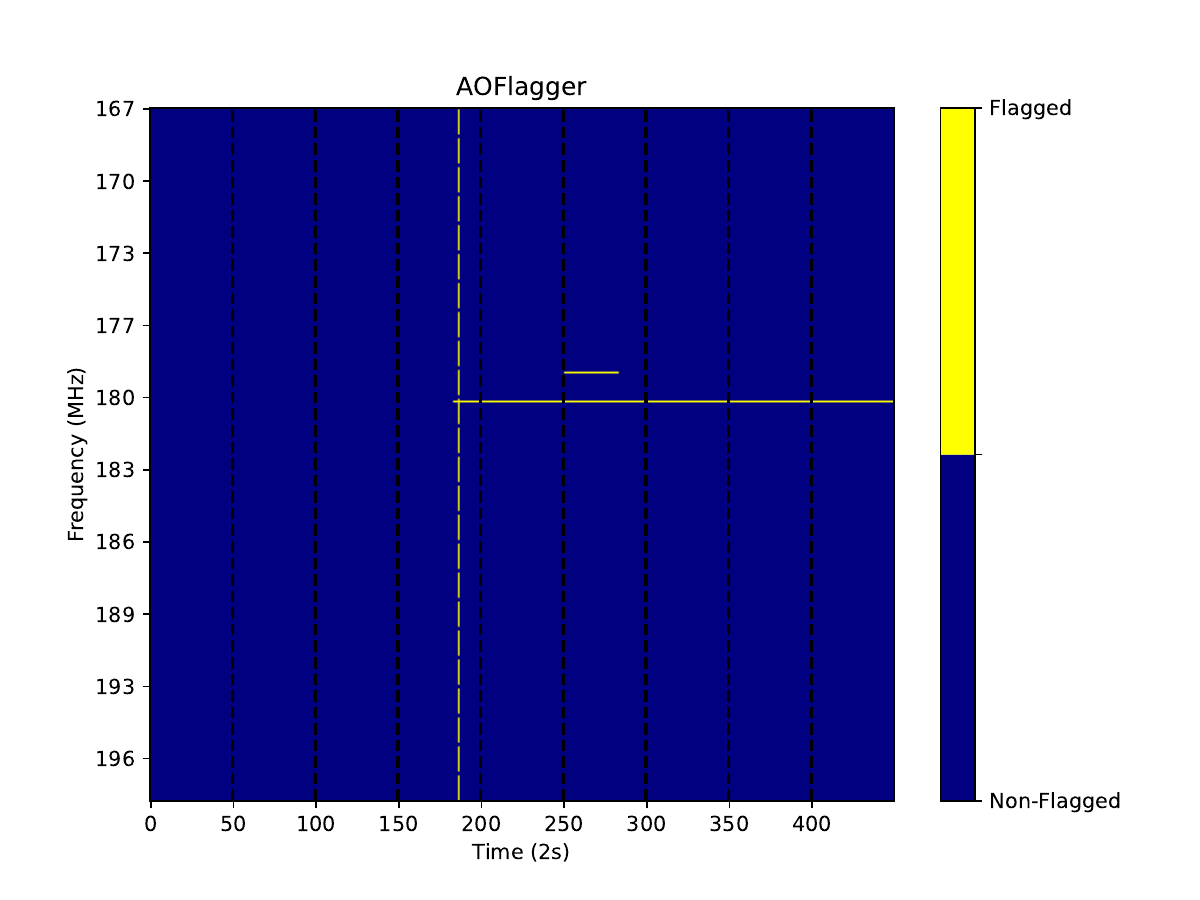}
\caption{\textsc{AOFlagger} results with coarse channels on the left and without on the right. The threshold was relaxed from 6.2 to 4, and already noise around time 190 is seen.}
\label{fig:AOFlaggerThresh9Datasets}
\end{figure*}

\section{Baseline-by-baseline}\label{sec:BaselineByBaseline}


\begin{figure*}[t!]
\centering
\includegraphics[scale=0.25]{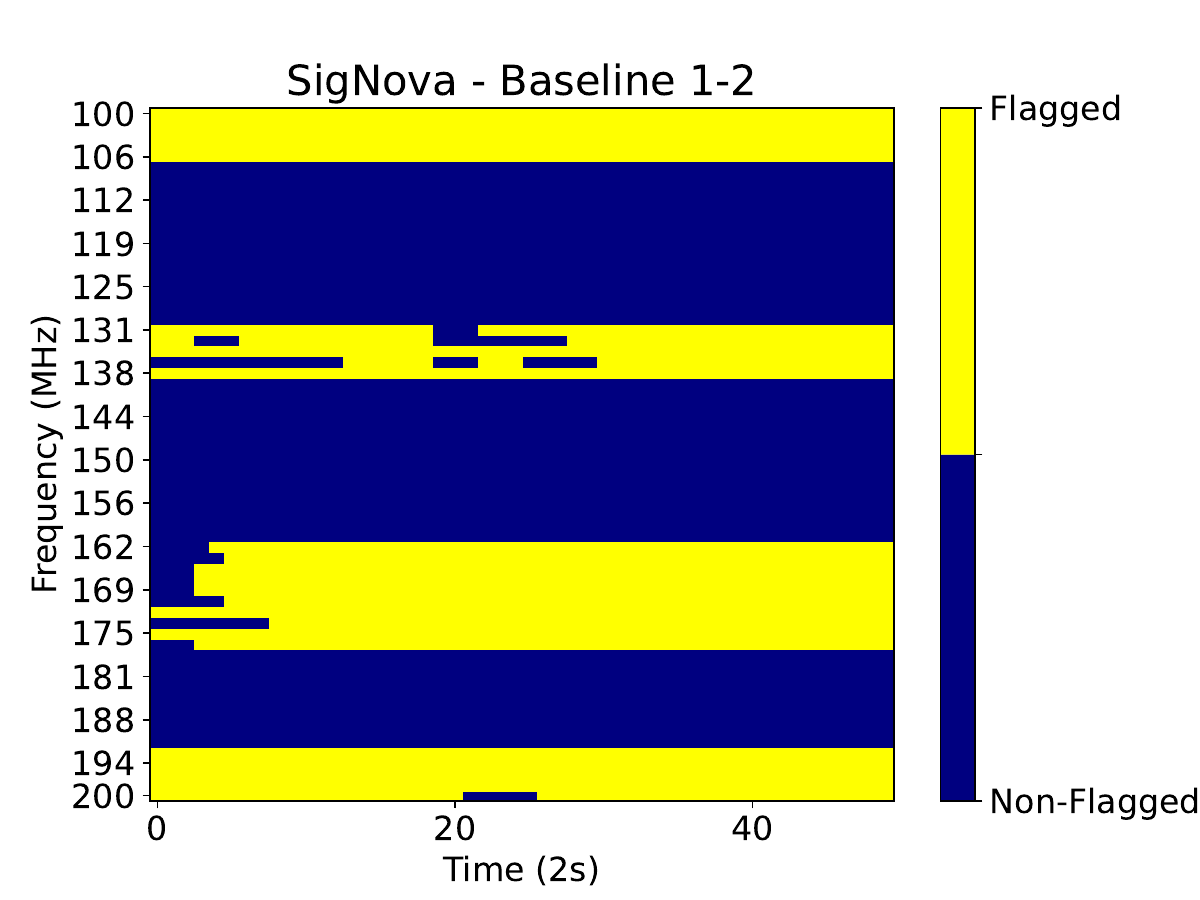}\includegraphics[scale=0.25]{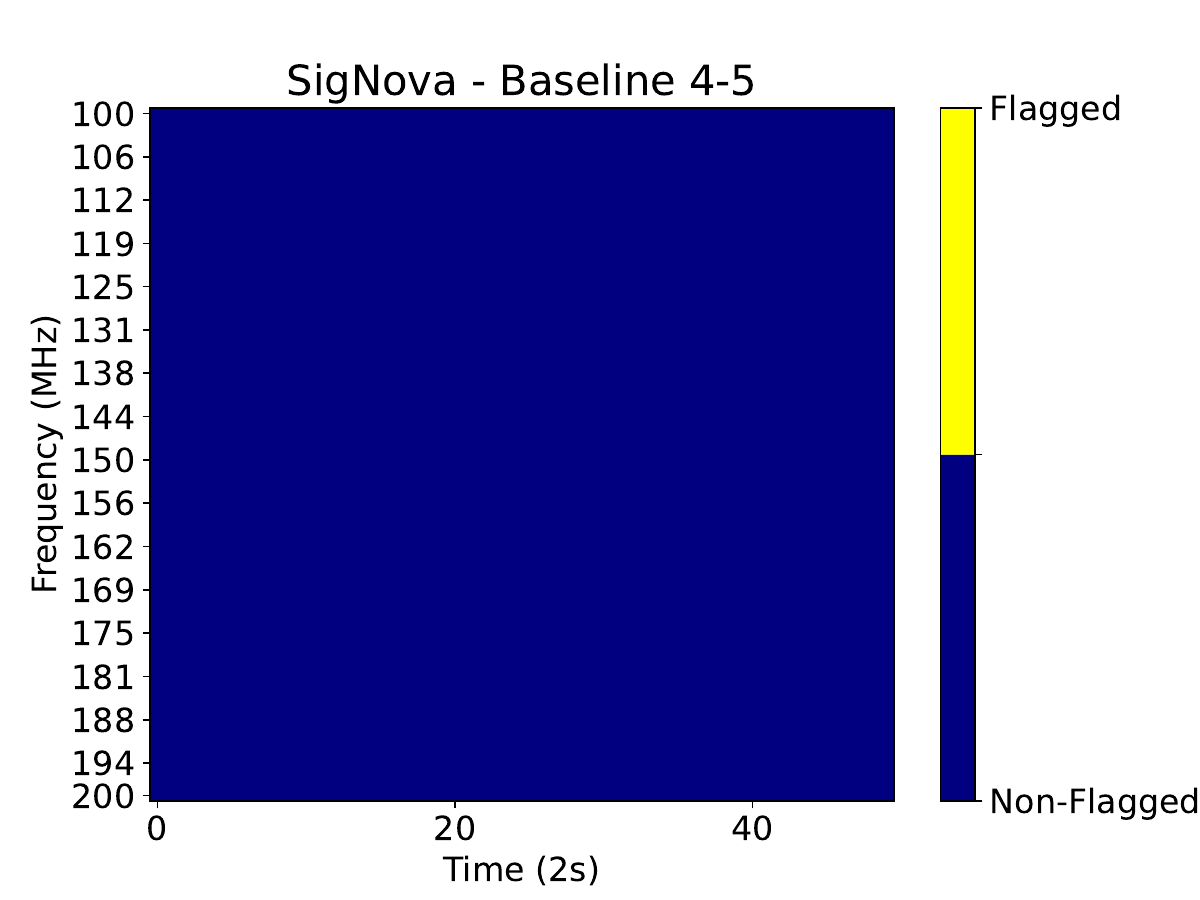}
\caption{The SigNova output for the Baseline-by-baseline approach is demonstrated using simulated data that only affects Antenna 1. The left plot displays the corrected flagged frequency channels, while the right plot shows the baseline corresponding to 4 and 5, which contains no RFI and is correctly identified.}
\label{fig:AOFlaggerThresh9Datasets2}
\end{figure*}

\end{document}